\documentclass[a4paper,fleqn]{cas-sc}

\usepackage[T1]{fontenc}
\usepackage[polutonikogreek,italian,english]{babel}
\usepackage[page,toc,titletoc,title]{appendix}
\usepackage[numbers,sort&compress]{natbib}
\usepackage{graphicx}
\usepackage{graphics}
\usepackage{braket}
\usepackage{bbold}
\usepackage{amssymb}
\usepackage{amsmath}
\usepackage{nicefrac}
\usepackage{dcolumn}
\usepackage{bm}
\usepackage{slashed}
\usepackage{datetime}
\usepackage{mciteplus}
\usepackage{multirow}
\usepackage{bbold}
\usepackage{color}
\usepackage{xcolor}
\usepackage{float}
\usepackage[utf8]{inputenc}
\usepackage[normalem]{ulem}
\usepackage{acro}
\usepackage{stfloats}
\usepackage{placeins}

\DeclareAcronym{AI}{
  short=AI,
  long=Artificial Intelligence,
}
\DeclareAcronym{BSM}{
  short=BSM,
  long=Beyond-the-Standard-Model,
}
\DeclareAcronym{QCD}{
  short=QCD,
  long=Quantum ChromoDynamics,
}
\DeclareAcronym{SM}{
  short=SM,
  long=Standard Model,
}
\DeclareAcronym{CP}{
  short=CP,
  long=Charge-Parity,
}
\DeclareAcronym{HF-NRevo}{
  short=HF-NRevo,
  long=Heavy-flavor NonRelativistic evolution,
}
\DeclareAcronym{HyF}{
  short=HyF,
  long=Hybrid collinear and high-energy Factorization,
}
\DeclareAcronym{SLAC}{
  short=SLAC,
  long=Stanford Linear Accelerator Center,
}
\DeclareAcronym{BNL}{
  short=BNL,
  long=Brookhaven National Laboratory,
}
\DeclareAcronym{FCNCs}{
  short=FCNCs,
  long=Flavor-Changing Neutral Currents,
}
\DeclareAcronym{GIM}{
  short=GIM,
  long=Glashow--Iliopoulos--Maiani,
}
\DeclareAcronym{CEM}{
  short=CEM,
  long= Color Evaporation Model,
}
\DeclareAcronym{CSM}{
  short=CSM,
  long= Color Singlet Mechanism,
}
\DeclareAcronym{CO}{
  short=CO,
  long=Color Octet,
}
\DeclareAcronym{NRQCD}{
  short=NRQCD,
  long=NonRelativistic QCD,
}
\DeclareAcronym{SDC}{
  short=SDC,
  long=Short-Distance Coefficient,
}
\DeclareAcronym{SDCs}{
  short=SDCs,
  long=Short-Distance Coefficients,
}
\DeclareAcronym{LDME}{
  short=LDME,
  long=Long-Distance Matrix-Element,
}
\DeclareAcronym{LDMEs}{
  short=LDMEs,
  long=Long-Distance Matrix Elements,
}
\DeclareAcronym{VSA}{
  short=VSA,
  long=Vacuum Saturation Approximation,
}
\DeclareAcronym{LO}{
  short=LO,
  long=Leading Order,
}
\DeclareAcronym{NLO}{
  short=NLO,
  long=Next-to-Leading-Order,
}
\DeclareAcronym{NNLO}{
  short=NNLO,
  long=Next-to-NLO,
}
\DeclareAcronym{MHOUs}{
  short=MHOUs,
  long=Missing Higher-Order Uncertainties,
}
\DeclareAcronym{F-MHOUs}{
  short=F-MHOUs,
  long=Fragmentation MHOUs,
}
\DeclareAcronym{H-MHOUs}{
  short=H-MHOUs,
  long=Hard-factor MHOUs,
}
\DeclareAcronym{DIS}{
  short=DIS,
  long=Deep Inelastic Scattering,
}
\DeclareAcronym{DGLAP}{
  short=DGLAP,
  long=Dokshitzer--Gribov--Lipatov--Altarelli--Parisi,
}
\DeclareAcronym{PDFs}{
  short=PDFs,
  long=Parton Distribution Functions,
}
\DeclareAcronym{FFs}{
  short=FFs,
  long=Fragmentation Functions,
}
\DeclareAcronym{MPI}{
  short=MPI,
  long=Multi-Parton-Interaction,
}
\DeclareAcronym{MPIs}{
  short=MPIs,
  long=Multi-Parton Interactions,
}
\DeclareAcronym{DPS}{
  short=DPS,
  long=Double-Parton Scattering,
}
\DeclareAcronym{SCET}{
  short=SCET,
  long=Soft and Collinear Effective Theory,
}
\DeclareAcronym{TM}{
  short=TM,
  long=Transverse-Momentum,
}
\DeclareAcronym{TMD}{
  short=TMD,
  long=Transverse-Momentum-Dependent,
}
\DeclareAcronym{FFNS}{
  short=FFNS,
  long=Fixed-Flavor Number Scheme,
}
\DeclareAcronym{VFNS}{
  short=VFNS,
  long=Variable-Flavor Number Scheme,
}
\DeclareAcronym{ZM-VFNS}{
  short=ZM-VFNS,
  long=Zero-Mass-Variable-Flavor Number Scheme,
}
\DeclareAcronym{GM-VFNS}{
  short=GM-VFNS,
  long=General-Mass Variable-Flavor Number Scheme,
}
\DeclareAcronym{HQSS}{
  short=HQSS,
  long=Heavy-Quark Spin Symmetry,
}
\DeclareAcronym{ABF}{
  short=ABF,
  long=Altarelli--Ball--Forte,
}
\DeclareAcronym{BFKL}{
  short=BFKL,
  long=Balitsky--Fadin--Kuraev--Lipatov,
}
\DeclareAcronym{LL}{
  short=LL,
  long=Leading-Logarithmic,
}
\DeclareAcronym{NLL}{
  short=NLL,
  long=Next-to-Leading-Logarithmic,
}
\DeclareAcronym{NNLL}{
  short=NNLL,
  long=Next-to-NLL,
}
\DeclareAcronym{LVM}{
  short=LVM,
  long=Light Vector Meson,
}
\DeclareAcronym{UGD}{
  short=UGD,
  long=Unintegrated Gluon Distribution,
}
\DeclareAcronym{LHC}{
  short=LHC,
  long=Large Hadron Collider,
}
\DeclareAcronym{HL-LHC}{
  short=HL-LHC,
  long=High-Luminosity Large Hadron Collider,
}
\DeclareAcronym{FCC}{
  short=FCC,
  long=Future Circular Collider,
}
\DeclareAcronym{EIC}{
  short=EIC,
  long=Electron-Ion Collider,
}
\DeclareAcronym{HFAG}{
  short=HFAG,
  long=Heavy Flavor Averaging Group,
}
\DeclareAcronym{SCA}{
  short=SCA,
  long=Small-Cone Algorithm
}
\DeclareAcronym{BLM}{
  short=BLM,
  long=Brodsky--Lepage--Mackenzie,
}
\DeclareAcronym{SNAJ}{
  short=SNAJ,
  long=Suzuki--Nejad--Amiri--Ji,
}
\DeclareAcronym{MSb}{
  short={$\boldsymbol{\overline{\rm MS}}$},
  long=Modified Minimal Subtraction,
}
\DeclareAcronym{MOM}{
  short=MOM,
  long=MOMentum,
}
\DeclareAcronym{HELL}{
  short={HELL},
  long=High Energy Large Logarithms,
}

\def\tsc#1{\csdef{#1}{\textsc{\lowercase{#1}}\xspace}}
\tsc{WGM}
\tsc{QE}

\newcommand{\drv}{{\rm d}}

\newcommand{\LQCD}{\Lambda_{\rm QCD}}
\newcommand{\MSb}{\overline{\rm MS}}

\newcommand{\NLO}{{\rm NLO}}

\newcommand{\LL}{{\rm LL/LO}}
\newcommand{\NLL}{{\rm NLL/NLO}}
\newcommand{\NLLp}{{\rm NLL/NLO^+}}

\newcommand{\HENLOp}{{\rm HE}\mbox{-}{\rm NLO^+}}

\newcommand{\CnLL}{{\cal C}_n^\LL}

\newcommand{\CnNLLp}{{\cal C}_n^\NLLp}

\newcommand{\CnHENLOp}{{\cal C}_n^{{\rm HE}\text{-}{\rm NLO}^+}}
\newcommand{\DY}{\Delta Y}

\newcommand{\R}{{\cal R}}

\newcommand{\Jpsi}{J/\psi}
\newcommand{\Yps}{\Upsilon}
\newcommand{\BCs}{B_c(^1S_0)}
\newcommand{\Bss}{B_c(^3S_1)}

\newcommand{\QXQq}{X_{Q\bar{Q}q\bar{q}}}

\newcommand{\TQQ}{T_{4Q}}
\newcommand{\TQc}{T_{4c}}
\newcommand{\TQcZpp}{T_{4c}(0^{++})}
\newcommand{\TQcOpm}{T_{4c}(1^{+-})}
\newcommand{\TQcTpp}{T_{4c}(2^{++})}
\newcommand{\TQb}{T_{4b}}
\newcommand{\TQbZpp}{T_{4b}(0^{++})}
\newcommand{\TQbOpm}{T_{4b}(1^{+-})}
\newcommand{\TQbTpp}{T_{4b}(2^{++})}

\newcommand{{\HFNRevo}}{\textsc{HF-NRevo}}

\newcommand{{\Jethad}}{\textsc{Jethad}}
\newcommand{{\symJethad}}{\textsc{symJethad}}
\newcommand{{\psymJethad}}{\textsc{(sym)Jethad}}
\newcommand{{\Hell}}{\textsc{Hell}}
\newcommand{{\RadISH}}{\textsc{RadISH}}
\newcommand{{\Pegasus}}{\textsc{QCD-PEGASUS}}
\newcommand{{\HOPPET}}{\textsc{HOPPET}}
\newcommand{{\QCDNUM}}{\textsc{QCDNUM}}
\newcommand{{\APFEL}}{\textsc{APFEL}}
\newcommand{{\APFELpp}}{\textsc{APFEL++}}
\newcommand{{\APFELppp}}{\textsc{APFEL(++)}}
\newcommand{{\EKO}}{\textsc{EKO}}
\newcommand{{\FeynCalc}}{\textsc{FeynCalc}}

\newcommand{{\HCFF}}{{\tt HCFF1.0}}
\newcommand{{\NRFF}}{{\tt NRFF1.0}}

\begin{document}
\let\WriteBookmarks\relax
\def\floatpagepagefraction{1}
\def\textpagefraction{.001}

\shorttitle{Tetraquark-Jet Systems at the High-Luminosity LHC}    

\shortauthors{Celiberto, Francesco Giovanni}  

\title []{\Huge Tetraquark-Jet Systems at the High-Luminosity LHC}  

\author[1]{Francesco Giovanni Celiberto}[orcid=0000-0003-3299-2203]

\cormark[1]


\ead{francesco.celiberto@uah.es}


\affiliation[1]{organization={Universidad de Alcal\'a (UAH), Departamento de F\'isica y Matem\'aticas},
            addressline={Campus Universitario}, 
            city={Alcal\'a de Henares},
            postcode={E-28805}, 
            state={Madrid},
            country={Spain}}




\begin{abstract}
We investigate the high-energy production of tetraquark-jet systems at the LHC and its forthcoming High-Luminosity upgrade.
In this review, we examine the leading-power fragmentation of fully heavy tetraquarks ($T_{4Q}$) in hadronic collisions, highlighting their relevance as novel probes of multiquark dynamics in QCD.
Our analysis relies on the hadron-structure-oriented {\tt TQ4Q1.1} fragmentation functions, built within a nonrelativistic QCD framework that incorporates both gluon- and heavy-quark-initiated channels.
Threshold-consistent DGLAP evolution is performed through the {\HFNRevo} scheme, enabling a unified treatment of mass thresholds and scale variations.
We also provide a systematic discussion of uncertainties arising from color-composite long-distance matrix elements (LDMEs) and from perturbative hard- and fragmentation-scale inputs (H- and F-MHOUs).
Phenomenological predictions are obtained using the \textsc{(sym)Jethad} framework at NLL/NLO$^+$ accuracy for semi-inclusive tetraquark-jet production at the LHC and beyond.
This review connects the emerging spectroscopy of fully heavy exotics with modern fragmentation-based approaches to hadron structure and high-energy QCD.
\end{abstract}



\begin{keywords}
 {\tt TQ4Q1.1} FF update \sep
 Hadronic structure \sep
 Precision QCD \sep
 Exotic matter \sep
 Tetraquarks \sep
 Heavy flavor \sep
 Fragmentation \sep
 Resummation \sep
 HL-LHC \sep
\end{keywords}

\maketitle

\newcounter{appcnt}


\tableofcontents
\clearpage

\setlength{\parskip}{3pt}%

\section{Introduction}
\label{sec:introduction}

Understanding how heavy quarks are produced and hadronize in high-energy collisions is essential for uncovering the inner workings of Quantum Chromodynamics (QCD).  
The study of heavy-flavored final states provides a dual benefit: it probes the dynamics of the strong interaction at short distances while offering sensitive channels to potential \ac{BSM} phenomena involving heavy-quark couplings.  
Because of their large masses, heavy quarks enable perturbative \ac{QCD} calculations with high precision, making them ideal tools for testing the Standard Model at current and future colliders, including the \ac{HL-LHC}~\cite{Apollinari:2015wtw,Apollinari:2015bam,Apollinari:2017lan,Chapon:2020heu}, the \ac{EIC}~\cite{AbdulKhalek:2021gbh,Khalek:2022bzd,Hentschinski:2022xnd,Amoroso:2022eow,Abir:2023fpo,Allaire:2023fgp}, and the \ac{FCC}~\cite{FCC:2018byv,FCC:2018evy,FCC:2018vvp,FCC:2018bvk,FCC:2025lpp,FCC:2025uan,FCC:2025jtd}.

\ac{QCD}, the non-Abelian gauge theory of the strong interaction, is a central component of the \ac{SM}.  
Formulated on the $SU(N_c)$ group with $N_c = 3$, it describes quark fermionic fields in the fundamental triplet representation, and gluons---the massless vector bosons mediating color interactions in the adjoint octet~\cite{Gell-Mann:1962yej,Gell-Mann:1964ewy,Zweig:1964jf,Fritzsch:1973pi}.  
Beyond its established framework within the SM, QCD also offers a natural testing ground for \ac{BSM} extensions, including axion models~\cite{Peccei:1977hh,Peccei:1977ur,Peccei:2006as,Duffy:2009ig}, dark non-Abelian forces~\cite{Forestell:2017wov,Huang:2020crf}, quarkyonic matter~\cite{McLerran:2007qj,Hidaka:2008yy,McLerran:2018hbz}, and higher-dimensional QCD operators~\cite{Buchmuller:1985jz,Witten:1979kh,Dudek:2010wm,Afonin:2019unu}.  
These ideas broaden the scope of strong-interaction studies, linking QCD phenomenology to the search for new fundamental physics.

A particularly rich subfield of QCD concerns hadrons that contain two or more heavy quarks.  
Among them, heavy quarkonia---bound $Q\bar Q$ mesons---played a defining role during the ``November Revolution'' of 1974, when the $\Jpsi$ meson was independently discovered at SLAC~\cite{SLAC-SP-017:1974ind} and BNL~\cite{E598:1974sol}, and later confirmed at Frascati~\cite{Bacci:1974za}.  
These discoveries provided direct evidence of quarks as physical degrees of freedom and offered key insights into confinement dynamics.

While quarkonia belong to the class of conventional hadrons, the color-neutrality of QCD allows for the formation of more intricate bound systems, collectively known as exotic hadrons.  
Such states possess quantum numbers that cannot be accommodated within the traditional $q\bar q$ or $qqq$ structures.  
Exotics can be grouped into two broad categories: states with explicit gluonic content (hybrids and glueballs)~\cite{Close:1991pf,Close:1997qda,Close:1998zz,Minkowski:1998mf,Close:2000yg,Mathieu:2008me,Hsiao:2013dta,D0:2020tig,Csorgo:2019ewn}, and multiquark configurations such as tetraquarks and pentaquarks~\cite{Gell-Mann:1964ewy,Jaffe:1976ig,Jaffe:1976ih,Ader:1981db}.  

The discovery of the $\Jpsi$ in November 1974~\cite{SLAC-SP-017:1974ind,E598:1974sol,Bacci:1974za} triggered the ``First Quarkonium Revolution,'' establishing heavy quarkonia as clean laboratories of QCD. 
Three decades later, the observation of the $X(3872)$ by Belle in 2003~\cite{Belle:2003nnu}, subsequently confirmed by several experiments~\cite{CDF:2003cab,LHCb:2013kgk,CMS:2021znk,Swanson:2006st}, inaugurated the so-called ``Second Quarkonium Revolution,'' marking the beginning of modern exotic-hadron spectroscopy.
More recently, the LHCb discovery of the $X(2900)$~\cite{LHCb:2020bls}, the first candidate with open-charm flavor, has consolidated what is now often referred to as the broader ``Exotic-matter Revolution,'' extending the field well beyond conventional quark-antiquark dynamics.

Recent progress in QCD factorization and all-order resummation techniques has substantially extended the predictive power of precision calculations for exotic-hadron observables~\cite{Feng:2020riv,Feng:2020qee,Feng:2023agq,Feng:2023ghc,Bai:2024ezn,Bai:2024flh,Nejad:2021mmp,Celiberto:2023rzw,Celiberto:2024mab,Celiberto:2024beg,Celiberto:2025dfe,Celiberto:2025ipt}.  
Accurate predictions for cross sections and kinematic distributions, systematically validated against collider data, now enable detailed investigations of the interplay between perturbative dynamics and nonperturbative hadronization mechanisms.

A coordinated research program that connects exotic spectroscopy with modern QCD methods can illuminate the effective degrees of freedom governing strongly coupled systems and clarify the organizing principles behind multiquark dynamics.  
Such an effort promises deeper insight into the nonperturbative regime of QCD and the processes that shape the formation of bound hadronic matter.

Exotic hadrons are broadly categorized into two classes: gluon-rich configurations, including quark-gluon hybrids~\cite{Kou:2005gt,Braaten:2013boa,Berwein:2015vca} and glueballs~\cite{Minkowski:1998mf,Mathieu:2008me,Chen:2021cjr,Csorgo:2019ewn,D0:2020tig}, and compact multiquark states such as tetraquarks, pentaquarks, and hexaquarks~\cite{Gell-Mann:1964ewy,Jaffe:1976ig,Jaffe:1976ih,Jaffe:1976yi,Ader:1981db,Rosner:1985yh,Pepin:1998ih,Vijande:2011im,Esposito:2016noz,Lebed:2016hpi,Guo:2017jvc,Lucha:2017mof,Ali:2019roi}.  
Although the former include explicit gluonic degrees of freedom, the latter are described as leading Fock-state combinations of four or more quarks.

Identified as a hidden-charm resonance with a $c\bar{c}$ core, the $X(3872)$ is widely regarded as the first clear candidate for a tetraquark state~\cite{Chen:2016qju,Liu:2019zoy}.
This milestone was later complemented by the LHCb observation of the $X(2900)$~\cite{LHCb:2020bls}, the first exotic structure exhibiting open-charm flavor.

Although the $X(3872)$ shares the quantum numbers of conventional quarkonia, its pronounced isospin-violating decays reveal a more complex internal structure, motivating various models within the tetraquark paradigm.  
Among these, the \textbf{compact diquark picture} describes the state as a tightly bound diquark-antidiquark system~\cite{Maiani:2004vq,tHooft:2008rus,Maiani:2013nmn,Maiani:2014aja,Maiani:2017kyi,Mutuk:2021hmi,Wang:2013vex,Wang:2013exa,Grinstein:2024rcu}.  
Alternatively, the \textbf{molecular interpretation} envisions a weakly bound meson-meson pair~\cite{Tornqvist:1993ng,Braaten:2003he,Guo:2013sya,Mutuk:2022ckn,Wang:2013daa,Wang:2014gwa,Esposito:2023mxw,Grinstein:2024rcu}, while the \textbf{hadroquarkonium scenario} posits a compact quarkonium core surrounded by light mesonic degrees of freedom~\cite{Dubynskiy:2008mq,Voloshin:2013dpa,Guo:2017jvc,Ferretti:2018ojb,Ferretti:2018tco,Ferretti:2020ewe}.  
Each approach captures different aspects of the $X(3872)$ dynamics and provides a framework for interpreting other exotic states with similar phenomenology.  
Additional information may emerge from heavy-ion and high-multiplicity proton collisions~\cite{Esposito:2020ywk} or from studies of its thermal evolution in hadronic media~\cite{Armesto:2024zad}.

The landscape evolved in 2021 with the LHCb discovery of the doubly charmed $T_{cc}^+$~\cite{LHCb:2021vvq,LHCb:2021auc}, interpreted as a near-threshold $|DD\rangle$ molecule and analyzed within the XEFT framework~\cite{Fleming:2021wmk,Dai:2023mxm,Hodges:2024awq,Fleming:2007rp,Fleming:2008yn,Braaten:2010mg,Fleming:2011xa,Mehen:2015efa,Braaten:2020iye}.  
Soon after, LHCb reported a broad resonance in the double $\Jpsi$ spectrum~\cite{LHCb:2020bwg}, the $X(6900)$, interpreted as either the scalar ($0^{++}$) or tensor ($2^{++}$) excitation of a fully charmed $\TQc$ tetraquark~\cite{Chen:2022asf}.  
This signal, later discussed in the LHCb review of exotic spectroscopy~\cite{Nogga:2025qcm}, has sparked renewed theoretical interest. 
Proposals range from Pomeron-driven and coupled-channel interpretations~\cite{Gong:2020bmg} to models based on Bethe--Salpeter and Regge analyses, where $X(6900)$ may correspond to a compact $|c\bar c c\bar c\rangle$ resonance accompanied by other fully heavy excitations.  
Recent suggestions also highlight photon-photon fusion in ultraperipheral collisions as a promising channel to determine the spin-parity structure of $\TQc$ states~\cite{Niu:2022cug}.

The CMS Collaboration has recently sharpened our picture of fully charmed tetraquarks by delivering the first spin-parity determination in the $\Jpsi$ and $(\Jpsi,\psi(2S))$ channels.  
An earlier report of three resonant structures in the $4m_c < M < 7\,{\rm GeV}$ region~\cite{CMS:2023owd} (see also~\cite{Zhu:2024swp}) has now been followed by an angular analysis~\cite{CMS:2025fpt} that favors a $J^{PC}=2^{++}$ assignment for the leading peak.  
This preference challenges loosely bound molecular scenarios and supports a compact diquark-antidiquark interpretation in which spin-1 diquarks naturally populate $J=2$ configurations.  
The same spin constraints need not apply to mixed-flavor exotics such as $X(3872)$ or $Z_c(3900)$, which remain compatible with lower spin and molecular descriptions.  
CMS thus provides crucial input for modeling the internal dynamics of fully heavy tetraquarks.

Complementary evidence on multi-parton dynamics comes from quarkonium-pair measurements.  
ATLAS extracted \ac{DPS} effective cross sections from prompt $\Jpsi$ pairs at $\sqrt{s}=8$~TeV~\cite{ATLAS:2016ydt}, and CMS measured total and differential double $\Jpsi$ cross sections at $\sqrt{s}=7$~TeV~\cite{CMS:2014cmt}.  
More recently, CMS observed double $\Jpsi$ production in $p$Pb at $8.16$~TeV and separated single- and double-parton contributions across phase space~\cite{CMS:2024wgu}.  
These results constrain \ac{MPI} effects that are directly relevant to multiquark production mechanisms.

From a theory standpoint, doubly and fully heavy states, $\QXQq$ and $\TQQ$, are clean laboratories for the strong force.  
In $\QXQq$ systems, nonrelativistic heavy quarks couple to light degrees of freedom, often through diquarklike substructures, providing a mixed-mass test of QCD.  
In contrast, $\TQQ$ hadrons are composed solely of heavy constituents in a $|Q\bar{Q}Q\bar{Q}\rangle$ configuration, with no valence light quarks or active gluon fields, and thus behave as exotic, doubled quarkonia. 
For fully stranged analogs of $\TQQ$, see Refs.~\cite{Ding:2006ya,Ho:2019org,Su:2022eun,Liu:2020lpw,Dong:2022otb,Xi:2023byo,Ma:2024vsi}.  
Given their nonrelativistic dynamics and Fock-space structure, these exotics can be analyzed using quarkonium-inspired tools: charmonia are often likened to hydrogenlike bound states~\cite{Pineda:2011dg}, whereas $\QXQq$ and $\TQQ$ may be viewed as QCD nuclei or molecules, depending on the modeling~\cite{Maiani:2019cwl}.

Despite major advances on spectra and decays since the $X(3872)$ discovery, production mechanisms remain poorly constrained. 
Only a few model-dependent approaches exist so far, based on color evaporation~\cite{Maciula:2020wri} or hadron-quark duality~\cite{Berezhnoy:2011xy,Karliner:2016zzc,Becchi:2020mjz}.  
Complementary lines of inquiry address the role of MPI in tetraquark production~\cite{Carvalho:2015nqf,Abreu:2023wwg} and potential high-energy signatures in exotic formation~\cite{Cisek:2022uqx}, alongside studies of exclusive radiative decays at $B$ factories~\cite{Feng:2020qee} and photoproduction in lepton-hadron scattering~\cite{Feng:2023ghc}.
A recent higher-order study of fully charmed tetraquark production, including gluon-radiation effects, was presented in Ref.~\cite{Wang:2025hex}.
Machine-learning strategies for multiquark binding have also been proposed~\cite{Wu:2025wvv}.

The bottom sector remains less explored.  
BELLE reported charged bottomoniumlike structures in $\Yps(5S)$ decays~\cite{Belle:2011aa}, hinting at exotic contributions, but no $|b\bar{b}b\bar{b}\rangle$ or $|b\bar{b}q\bar{q}\rangle$ states have been confirmed so far.  
An ANDY observation at RHIC of a resonance near $18.15$~GeV in Cu-Au collisions~\cite{ANDY:2019bfn} is consistent with predictions for $\TQb$~\cite{Vogt:2021lei}.  
Lattice studies have probed bottom-charm and doubly bottomed tetraquarks~\cite{Francis:2018jyb,Padmanath:2023rdu,Bicudo:2015vta,Leskovec:2019ioa,Alexandrou:2024iwi}.

Large-transverse-momentum measurements of $X(3872)$ by ATLAS, CMS, and LHCb~\cite{CMS:2013fpt,ATLAS:2016kwu,LHCb:2021ten} point to fragmentation-dominated production, offering a clean lever arm to test high-energy QCD descriptions.  
Motivated by this, Ref.~\cite{Celiberto:2024mab} introduced the first-generation \ac{FFs} for fully charmed tetraquarks, {\tt TQ4Q1.0}, targeting scalar ($J^{PC}=0^{++}$) and tensor ($J^{PC}=2^{++}$) channels.  
Initial-scale inputs include both gluon and charm fragmentation, modeled via \ac{NRQCD} potential methods~\cite{Caswell:1985ui,Thacker:1990bm,Bodwin:1994jh,Cho:1995vh,Cho:1995ce,Leibovich:1996pa,Bodwin:2005hm} and spin-inspired prescriptions~\cite{Suzuki:1977km,Suzuki:1985up,Amiri:1986zv,Nejad:2021mmp}. 
Evolution to experimental scales is performed with the \ac{HF-NRevo} scheme~\cite{Celiberto:2025euy,Celiberto:2024mex,Celiberto:2024bxu,Celiberto:2024rxa,Celiberto:2025xvy}, enabling a consistent \ac{DGLAP}~\cite{Dokshitzer:1977sg,Gribov:1972ri,Altarelli:1977zs} treatment of multistep heavy-flavor thresholds, including gluon and charm activation.

Scalar and tensor channels thus provide robust benchmarks for fully heavy spectroscopy. 
In particular, all latest CMS results~\cite{CMS:2023owd,CMS:2025fpt} favor a $2^{++}$ assignment for the dominant all-charm signal, consistent with a compact configuration of spin-1 diquarks and making the tensor channel a natural ground-state candidate on symmetry and dynamical grounds.  
This motivates a targeted phenomenology and dedicated experimental comparisons.

Subsequent work extended the framework to {\tt TQ4Q1.1}~\cite{Celiberto:2024beg}, updating initial inputs from NRQCD for both gluon~\cite{Feng:2020riv} and charm~\cite{Bai:2024ezn} channels and covering fully charmed and fully bottomed states, thereby generalizing the fragmentation picture across heavy flavors.  
A further advance addressed axial-vector ($J^{PC}=1^{+-}$) tetraquarks~\cite{Celiberto:2025dfe}, for which \ac{LDME} uncertainties were propagated consistently into the FFs for the first time.  
These states are particularly diagnostic: in light systems, chiral dynamics may induce strong mixing with conventional mesons~\cite{Kim:2017yvd,Kim:2018zob}, whereas in the heavy sector \ac{HQSS}~\cite{Isgur:1991wq,Neubert:1993mb} suppresses $1^{++}\leftrightarrow 1^{+-}$ mixing~\cite{Weng:2020jao,An:2022qpt}, yielding nearly pure spin eigenstates.  
Production is also selective: Landau--Yang constraints~\cite{Landau:1948kw,Yang:1950rg} and $C$-parity rules suppress \ac{LO} gluon fusion, making axial-vector channels sensitive probes of nonperturbative dynamics~\cite{Karliner:2020dta,Becchi:2020mjz}.  
Together, the scalar, tensor, and axial-vector components of {\tt TQ4Q1.1} provide a unified, evolution-consistent infrastructure for modeling fragmentation into fully heavy tetraquarks and for systematic, flavor-spanning comparisons with data.

In this review, we extend and complete the phenomenological study initiated in Ref.~\cite{Celiberto:2025ziy}, providing a comprehensive analysis of collinear fragmentation into fully heavy tetraquarks through the {\tt TQ4Q1.1} family of functions.  

A key novelty of this study is the quantitative propagation of both perturbative and nonperturbative uncertainties into the FFs.  
For the first time, LDMEs are treated as independent sources of theoretical error and systematically incorporated into the final functions, together with a replica-based uncertainty-quantification strategy derived from \ac{MHOUs}.  
The LDMEs are defined on a color-composite basis consistent with diquark-antidiquark clustering and are extracted from potential NRQCD calculations for the scalar and tensor channels, while the axial-vector case employs model-averaged inputs.  
This combined treatment yields a realistic estimate of the total theoretical uncertainty, linking the perturbative and nonperturbative components of the fragmentation mechanism.  

The resulting {\tt TQ4Q1.1} framework thus becomes a dual-purpose instrument: not only a predictive tool for collider phenomenology, but also a diagnostic platform to assess model dependence in exotic-hadron formation.  
As a reference case within a broader research program, it establishes the basis for future \emph{multimodal} fragmentation studies, in which perturbative and nonperturbative ingredients of different nature can be explored and validated in a unified, uncertainty-aware setting.

As a test ground for this fragmentation framework, we investigate the inclusive production of fully heavy tetraquarks in association with a jet in proton-proton collisions at energies relevant to the HL-LHC and the FCC.  
The process is studied within the \ac{HyF} at $\NLLp$ that merges \ac{NLO} collinear factorization with the resummation of high-energy logarithms beyond \ac{NLL} accuracy.  
This setup provides access to observables particularly sensitive to high-energy QCD dynamics, including rapidity intervals and azimuthal-angle multiplicities, the latter introduced here as novel discriminators of energy-flow patterns---resummed versus fixed order---in QCD radiation.  
All numerical predictions are obtained using the {\Jethad} computational framework, supplemented by its symbolic extension {\symJethad}~\cite{Celiberto:2020wpk,Celiberto:2022rfj,Celiberto:2023fzz,Celiberto:2024mrq,Celiberto:2024swu,Celiberto:2025_P5Q_review,Celiberto:2025csa}.

Overall, this review contributes to establishing a systematic, evolution-consistent description of fully heavy tetraquark fragmentation.  
By integrating multiple spin-parity channels and both charm and bottom flavors, it delivers a unified reference for compact four-quark systems and a consistent bridge between QCD-based modeling, lattice results, and potential-theory predictions.  
This framework sets the stage for quantitative cross-validation of heavy-flavor observables at current and next-generation hadron colliders.

The remainder of this review is organized as follows.  
Section~\ref{sec:FFs} outlines the derivation of the {\tt TQ4Q1.1} FFs within the {\HFNRevo} evolution framework.  
The principles and implementation of the HyF scheme are presented in Section~\ref{sec:HE-resummation}.  
Section~\ref{sec:phenomenology} is devoted to numerical predictions and phenomenological interpretations, while closing remarks and future perspectives are summarized in Section~\ref{sec:conclusions}.

\section{{\HFNRevo} fragmentation for $\TQQ$ states}
\label{sec:FFs}

We first summarize the essential aspects of heavy-flavor fragmentation, emphasizing its realization in heavy-light hadrons, quarkonia, and exotic multiquark configurations (Section~\ref{ssec:FFs-HF}).
We then illustrate how the NRQCD formalism provides a consistent framework for constructing initial-scale inputs in both gluon- and heavy-quark-initiated fragmentation channels leading to $\TQQ$ production (Sections~\ref{ssec:FFs-NRQCD} and~\ref{ssec:FFs-inputs}).
The final part of the section (Section~\ref{ssec:FFs-HFNRevo}) is devoted to the timelike DGLAP evolution of the full {\tt TQ4Q1.1} set within the {\HFNRevo} scheme, presented as a prototype of a unified, modular framework for describing the fragmentation of fully heavy tetraquarks.

\subsection{Heavy-flavor fragmentation framework}
\label{ssec:FFs-HF}

Heavy-flavor hadrons differ fundamentally from their light counterparts because their fragmentation dynamics partly emerge from perturbative QCD.
The large quark masses make the coupling small enough at the production scale that short-distance components must explicitly enter the FF definition, whereas light-hadron FFs remain fully nonperturbative.

For singly heavy hadrons such as $D$, $B$, and $\Lambda_{c,b}$, the process can be modeled through a two-step mechanism~\cite{Cacciari:1996wr,Cacciari:1993mq,Jaffe:1993ie,Kniehl:2005mk,Helenius:2018uul,Helenius:2023wkn,Generet:2023vte}.
A high-energy parton $j$ first produces a heavy quark $Q$ via a \ac{SDC} calculable in perturbative QCD~\cite{Mele:1990yq,Mele:1990cw,Rijken:1996vr,Mitov:2006wy,Blumlein:2006rr,Melnikov:2004bm,Mitov:2004du,Biello:2024zti}.
The heavy quark then undergoes nonperturbative hadronization, described through phenomenological parametrizations~\cite{Kartvelishvili:1977pi,Bowler:1981sb,Peterson:1982ak,Andersson:1983jt,Collins:1984ms,Colangelo:1992kh} or effective-theory approaches~\cite{Georgi:1990um,Eichten:1989zv,Grinstein:1992ss,Neubert:1993mb,Jaffe:1993ie}.
To obtain a complete \ac{VFNS} description, these inputs are evolved through DGLAP equations, introducing scaling violations order by order in $\alpha_s$.

A similar two-step picture applies to quarkonium production, though the presence of a $Q \bar Q$ pair requires the NRQCD framework~\cite{Caswell:1985ui,Thacker:1990bm,Bodwin:1994jh,Cho:1995vh,Cho:1995ce,Leibovich:1996pa,Bodwin:2005hm}.
NRQCD factorizes short-distance production from nonperturbative hadronization through LDMEs, expanding each state in powers of $\alpha_s$ and the heavy-quark velocity $v_{\cal Q}$~\cite{Grinstein:1998xb,Kramer:2001hh,QuarkoniumWorkingGroup:2004kpm,Lansberg:2005aw,Lansberg:2019adr}.
Low-transverse-momentum production is dominated by direct $Q \bar Q$ creation, while at large transverse momenta single-parton fragmentation governed by DGLAP evolution becomes dominant, effectively defining a VFNS.
The \ac{FFNS} approach~\cite{Alekhin:2009ni} remains relevant at lower scales, where multi-parton fragmentation and higher-twist effects contribute~\cite{Fleming:2012wy,Kang:2014tta,Echevarria:2019ynx,Boer:2023zit,Celiberto:2024mex,Celiberto:2024bxu,Celiberto:2024rxa,Celiberto:2025xvy}.

Early LO predictions for gluon and heavy-quark fragmentation into $S$-wave vector quarkonia~\cite{Braaten:1993rw,Braaten:1993mp} were later extended to NLO~\cite{Zheng:2019gnb,Zheng:2021sdo}, enabling the first VFNS-evolved FFs, {\tt ZCW19$^+$}~\cite{Celiberto:2022dyf,Celiberto:2023fzz}, and their $\BCs$ and $\Bss$ counterparts, {\tt ZCFW22}~\cite{Celiberto:2022keu,Celiberto:2024omj}.
Their predictions matched LHCb data~\cite{LHCb:2014iah,LHCb:2016qpe,Celiberto:2024omj}, confirming the expected $\BCs$-to-$B$ suppression at the 0.1\% level~\cite{Celiberto:2024omj} and validating VFNS applicability at large transverse momenta.

NRQCD-inspired approaches have since been extended to exotic hadrons, such as the double $\Jpsi$ signals~\cite{LHCb:2020bwg,ATLAS:2023bft,CMS:2023owd}, interpreted as compact tetraquarks~\cite{Zhang:2020hoh,Zhu:2020xni}.
Their formation originates from the short-distance production of two heavy-quark pairs, $\sim 1/m_Q$, with $m_Q$ being the heavy-quark mass, followed by nonperturbative binding.
The first NRQCD-based input for gluon fragmentation into color-singlet $S$-wave $\TQc$ states was provided in~\cite{Feng:2020riv}.
Building upon this, the {\tt TQHL1.0} FFs~\cite{Celiberto:2023rzw,Celiberto:2024mrq} established the first VFNS framework for heavy-light tetraquarks.
The subsequent {\tt TQ4Q1.1}~\cite{Celiberto:2025ziy} and {\tt TQHL1.1}~\cite{Celiberto:2024beg} families incorporated NRQCD-based modeling for both gluon~\cite{Feng:2020riv} and heavy-quark~\cite{Bai:2024ezn} channels, improved treatment of doubly-heavy states, and extended coverage to bottomoniumlike systems.

A detailed study of axial-vector ($1^{+-}$) tetraquarks followed in~\cite{Celiberto:2025dfe}, where LDME uncertainties were first propagated into the FFs.
The {\tt TQ4Q1.1} set was subsequently applied to investigate charmed-tetraquark signatures in Higgs and electroweak decays~\cite{Ma:2025ryo}.
Parallel developments produced FFs for fully charmed pentaquarks ({\tt PQ5Q1.0}) and triply heavy $\Omega$ baryons ({\tt OMG3Q1.0})~\cite{Celiberto:2025ipt,Celiberto:2025ogy}, advancing the systematic exploration of heavy exotic fragmentation in QCD.

\subsection{Extending NRQCD factorization from quarkonia to tetraquarks}
\label{ssec:FFs-NRQCD}

We briefly present the NRQCD-based framework for collinear fragmentation, outlining how it provides a consistent and hierarchical description of the transition from a perturbative partonic configuration to a physical hadron.
For clarity, we first summarize the NRQCD construction for standard quarkonium production, which serves as a conceptual baseline before extending the discussion to exotic multiquark systems.

The derivation of tetraquark FFs follows the formalism established in~\cite{Feng:2020riv,Bai:2024ezn}, adapted here to the physical context of our study.
For the full analytical details of the NRQCD-based calculation and the treatment of short- and long-distance terms, the reader is referred to those works.

Within NRQCD, the FF describing the transition of a parton $j$ into a quarkonium $\cal Q$ carrying momentum fraction $z$ at the initial scale $\mu_{F,0}$ reads
\begin{equation}
\label{FFs_NRQCD_onium_rfrm}
 D_j^{\cal Q}(z,\mu_{F,0}) 
 \,= \sum_{[n]} {\cal D}_j^{Q\bar Q}(z,[n])
 \,
 \langle {\cal O}^{\cal Q}([n]) \rangle \;.
\end{equation}
Here ${\cal D}_j^{Q\bar Q}(z,[n])$ represents the perturbative SDC for producing a $Q \bar Q$ pair with quantum numbers $[n]$, while $\langle {\cal O}^{\cal Q}([n]) \rangle$ encodes the nonperturbative transition probability governed by the LDME.
The label $[n] = {}^{2S+1}L_J^{(c)}$ specifies spin, orbital, and color structure, with $(c)$ = 1 or 8 for singlet or octet states, respectively.
Equation~\eqref{FFs_NRQCD_onium_rfrm} embodies the two NRQCD postulates: the expansion of the quarkonium state over all Fock components and the ordering of their contributions in powers of $\alpha_s$ and $v_{\cal Q}$.

Fragmentation into singly heavy hadrons $H_Q$ exhibits partial analogies but crucial differences.
There, the $j \to H_Q$ FF at the initial scale results from the convolution of a perturbative $j \to Q$ coefficient with a $z$-dependent hadronization function describing the probability for $Q$ to form $H_Q$~\cite{Cacciari:1996wr,Cacciari:1993mq,Jaffe:1993ie}.
In quarkonium, by contrast, hadronization occurs without momentum redistribution since both heavy constituents are created perturbatively.
Thus LDMEs act as constant normalization factors, independent of $z$, and the overall FF is expressed as a linear combination of perturbative terms weighted by scalar coefficients.

The nonperturbative LDMEs cannot be derived from first principles but can be extracted from fits or estimated via lattice and potential models.
A useful simplification comes from the \ac{VSA}~\cite{Shifman:1978bx,Gilman:1979bc,Bodwin:1994jh}, which assumes that intermediate states other than the vacuum are suppressed by powers of $v_{\cal Q}$.
In this limit,
\begin{equation}
\label{VSA_rfrm}
\langle 0 | \chi^\dagger \Pi_n \psi \, \mathcal{P}_{\cal Q} \, \psi^\dagger \Pi_n^\prime \chi | 0 \rangle
\,\simeq\,
\langle 0 | \chi^\dagger \Pi_n \psi | {\cal Q} \rangle
\,
\langle {\cal Q} | \psi^\dagger \Pi_n^\prime \chi | 0 \rangle \;,
\end{equation}
where $\Pi_n$ and $\Pi_n^\prime$ are color-spin projectors and $\mathcal{P}_{\cal Q}$ selects the physical quarkonium state.
For the same reason, the FF for a parton $j$ fragmenting into a fully heavy tetraquark $\TQQ$ can be expressed as
\begin{equation}
\label{FFs_TQc_general_rfrm}
 D_j^{\TQQ}(z,\mu_{F,0})
 \,= \sum_{[n]} {\cal D}_j^{4Q}(z,[n]) 
 \,
 \langle {\cal O}^{\TQQ}([n]) \rangle \;,
\end{equation}
where ${\cal D}_j^{4Q}$ describes the short-distance formation of a compact four-quark system and $\langle {\cal O}^{\TQQ}([n]) \rangle$ represents its hadronization probability.
Compared to quarkonia, tetraquarks involve richer internal structures: spin-color couplings among four heavy constituents yield a broader Fock-state spectrum and a more complex pattern of LDMEs.

Although the NRQCD expansion remains applicable thanks to $m_Q \gg \Lambda_{\rm QCD}$, the weaker binding and coexistence of compact and extended configurations may affect convergence in $v_Q$.
Still, for low-lying $S$-wave states, the factorized form of Eq.~\eqref{FFs_TQc_general_rfrm} provides a sound starting point for building initial FFs and connecting them to higher scales through DGLAP evolution.

Restricting to $\TQQ(J^{PC})$ states with $J^{PC} = 0^{++}$, $1^{+-}$, and $2^{++}$, and retaining the leading terms in $v_Q$, one obtains
\begin{equation}
\label{TQQ_FF_initial-scale_rfrm}
 D^{\TQQ(J^{PC})}_j(z,\mu{F,0}) 
 \,=\,
 \frac{1}{m_Q^9}
 \sum_{[n]}
 \tilde{\cal D}^{(J^{PC})}_j(z,[n])
 \,\langle {\cal O}^{\TQQ(J^{PC})}([n]) \rangle \;,
\end{equation}
where $m_Q = 1.5$~GeV for charm and $4.9$~GeV for bottom, and
\begin{equation}
\tilde{\cal D}^{(J^{PC})}_j(z,[n]) \equiv m_Q^9 {\cal D}^{(J^{PC})}_j(z,[n]) \;,
\end{equation}
with symmetry relations
\begin{equation}
 \tilde{\cal D}^{(J^{PC})}_j(z,[3,6]) 
 \,=\, 
 \tilde{\cal D}^{(J^{PC})}_j(z,[6,3]) \;, \qquad
 \langle {\cal O}^{\TQQ(J^{PC})}([3,6])  \rangle 
 \,=\,
 \langle {\cal O}^{\TQQ(J^{PC})}([6,3]) \rangle^* \;.
\end{equation}

The fragmentation regime is expected to dominate for transverse momenta $|\vec \kappa| \gtrsim 3 M_{\TQQ}$, ensuring $\ln(|\vec \kappa|^2/M_{\TQQ}^2)$ enhancement and suppression of $\mathcal{O}(M_{\TQQ}^2/|\vec \kappa|^2)$ corrections~\cite{Cacciari:1995yt,Artoisenet:2007xi,Ma:2014svb}.
For $\TQc$ ($M_{\TQc} \!\sim\! 6.5$~GeV) and $\TQb$ ($M_{\TQb} \!\sim\! 18$~GeV), this translates into $|\vec \kappa| \gtrsim 20$~GeV and $|\vec \kappa| \gtrsim 50$~GeV, respectively, defining the kinematic domain where our FFs are reliable.

At lower transverse momenta, DPS effects, where two independent hard interactions occur in a single $pp$ collision, may contribute significantly.
DPS has been studied across multiple channels, including double jets~\cite{Ducloue:2015jba}, double and mixed quarkonia~\cite{Lansberg:2014swa,Lansberg:2020rft,Lansberg:2016rcx,Lansberg:2017chq}, and triple $\Jpsi$ production~\cite{dEnterria:2016ids,Shao:2019qob}.
These analyses confirm its relevance at low to moderate transverse momenta, where it can even exceed single-parton scattering contributions~\cite{Maciula:2020wri,Carvalho:2015nqf,Abreu:2023wwg}.
However, DPS lacks logarithmic enhancement and falls off rapidly with $|\vec \kappa|$; thus, for $|\vec \kappa| \gtrsim 3 M_{\TQQ}$, fragmentation remains the leading mechanism within our framework, while DPS effects can be safely neglected at leading power.

\subsection{Modeling and initialization of fragmentation inputs}
\label{ssec:FFs-inputs}

\begin{figure}[!t]
\centering
\includegraphics[width=0.470\textwidth]{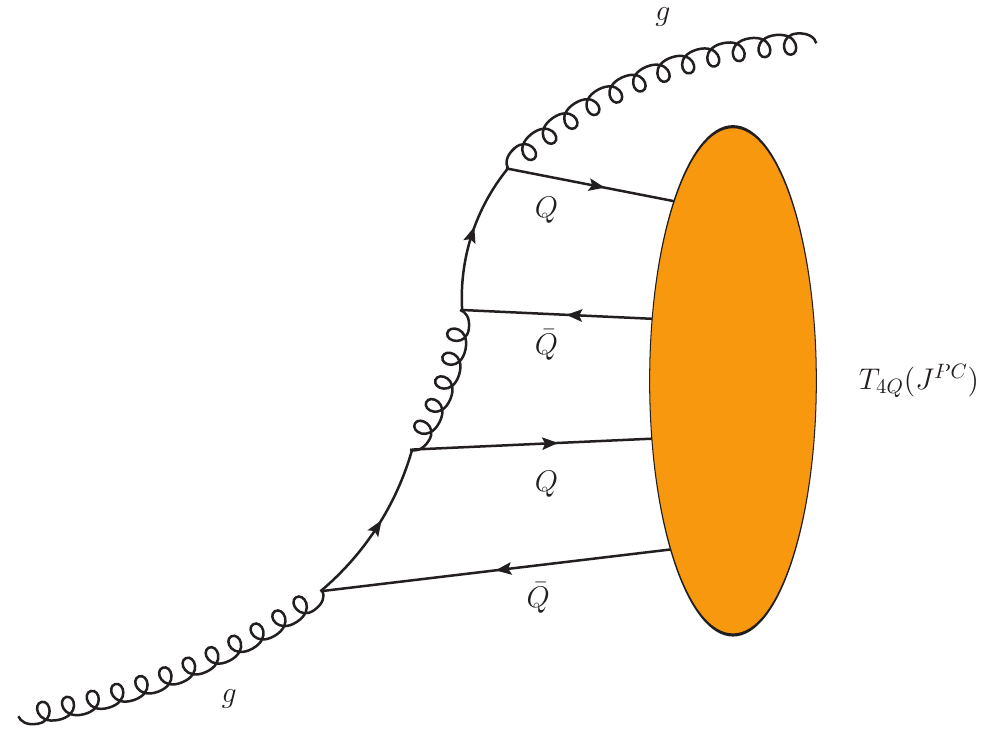}
\hspace{0.375cm}
\includegraphics[width=0.470\textwidth]{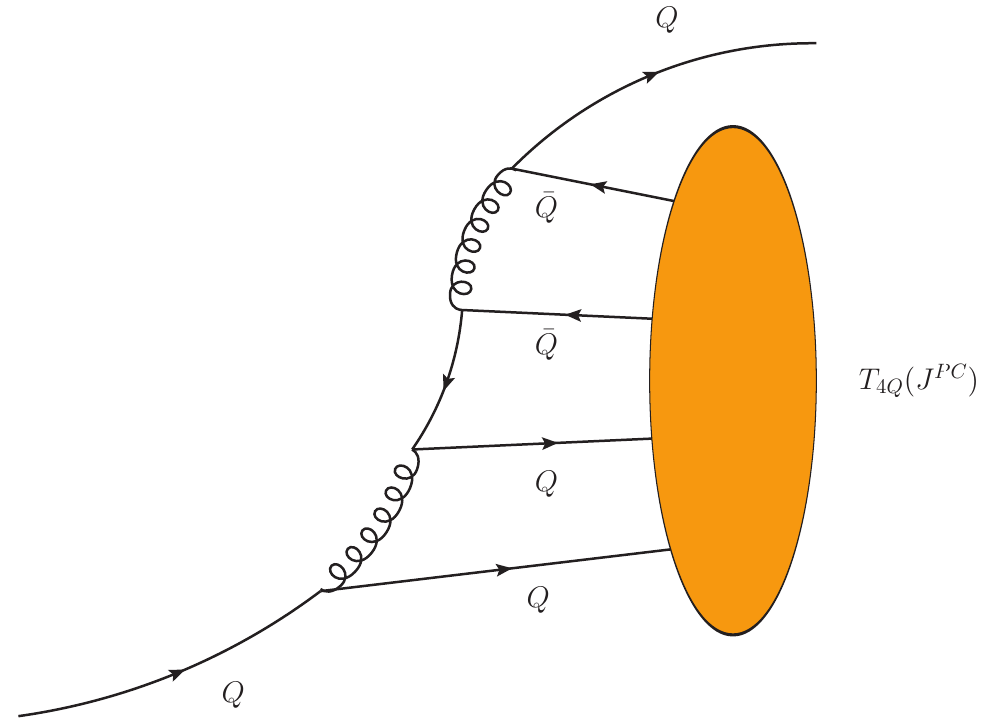}

\caption{
LO diagrams illustrating the collinear fragmentation of a gluon (left) and a heavy quark (right) into a fully heavy $S$-wave tetraquark in a color-singlet state.
The partonic interactions shown on the left of each panel represent the perturbative SDCs, whereas the orange blobs on the right denote the nonperturbative LDMEs responsible for the hadronization into the physical tetraquark.
}
\label{fig:FF-diagrams}
\end{figure}

The SDCs in Eq.~\eqref{TQQ_FF_initial-scale_rfrm} are obtained through the standard perturbative matching between QCD and NRQCD matrix elements, following the procedure outlined in Refs.~\cite{Feng:2020riv,Bai:2024ezn}.
Within NRQCD, these coefficients are determined by equating perturbative QCD and NRQCD amplitudes computed with hypothetical, nonbound multiquark states carrying the same quantum numbers as the target hadron~\cite{Bodwin:1994jh,Petrelli:1997ge}.
At present, analytical results for the initial-scale inputs of $\TQQ$ fragmentation exist only at LO in both $\alpha_s$ and the relative velocity $v_{\cal Q}$.

Analogous to quarkonium, SDCs for tetraquark production are insensitive to long-distance hadronization and can be evaluated by substituting the physical $\TQQ$ with a fictitious color-neutral configuration $|[QQ][\bar{Q}\bar{Q}]\rangle$.
The LO matching retains only the lowest NRQCD operators, neglecting derivative and higher-$v_{\cal Q}$ corrections.
The tetraquark is modeled as a diquark-antidiquark pair, and the projection onto definite spin, parity, and color is realized through corresponding four-quark NRQCD operators.
Both $[3\otimes\bar{3}]$ and $[6\otimes\bar{6}]$ color combinations are included.

Fermi--Dirac statistics and $S$-wave symmetry restrict the allowed quantum numbers.
For scalar ($0^{++}$) and tensor ($2^{++}$) states, both color configurations can contribute, although the $[6\otimes\bar{6}]$ term is relevant only to spin-0 channels.
In the axial-vector case ($1^{+-}$), antisymmetry forbids the $[6\otimes\bar{6}]$ component, leaving only the $[\bar{3}\otimes3]$ one.
Moreover, gluon fragmentation into $1^{+-}$ states is prohibited at LO by the Landau--Yang theorem, which prevents a single on-shell gluon from coupling to this configuration.
Consequently, only heavy-quark initiated fragmentation is retained for this channel.

The resulting dimensionless SDCs can be found in the first appendix of Ref.~\cite{Celiberto:2025ziy}.
Their explicit analytic forms, first derived in Refs.~\cite{Feng:2020riv,Bai:2024ezn} and independently confirmed using the {\psymJethad} symbolic environment~\cite{Celiberto:2020wpk,Celiberto:2022rfj,Celiberto:2023fzz,Celiberto:2024mrq,Celiberto:2024swu,Celiberto:2025_P5Q_review,Celiberto:2025csa}, apply to LO color-singlet configurations (see Fig.~\ref{fig:FF-diagrams}).

As discussed earlier, LDMEs encode the nonperturbative conversion of the perturbatively produced $|[QQ][\bar{Q}\bar{Q}]\rangle$ configuration into the physical tetraquark.
They must be either fitted to data or modeled phenomenologically.
A common simplification, paralleling the quarkonium case, is the VSA, which factorizes the LDME into a product of matrix elements between the vacuum and the lowest Fock component of the hadron.
For tetraquarks, however, the presence of color-correlated diquark and antidiquark subsystems requires a more elaborate construction.
Following Refs.~\cite{Feng:2020riv,Bai:2024ezn}, we adopt the color-composite LDME approach, expanding the physical state in a diquark-antidiquark basis and projecting it onto color-singlet operators.

This decomposition organizes the long-distance matrix elements according to the color configuration of the constituent clusters ($[3,3]$, $[6,6]$, and mixed $[3,6]$/$[6,3]$ terms) as defined in Eq.~\eqref{TQQ_FF_initial-scale_rfrm}.
Each LDME quantifies the probability amplitude for a four-quark configuration to hadronize into a bound tetraquark with specified spin and color quantum numbers.
For completeness, the color-composite operators entering the $\TQQ$ LDMEs are written as
\begin{align}
\label{TQQ_FF_LDMEs_operators_rephrased}
 {\cal O}^{(0)}_{\bar{3} \otimes 3} &= - 1/\sqrt{3} \, 
 \big[\psi^T_a (i\sigma^2)\sigma^i \psi_b\big]
 \big[\chi^\dagger_c \sigma^i (i\sigma^2)\chi^*d\big]
 {\cal C}^{ab;cd}_{\bar{3}\otimes3}\,, \nonumber \\[0.2cm]
 {\cal O}^{\alpha\beta;(2)}_{\bar{3}\otimes3} &=
 \big[\psi^T_a (i\sigma^2)\sigma^m\psi_b\big]
 \big[\chi^\dagger_c \sigma^n(i\sigma^2)\chi^d\big]
 \Gamma^{\alpha\beta;mn}{\cal C}^{ab;cd}_{\bar{3}\otimes3}\,, \\[0.2cm] \nonumber
 {\cal O}^{(0)}_{6\otimes\bar{6}} &=
 \big[\psi^T_a(i\sigma^2)\psi_b\big]
 \big[\chi^\dagger_c(i\sigma^2)\chi^d\big]
 {\cal C}^{ab;cd}_{6\otimes\bar{6}}\;,
\end{align}
where $\psi$ and $\chi$ denote the NRQCD Pauli fields and $\sigma^2$ the second Pauli matrix, consistent with Eq.~\eqref{VSA_rfrm}.
The rank-4 Lorentz tensor reads
\begin{equation}
\label{Gamma_rank4_short}
 \Gamma^{kl;mn}=\tfrac{1}{2}\left(\delta^{km}\delta^{ln}+\delta^{kn}\delta^{lm}-\tfrac{2}{3}\delta^{kl}\delta^{mn}\right)\,,
\end{equation}
and the color tensors are
\begin{equation}
\begin{split}
\label{C_abcd_short}
 C_{\bar{3}\otimes3}^{ab;cd}&=\tfrac{1}{2\sqrt{3}}(\delta^{ac}\delta^{bd}-\delta^{ad}\delta^{bc}),\\[0.15cm]
 C_{6\otimes\bar{6}}^{ab;cd}&=\tfrac{1}{2\sqrt{6}}(\delta^{ac}\delta^{bd}+\delta^{ad}\delta^{bc})\,,
\end{split}
\end{equation}
with $\epsilon^{ijk}$ and $\delta^{mn}$ used to form color-singlet combinations.

Although the diquark-antidiquark basis can be recast, through Fierz transformations, into a molecularlike expansion with $[1\otimes1]$ or $[8\otimes8]$ color topology, it does not imply a literal molecular interpretation.
Instead, it captures dominant spin-color correlations between compact diquark and antidiquark clusters---an appropriate representation for fragmentation-driven production of tightly bound multiquark states.

The treatment of color-octet channels within NRQCD remains one of the most intricate aspects of initial-scale fragmentation modeling.
In this work, as in Refs.~\cite{Feng:2020riv,Bai:2024ezn}, we confine our analysis to color-singlet configurations, consistent with the limited knowledge of the corresponding color-octet LDMEs.
Reliable determinations of such matrix elements are not yet available, and any attempt to include them would require ad-hoc assumptions.
A detailed theoretical discussion of their possible role in $\TQQ$ fragmentation was first presented in Ref.~\cite{Celiberto:2025ziy}, which emphasized the need for future lattice or potential-model estimates to achieve quantitative accuracy.

The relevance of color-octet mechanisms can be appreciated by recalling the well-known quarkonium production puzzle~\cite{Brambilla:2010cs,Andronic:2015wma}.
For the $\Jpsi$, singlet channels such as ${}^3S_1^{(1)}$ alone underestimate high-$p_T$ cross sections, while octet transitions ${}^3S_1^{(8)}$, ${}^1S_0^{(8)}$, and ${}^3P_J^{(8)}$~\cite{Braaten:1993rw,Cho:1995vh,Cho:1995ce,Beneke:1996tk,Butenschoen:2010rq,Chao:2012iv,Gong:2012ug} are essential to reproduce data.
Conversely, for pseudoscalar quarkonia such as $\eta_c$ and $\eta_b$, the ${}^1S_0^{(1)}$ singlet term dominates, and existing measurements~\cite{LHCb:2014oii,LHCb:2019zaj} confirm that octet effects are strongly suppressed~\cite{Han:2014jya}.
A similar pattern holds for bottomonia~\cite{LHCb:2022byt,Artoisenet:2008fc,Gong:2010bk}.

For fully heavy tetraquarks, the importance of color-octet channels depends on the quantum numbers $J^{PC}$.
In the scalar sector ($0^{++}$), multiple color combinations such as $[3,3]$ and $[6,6]$ can interfere, making fragmentation predictions model-dependent and potentially sensitive to octet admixtures.
The tensor case ($2^{++}$) is more constrained: only the $[3,3]$ component contributes, yet uncertainties in the wave-function normalization leave room for small octet effects.
The axial-vector state ($1^{+-}$), by contrast, features a unique antisymmetric $[3,3]$ color structure that forbids interference and yields stable, model-independent predictions~\cite{Bai:2024ezn,Celiberto:2025dfe}.
Unlike the $\Jpsi$, its singlet channel alone provides a consistent description, making it the cleanest benchmark for fragmentation studies.

From an NRQCD-scaling viewpoint, color-octet terms are expected to be heavily suppressed.
If the singlet FF scales as $v_{\cal Q}^3$, octet terms scale as $\alpha_s v_{\cal Q}^7$, giving a ratio $\text{octet/singlet}\!\sim\!\alpha_s v_{\cal Q}^4$.
Using $v_{{\cal Q}c}^2\!\sim\!0.3$ and $\alpha_s\!\sim\!0.2$ yields $\alpha_s v_{{\cal Q}c}^4\!\simeq\!0.02$, \emph{i.e.} two orders of magnitude suppression, while for bottom systems, $v_{{\cal Q}b}^2\!\sim\!0.1$ gives $\alpha_s v_{{\cal Q}_b}^4\!\simeq\!1.5\times10^{-3}$, implying an even stronger suppression.
Reduced phase space for soft-gluon emission in heavier systems further weakens octet transitions.

These scaling estimates justify the singlet-only approximation adopted here, especially for the $1^{+-}$ channel, while delineating the next step: once robust lattice or potential-model determinations of octet LDMEs become available, they can be incorporated to achieve a comprehensive and quantitatively refined description of $\TQQ$ fragmentation across all $J^{PC}$ sectors.

Following our previous works~\cite{Celiberto:2024mab,Celiberto:2024beg}, and in the absence of experimental or lattice determinations, we estimate the $\langle {\cal O}^{\TQQ(J^{PC})}([n]) \rangle$ matrix elements using potential-model calculations of the tetraquark wave function at the origin.

This approach builds on the framework introduced in~\cite{Feng:2020riv}, where the Schr{\"o}dinger equation for fully charmed systems was solved using a Cornell-type potential~\cite{Eichten:1974af,Eichten:1978tg}.
The resulting radial wave functions at the origin were then related to the LDMEs via the vacuum-saturation approximation.
Three potential models were originally proposed~\cite{Zhao:2020nwy,Lu:2020cns,liu:2020eha}, differing in the inclusion of relativistic effects.
The first and third models were later found to produce unstable or unrealistically large cross sections, while the second one~\cite{Lu:2020cns} provided numerically stable results consistent with existing quarkonium data~\cite{CMS:2017dju}.
For this reason, both Refs.~\cite{Celiberto:2024mab,Celiberto:2024beg} and the present work adopt it as the baseline reference.

A more refined treatment was proposed in~\cite{Bai:2024ezn}, who introduced two additional potential models (denoted IV and V).
Model IV~\cite{Yu:2022lak} predicts LDME values for the axial-vector state that are close to those from Ref.~\cite{Lu:2020cns}, whereas Model V~\cite{Wang:2019rdo} produces significantly smaller numbers.
In Ref.~\cite{Celiberto:2025dfe}, we therefore adopted the average of Models II and IV as our default input, interpreting their spread as an intrinsic theoretical uncertainty.
This averaging strategy offers a pragmatic balance between model diversity and physical realism.

Among all states, the axial-vector configuration provides the cleanest testing ground.
Because of Fermi--Dirac statistics and its $S$-wave symmetry, it admits only one color-spin component, $[3,3]$, with no internal interference.
This single-structure nature stabilizes the predictions and reduces model dependence.
In fact, as emphasized in Ref.~\cite{Celiberto:2025dfe}, results for the $1^{+-}$ state remain remarkably consistent across multiple LDME models, unlike the scalar and tensor cases.

\begin{table}[t]
\centering
\begin{tabular}{c|ccc|ccc}
\toprule
\multirow{2}{*}{$[n]$} &
\multicolumn{3}{c|}{\textbf{Charm sector} ($\TQc$) [GeV$^9$]} &
\multicolumn{3}{c}{\textbf{Bottom sector} ($\TQb$) [GeV$^9$]} \\
\cmidrule(lr){2-4}\cmidrule(lr){5-7}
 & $\TQcZpp$ & $\TQcOpm$ & $\TQcTpp$ & $\TQbZpp$ & $\TQbOpm$ & $\TQbTpp$ \\
\midrule
$[3,3]$ & $0.0347 \pm 0.0076$ & $0.0878 \pm 0.0098$ & $0.0720 \pm 0.0158$ 
        & $13.88 \pm 3.05$ & $35.1 \pm 3.9$ & $28.80 \pm 6.34$ \\
$[6,6]$ & $0.0128 \pm 0.0028$ & $0$ & $0$
        & $5.12 \pm 1.13$ & $0$ & $0$ \\
$[3,6]$ & $0.0211 \pm 0.0046$ & $0$ & $0$
        & $8.44 \pm 1.86$ & $0$ & $0$ \\
\bottomrule
\end{tabular}
\caption{Color-composite LDMEs $\langle {\cal O}^{\TQQ(J^{PC})}([n]) \rangle$ for the $\TQc$ and $\TQb$ tetraquark families.
Bottom-sector entries are obtained via color-Coulomb scaling from the corresponding $\TQc$ values, as described in the text.}
\label{tab:T4Q_LDMEs}
\end{table}

The scalar ($0^{++}$) and tensor ($2^{++}$) channels, by contrast, display greater sensitivity to the underlying potential.
The scalar configuration receives contributions from several color combinations ($[3,3]$, $[6,6]$, and mixed $[3,6]$), which can interfere constructively or destructively, amplifying model uncertainty.
Although the tensor state involves only a single color structure, its predicted LDME still depends strongly on the normalization of the wave function at the origin.
To account for these effects, we adopt a conservative prescription: the relative uncertainty obtained for the axial-vector LDME in Ref.~\cite{Celiberto:2025dfe} is doubled for the scalar and tensor channels to reflect their increased model dependence.

Since no first-principles LDME determinations exist for the fully bottomed sector, we estimate them using a scaling argument inspired by color-Coulomb dynamics~\cite{Feng:2023agq}.
Treating the tetraquark as a pair of tightly bound diquarks, we assume that the strength of the wave function at the origin scales with the heavy-quark mass and the corresponding strong coupling, approximately as $(\alpha_s m_Q)^9$.
This implies that the bottom-sector matrix elements are roughly four hundred times larger than their charmed counterparts, consistent with the expectation that heavier quarks form more compact and strongly bound configurations.

Possible corrections from confining potentials such as Cornell~\cite{Eichten:1978tg} or Buchm{\"u}ller--Tye~\cite{Buchmuller:1980su} were also explored.
These effects, which tend to decrease the wave function at the origin by about 25\%, would alter the LDMEs by less than 1\%, well below the 20--25\% uncertainty already inherited from the charmed sector.
Consequently, we treat this contribution as negligible and retain only the propagated uncertainty from the $\TQc$ analysis.

Table~\ref{tab:T4Q_LDMEs} summarizes the color-composite LDMEs adopted in this work for both the charm and bottom tetraquark families, together with their associated theoretical uncertainties.

\subsection{Generation and evolution of {\tt TQ4Q1.1} via {\HFNRevo}}
\label{ssec:FFs-HFNRevo}

In this subsection, we present the DGLAP evolution of the initial NRQCD inputs leading to the {\tt TQ4Q1.1} collinear FFs for fully heavy tetraquarks.
Unlike light-hadron fragmentation, both the gluon and heavy-quark channels feature distinct evolution thresholds, determined by the kinematics of the perturbative splittings $g \to (Q\bar{Q}Q\bar{Q})$ and $Q,\bar{Q} \to (Q\bar{Q}Q\bar{Q})+Q,\bar{Q}$ shown in Fig.~\ref{fig:FF-diagrams}.
These fragmentation splittings set $\mu_{F,0}(g\!\to\!\TQQ)=4m_Q$ and $\mu_{F,0}(Q\!\to\!\TQQ)=5m_Q$ as the starting scales for gluon and (anti)quark fragmentation, respectively.

To consistently include both thresholds, we adopt the {\HFNRevo} scheme~\cite{Celiberto:2025euy,Celiberto:2024mex,Celiberto:2024bxu,Celiberto:2024rxa,Celiberto:2025xvy}, a framework specifically developed for evolving heavy-hadron FFs derived from nonrelativistic inputs.
It provides a smooth connection between FFNS and VFNS regimes while allowing a systematic propagation of MHOUs through controlled threshold variations.
Originally tailored for quarkonium studies, {\HFNRevo} has been successfully extended to exotic-hadron applications, proving effective in handling multi-threshold evolution.

For fully heavy tetraquarks, the evolution proceeds in two stages.
First, the gluon FF, initialized at $\mu_{F,0}(g\!\to\!\TQQ)=4m_Q$, evolves via the LO $P_{gg}$ kernel up to $\mu_{F,0}(Q\!\to\!\TQQ)=5m_Q$, generating only collinear gluons.
This step, involving a single active channel, is performed analytically through the {\symJethad} plugin.
At the heavy-quark threshold ($Q_0=5m_Q$), the evolved gluon FF is matched to the NRQCD input of the (anti)quark channel.
From this ``evolution-ready'' scale, a full NLO DGLAP evolution including all active species is carried out using {\tt APFEL++}~\cite{Bertone:2013vaa,Carrazza:2014gfa,Bertone:2017gds}, yielding the final {\tt TQ4Q1.1} sets in {\tt LHAPDF} format~\cite{Buckley:2014ana}.
Future versions will also interface {\HFNRevo} with {\tt EKO}~\cite{Candido:2022tld,Hekhorn:2023gul}.

Uncertainties from the variation of energy scales inside perturbative SDC, labeled as \ac{F-MHOUs}, are estimated by varying $Q_0$ by a factor two around its central value and evolving each replica independently.
The resulting envelope quantifies the FF sensitivity to $Q_0$ and can be propagated to collider-level observables, complementing the uncertainty budget from perturbative \ac{H-MHOUs} and nonperturbative LDMEs (see Section~\ref{ssec:uncertainty}).

Light- and nonconstituent-heavy-quark channels are excluded from this release, as their FFs are suppressed by one or two orders of magnitude compared to gluon fragmentation~\cite{Bai:2024flh,Braaten:1993rw,Braaten:1993mp,Artoisenet:2014lpa,Zhang:2018mlo,Zheng:2021mqr,Zheng:2021ylc,Zheng:2019dfk}.
Moreover, in the LHC kinematic region $10^{-4}\!\lesssim\!x\!\lesssim\!10^{-2}$~\cite{Celiberto:2021dzy,Celiberto:2021fdp,Celiberto:2022dyf}, gluon PDFs exceed quark ones by over an order of magnitude, further reinforcing the dominance of gluon-initiated channels.
Consequently, the present {\tt TQ4Q1.1} release, restricted to gluon and constituent-heavy-quark inputs, provides a complete and phenomenologically reliable description of fully heavy tetraquark fragmentation.

\begin{figure}[!t]
\centering

   \hspace{-0.00cm}
   \includegraphics[scale=0.400,clip]{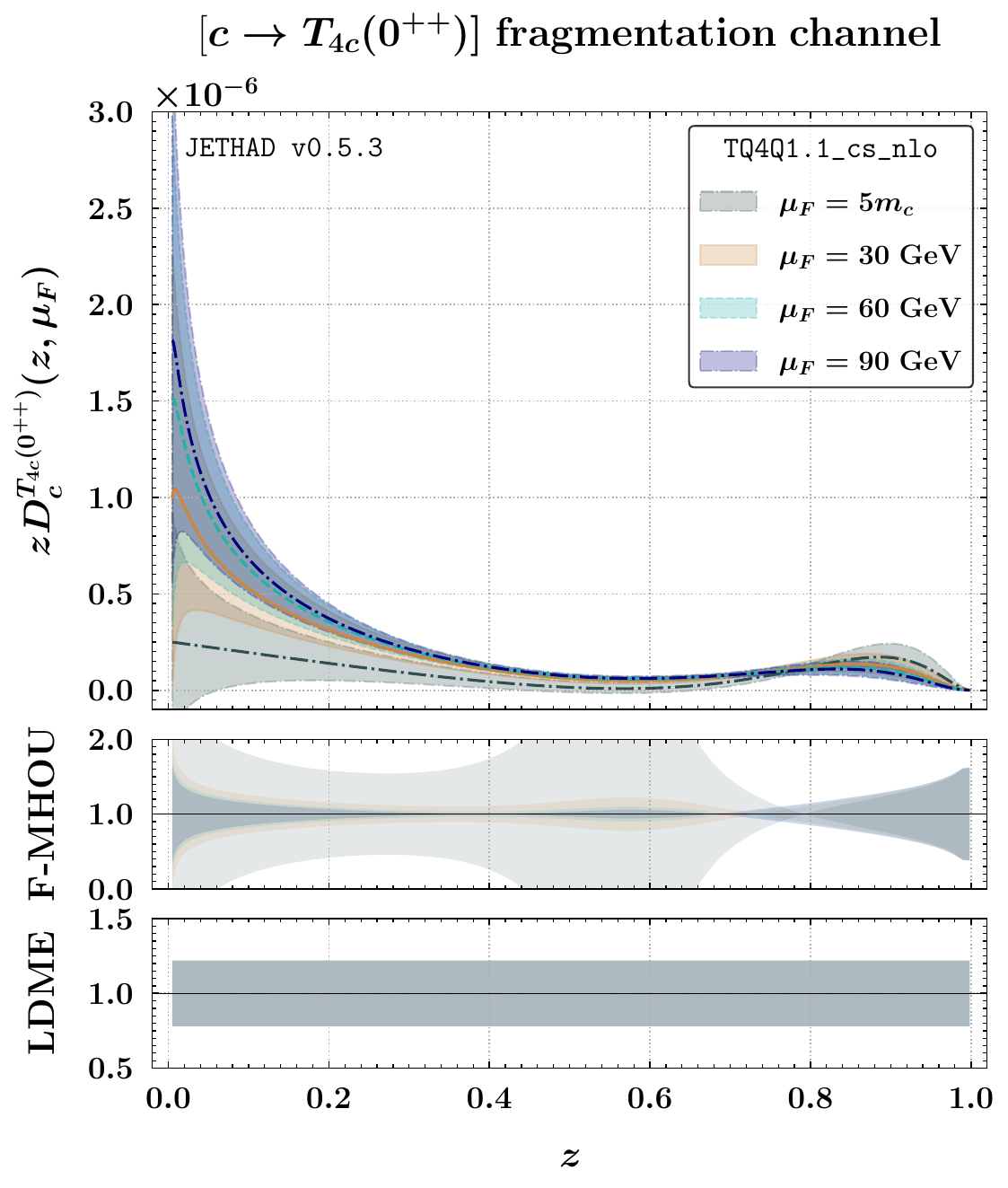}
   \hspace{0.90cm}
   \includegraphics[scale=0.400,clip]{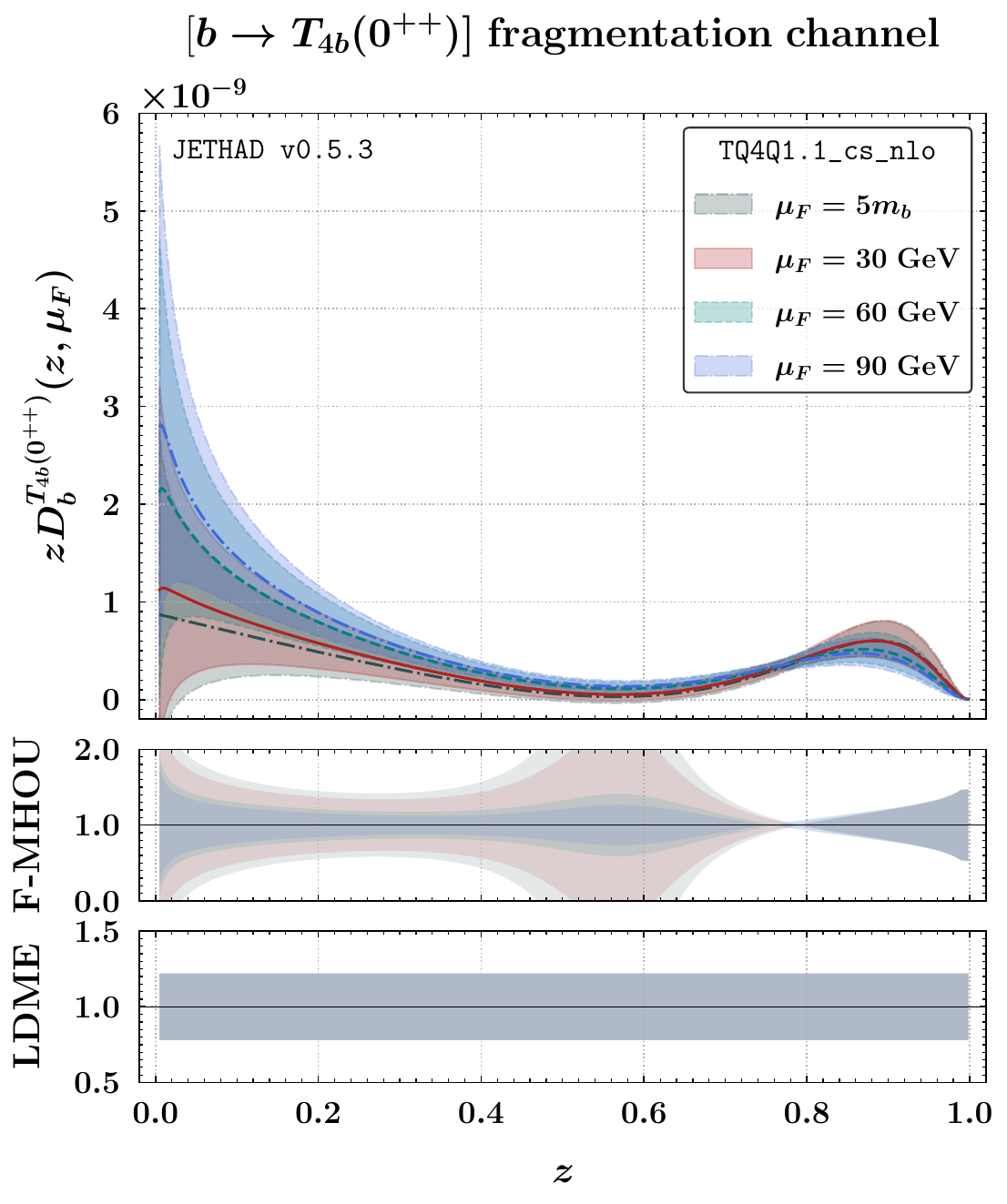}

   \vspace{0.35cm}

   \hspace{-0.00cm}
   \includegraphics[scale=0.400,clip]{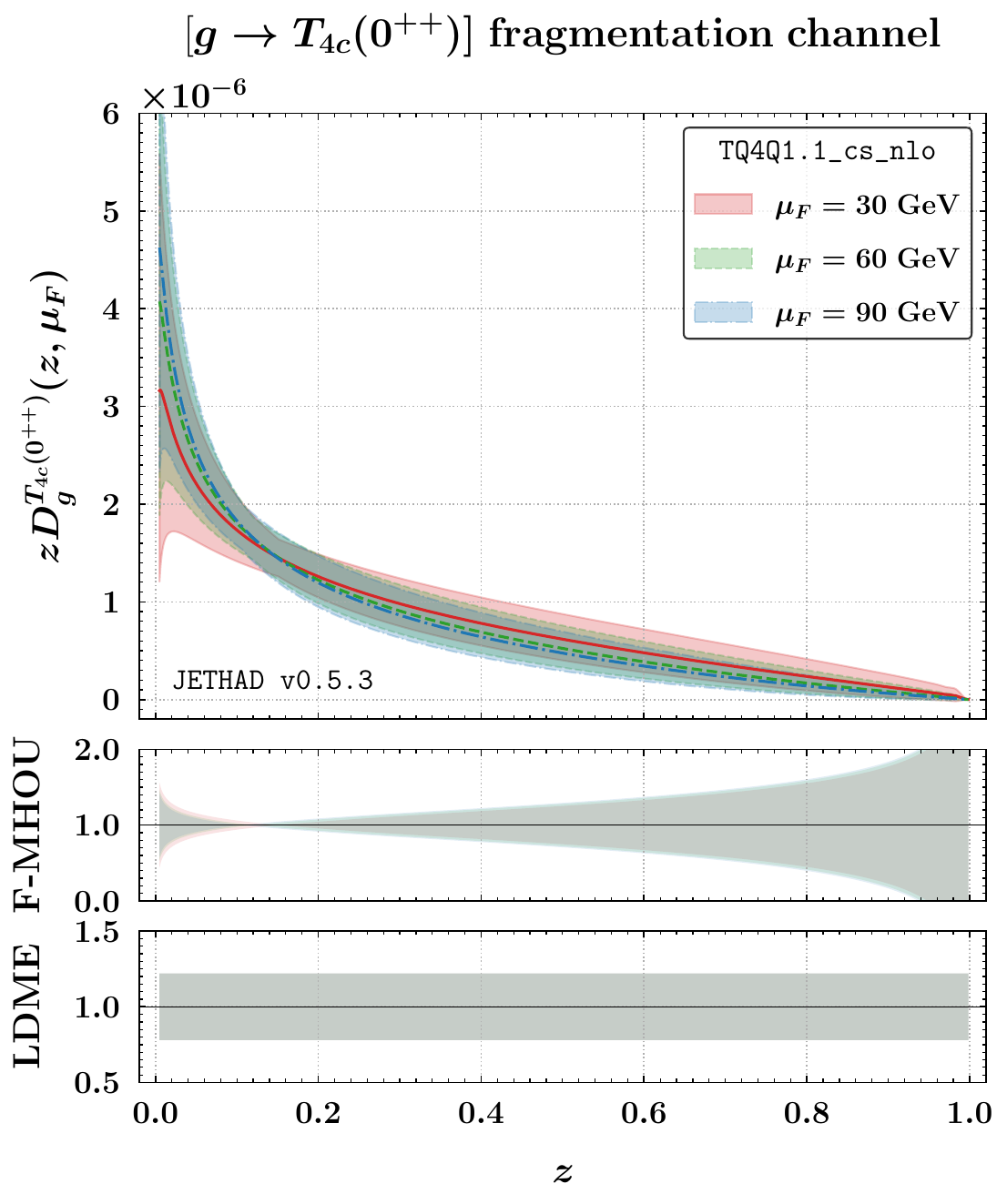}
   \hspace{0.90cm}
   \includegraphics[scale=0.400,clip]{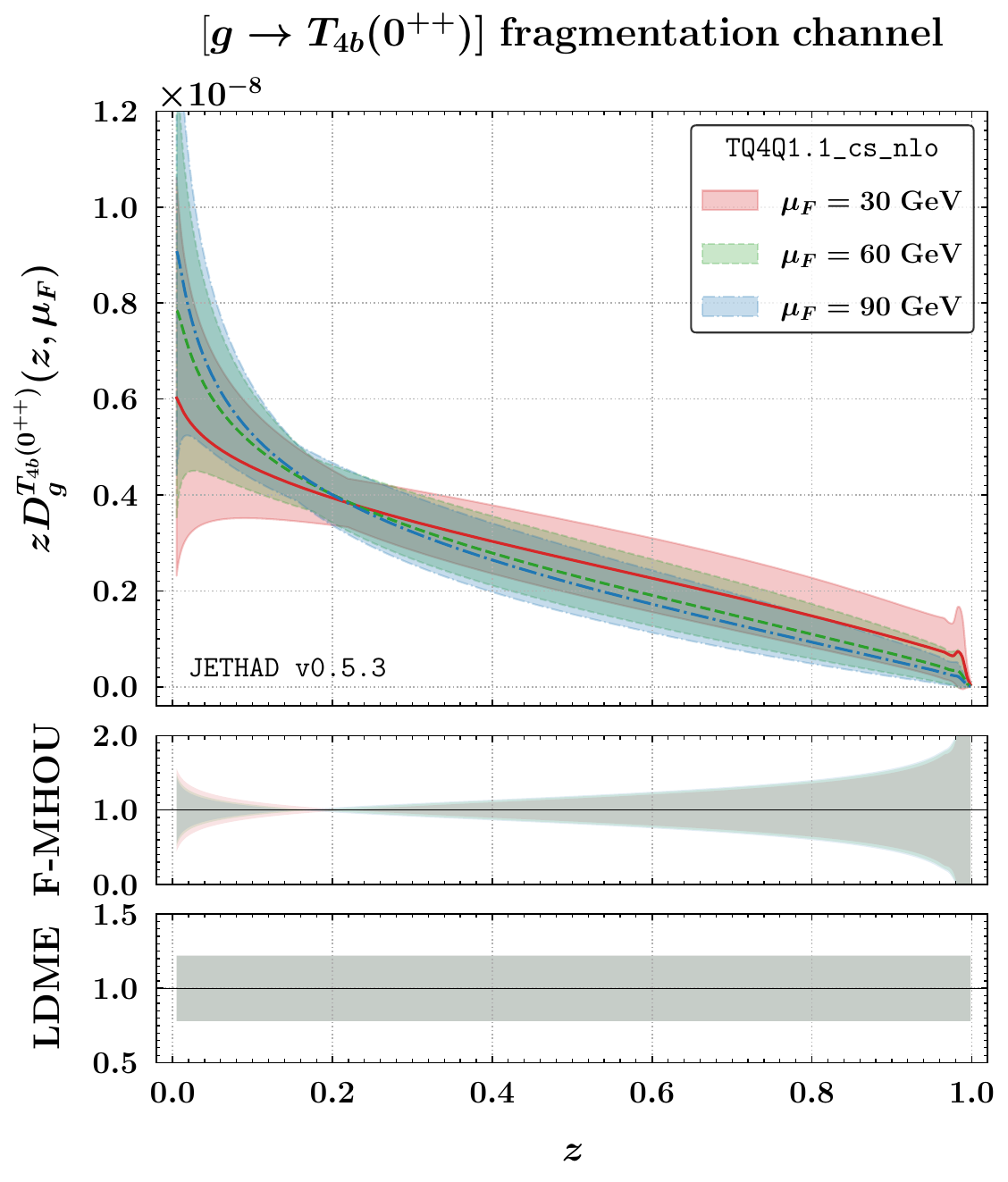}

\caption{
$z$-profiles of the {\tt TQ4Q1.1} FFs for scalar tetraquarks $\TQcZpp$ (left) and $\TQbZpp$ (right) at various scales.
Upper (lower) panels correspond to heavy-quark (gluon) channels.
Main-panel bands combine F-MHOU and LDME uncertainties; lower panels display, respectively, F-MHOUs as replica envelopes and LDME effects as ratios to the central curve.
}
\label{fig:FFs-z_TQ0}
\end{figure}

\begin{figure}[!t]
\centering

   \hspace{-0.00cm}
   \includegraphics[scale=0.400,clip]{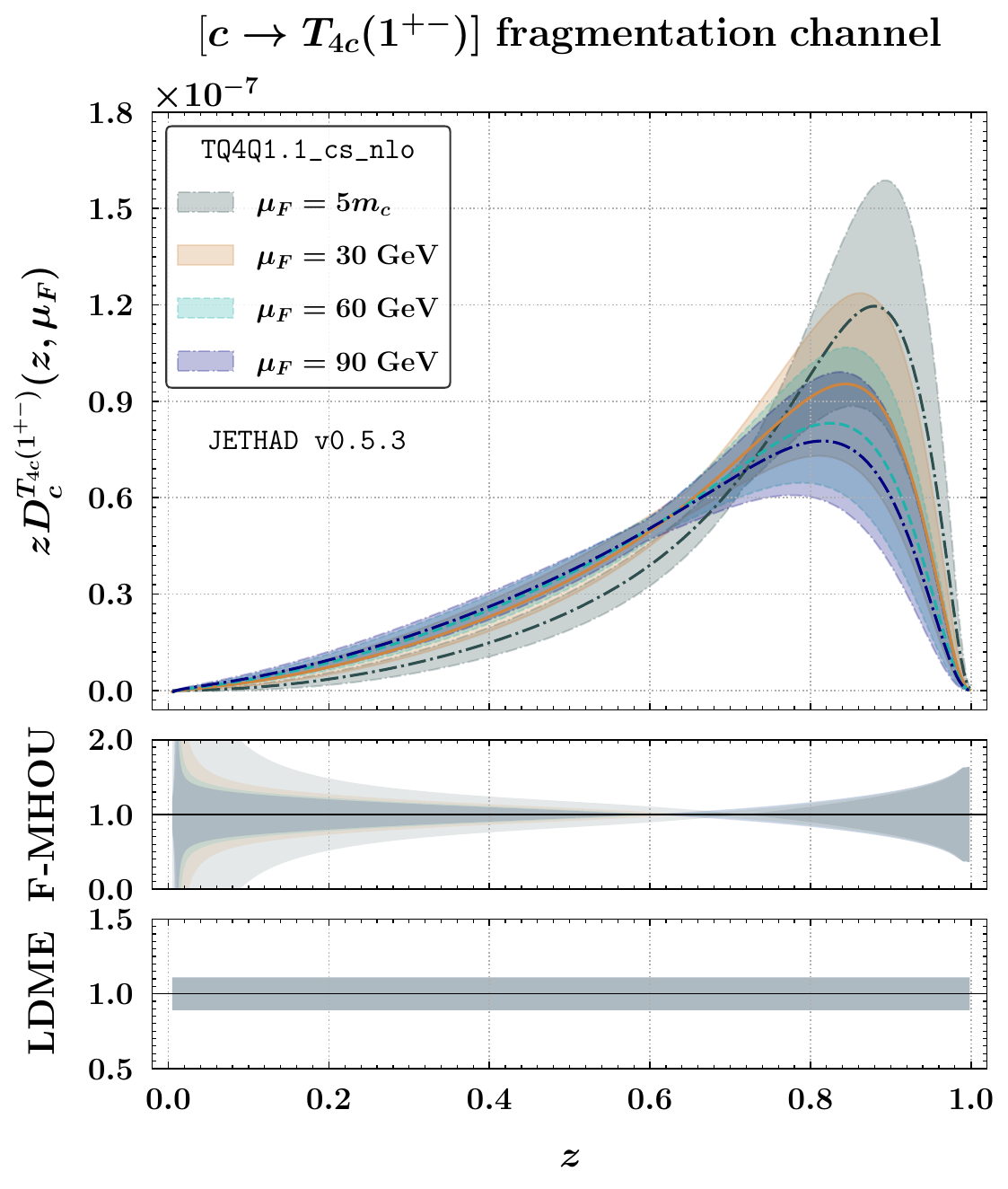}
   \hspace{0.90cm}
   \includegraphics[scale=0.400,clip]{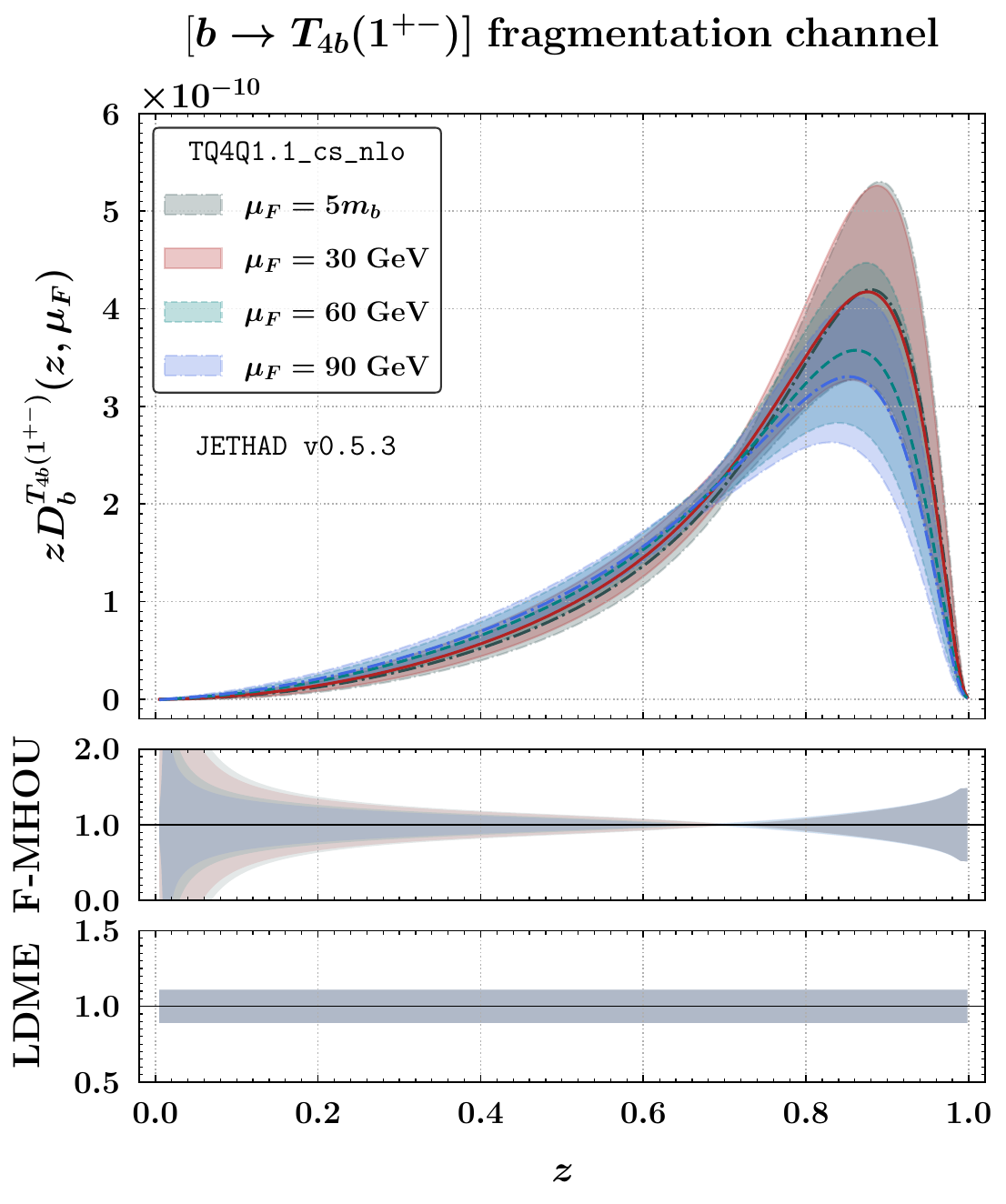}

   \vspace{0.35cm}

   \hspace{-0.00cm}
   \includegraphics[scale=0.400,clip]{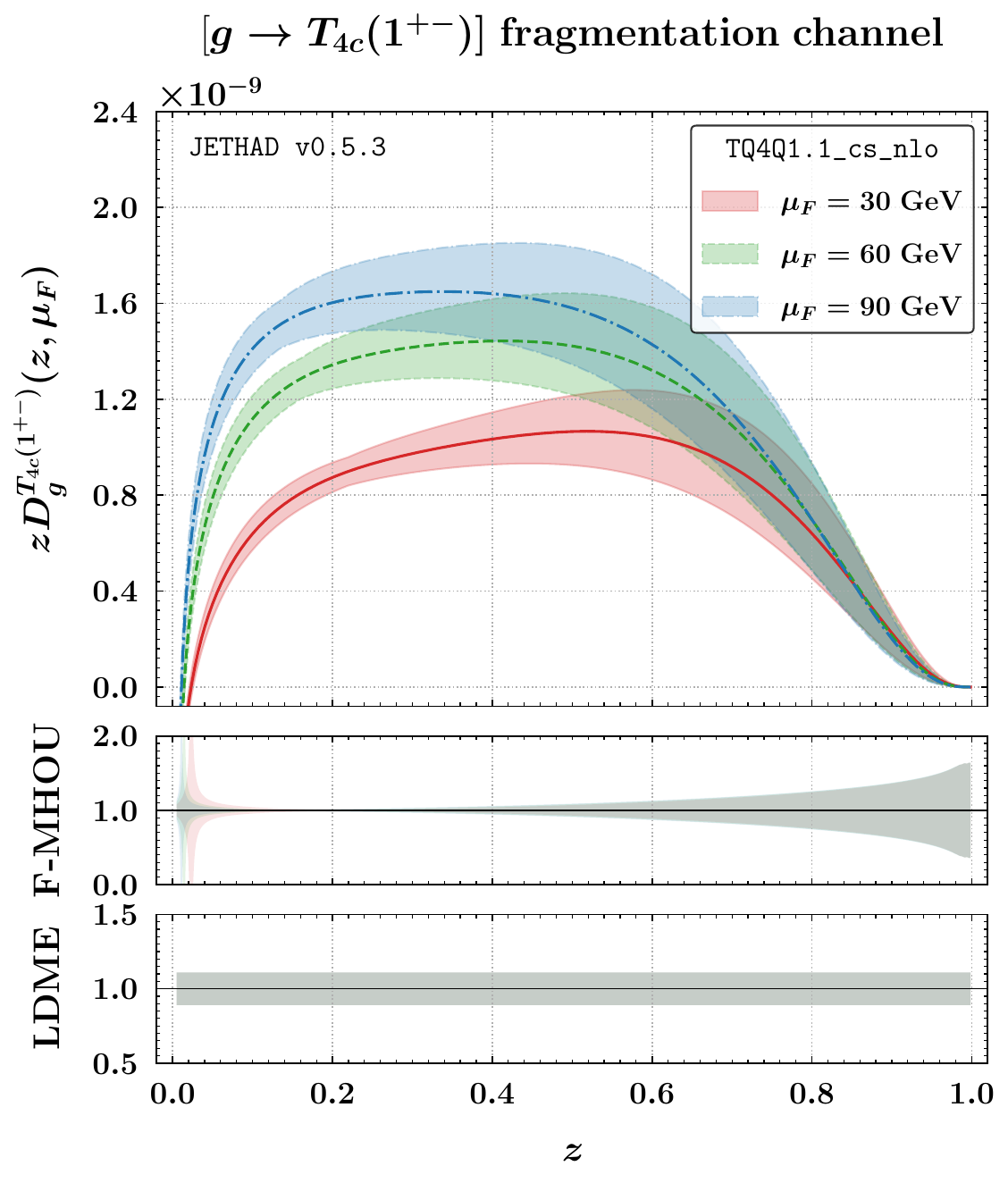}
   \hspace{0.90cm}
   \includegraphics[scale=0.400,clip]{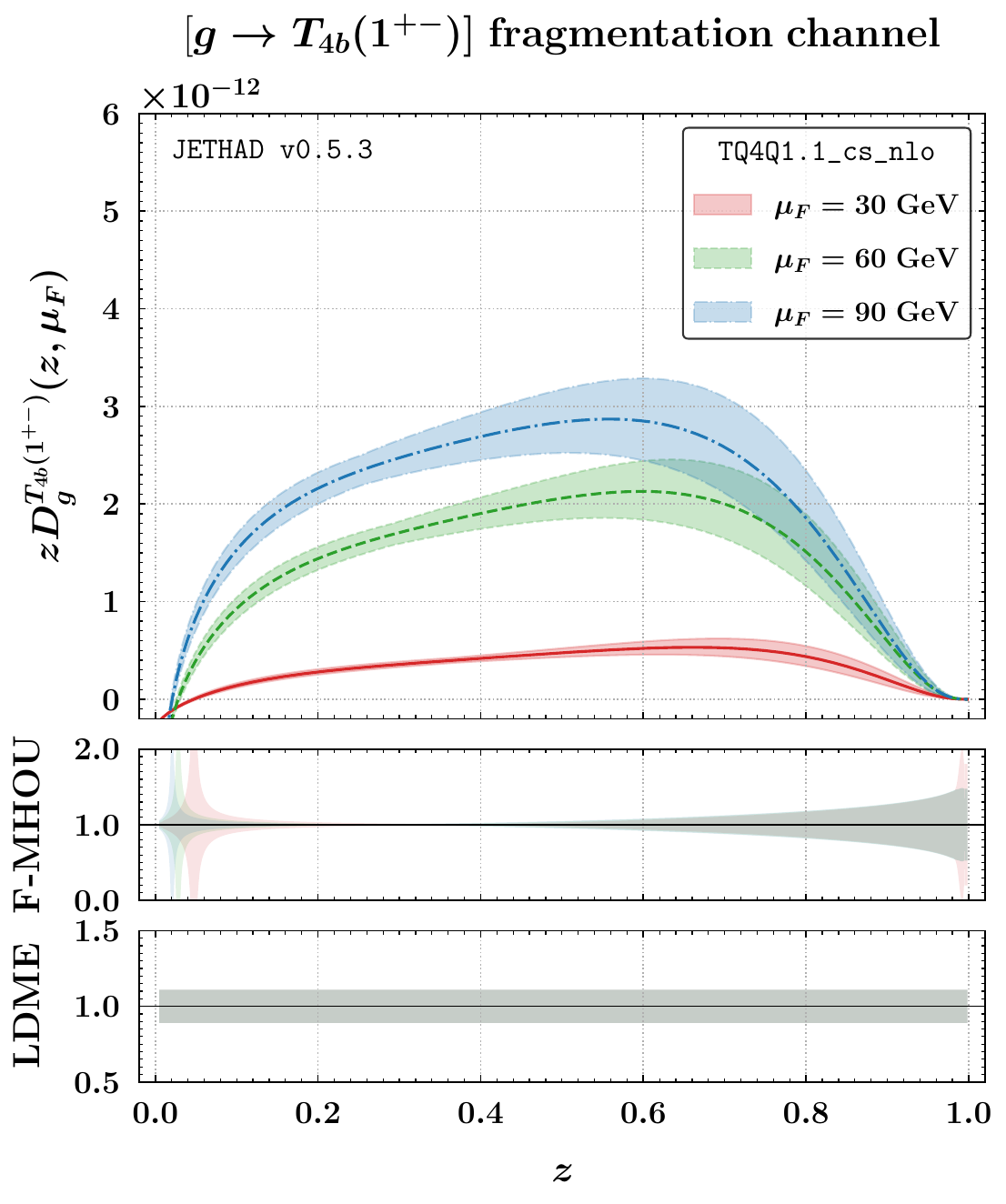}

\caption{
$z$-profiles of the {\tt TQ4Q1.1} FFs for axial-vector tetraquarks $\TQcOpm$ (left) and $\TQbOpm$ (right) at various scales.
Upper (lower) panels correspond to heavy-quark (gluon) channels.
Main-panel bands combine F-MHOU and LDME uncertainties; lower panels display, respectively, F-MHOUs as replica envelopes and LDME effects as ratios to the central curve.
}
\label{fig:FFs-z_TQ1}
\end{figure}

\begin{figure}[!t]
\centering

   \hspace{-0.00cm}
   \includegraphics[scale=0.400,clip]{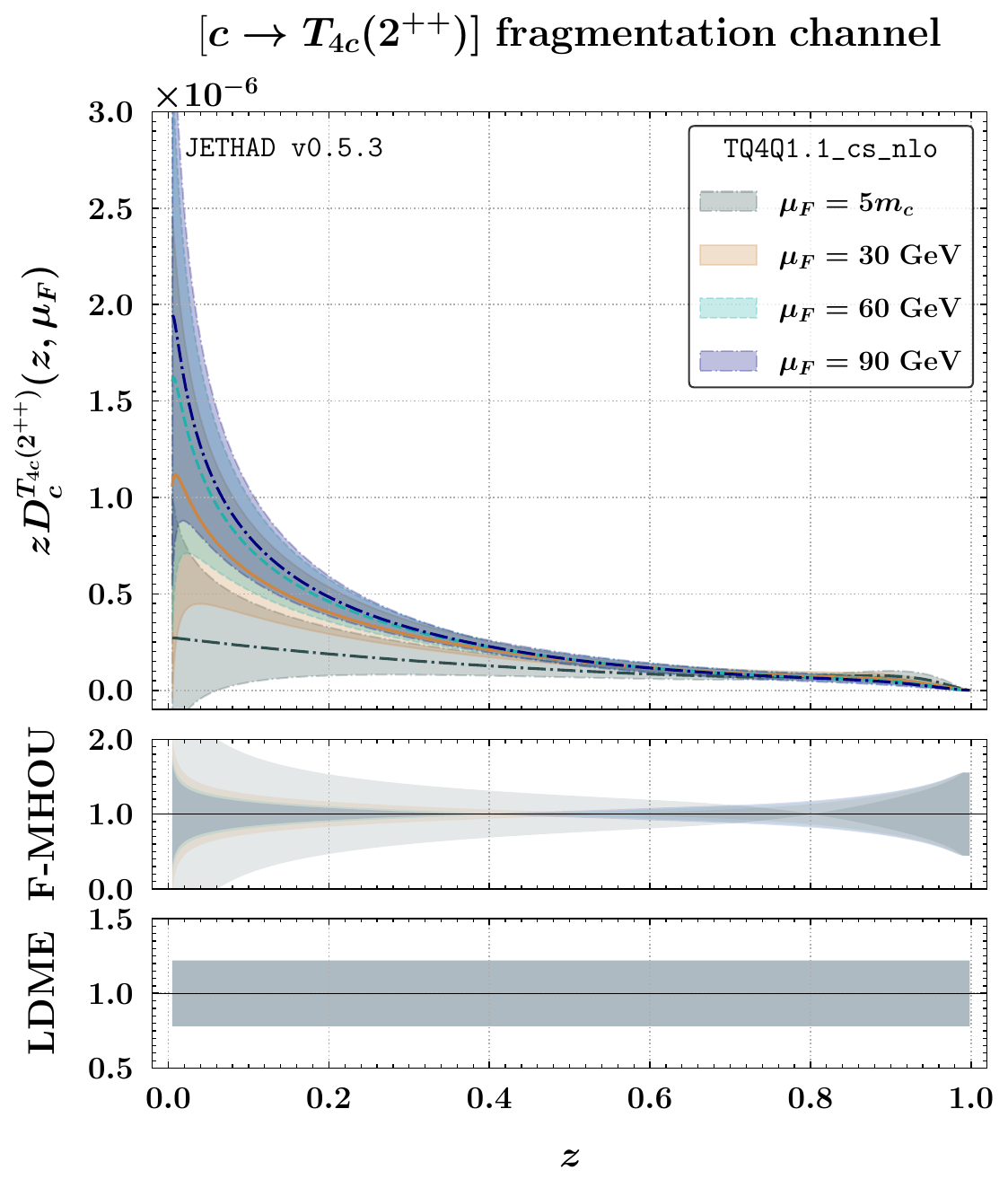}
   \hspace{0.90cm}
   \includegraphics[scale=0.400,clip]{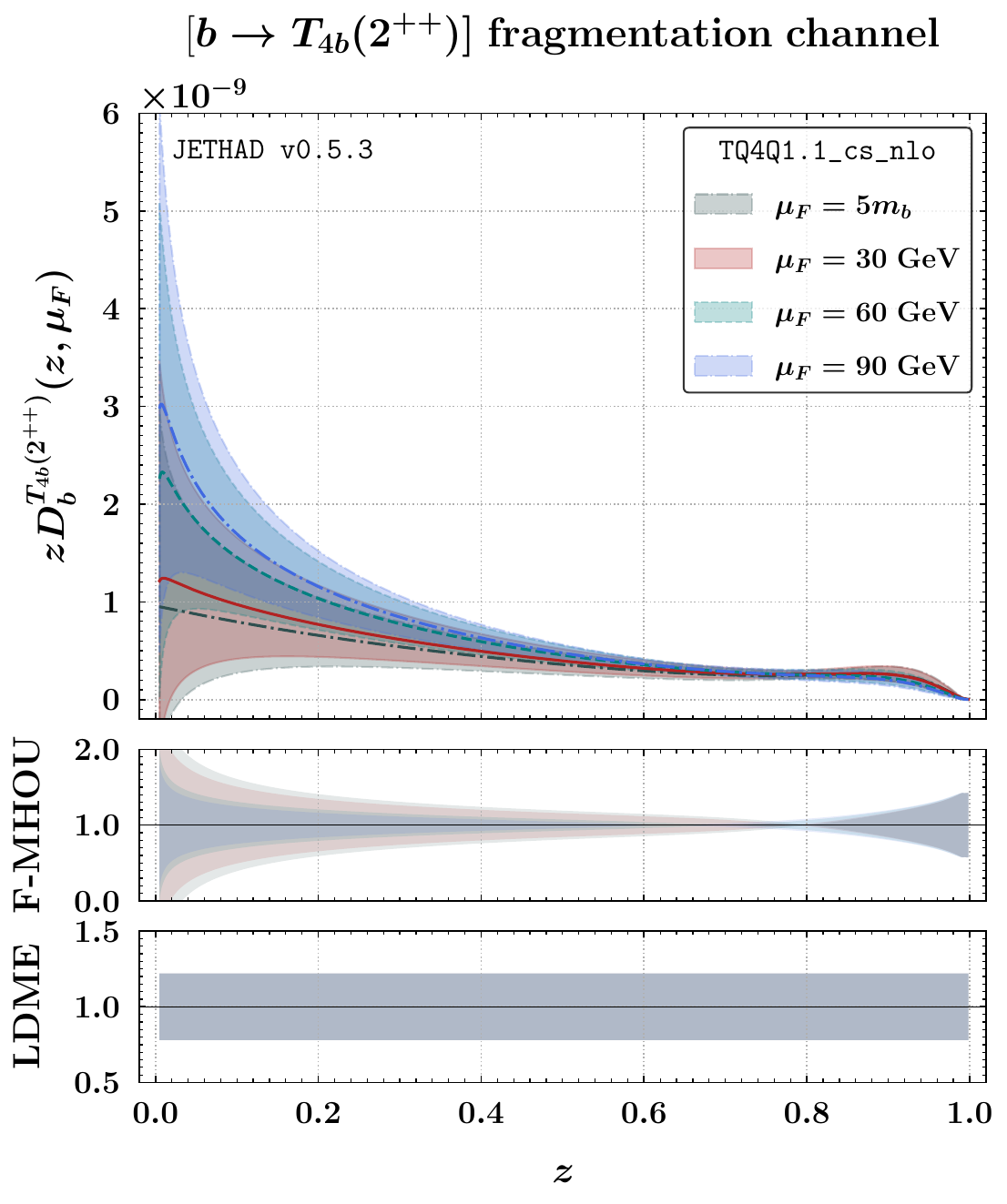}

   \vspace{0.35cm}

   \hspace{-0.00cm}
   \includegraphics[scale=0.400,clip]{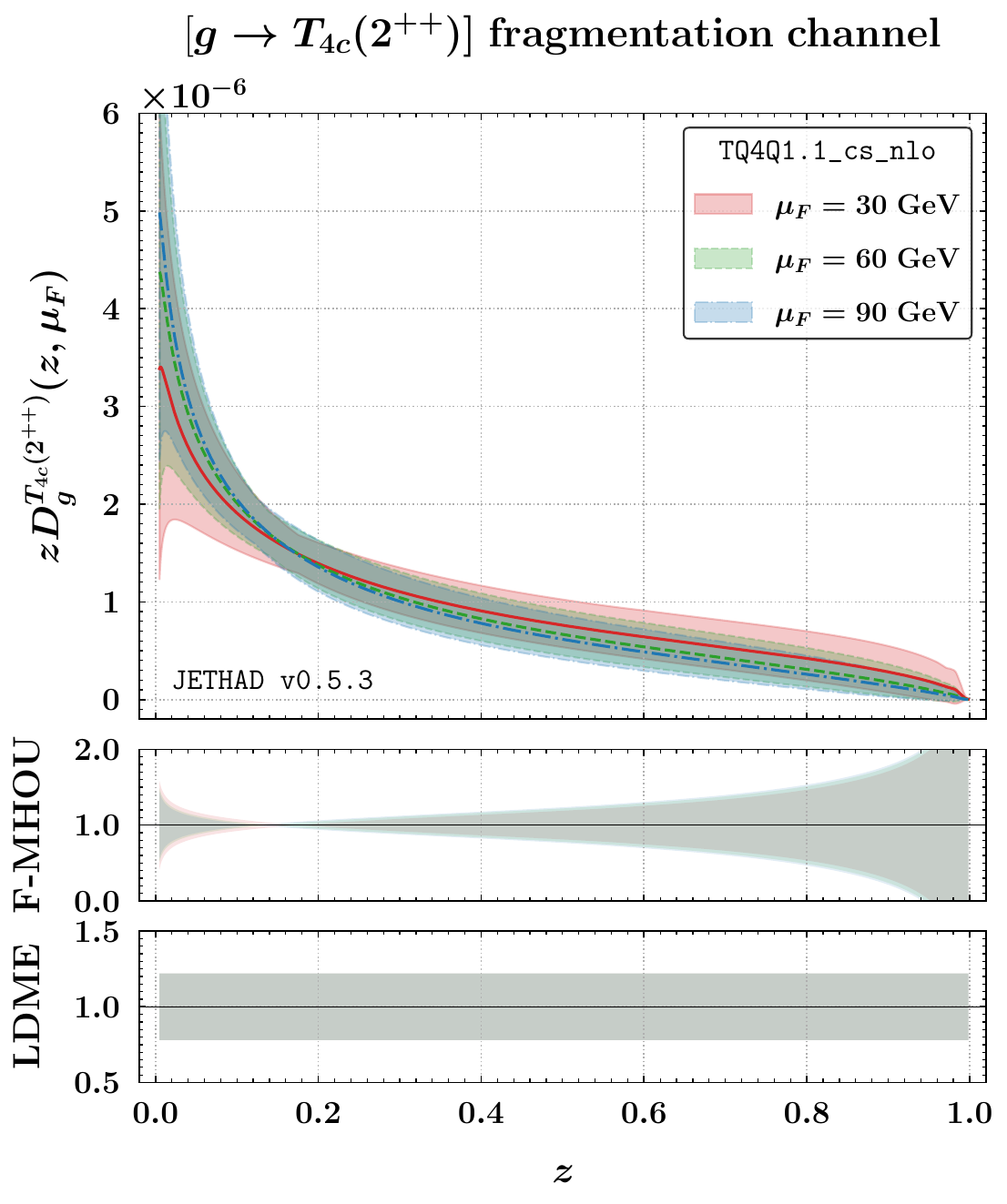}
   \hspace{0.90cm}
   \includegraphics[scale=0.400,clip]{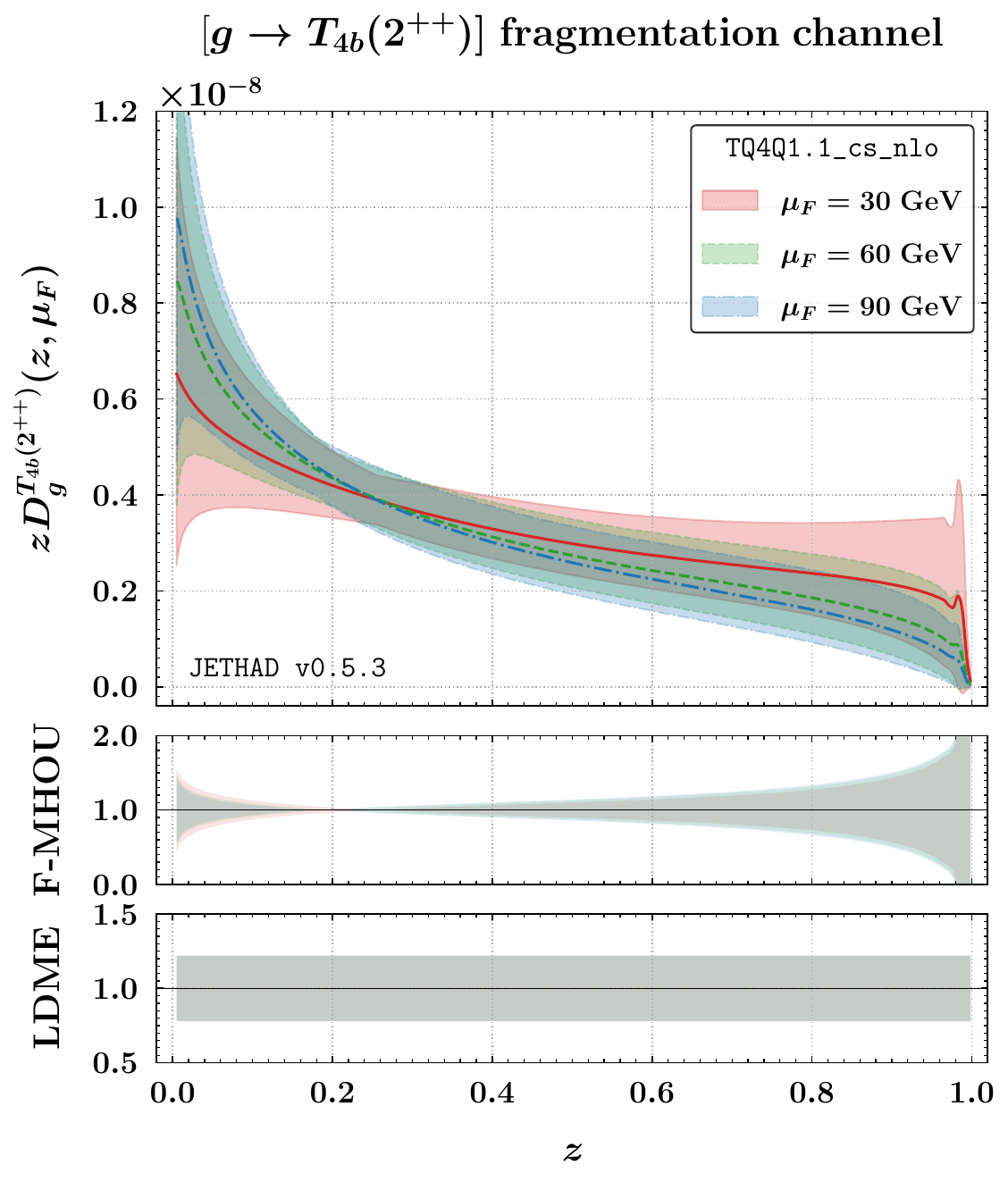}

\caption{
$z$-profiles of the {\tt TQ4Q1.1} FFs for tensor tetraquarks $\TQcTpp$ (left) and $\TQbTpp$ (right) at various scales.
Upper (lower) panels correspond to heavy-quark (gluon) channels.
Main-panel bands combine F-MHOU and LDME uncertainties; lower panels display, respectively, F-MHOUs as replica envelopes and LDME effects as ratios to the central curve.
}
\label{fig:FFs-z_TQ2}
\end{figure}

We next examine the momentum-fraction dependence of the evolved {\tt TQ4Q1.1} sets, focusing on their behavior as a function of $z$ for scalar ($0^{++}$, Fig.~\ref{fig:FFs-z_TQ0}), axial-vector ($1^{+-}$, Fig.~\ref{fig:FFs-z_TQ1}), and tensor ($2^{++}$, Fig.~\ref{fig:FFs-z_TQ2}) tetraquarks. 
Each plot features charm (left) and bottom (right) sectors, with upper and lower panels displaying, respectively, the heavy-quark and gluon fragmentation channels. 
Uncertainty bands in the main panels combine F-MHOUs and LDME variations, while the ancillary panels disentangle the two sources by showing their individual ratios with respect to the central prediction.

Four representative scales are displayed: the evolution-ready threshold $Q_0 = 5 m_Q$ and the evolved values $\mu_F = 30$, $60$, and $90$~GeV. Since the $g \to \TQQ(1^{+-})$ input vanishes by construction, it appears only at $\mu_F > Q_0$, whereas the scalar and tensor gluon FFs are nonzero already at threshold. 
For consistency, all gluon results are shown starting from $30$~GeV upward.

Heavy-quark FFs exhibit distinctive shapes across spin channels. 
In the scalar and tensor cases, they start from finite values at small $z$, decrease rapidly, and develop a secondary enhancement near $z \simeq 0.8 \div 0.9$, slightly shifted to lower $z$ for bottomed states. 
The small-$z$ slope steepens with increasing $\mu_F$, signaling stronger soft-gluon radiation under DGLAP evolution. 
Yet, a residual high-$z$ hump persists, reflecting a hard-fragmentation component characteristic of heavy-flavor dynamics~\cite{Suzuki:1977km,Bjorken:1977md}. 
This dual behavior stems from the coexistence of soft and hard regimes, where the fragmenting quark may retain a large share of the hadron momentum.

Earlier {\tt TQ4Q1.0} determinations~\cite{Celiberto:2024mab} used a purely kinematic model inspired by Suzuki~\cite{Suzuki:1977km}, predicting $\langle z \rangle \!\simeq\! 1-\Lambda_q/m_Q$ for heavy-light hadrons, but lacking justification for multiquark states. 
The new results confirm that even without light constituents, scalar and tensor channels naturally generate large-$z$ peaks, consistent with quarkonium fragmentation~\cite{Braaten:1993mp}. 
This feature therefore emerges dynamically, not kinematically, as a manifestation of the compactness of fully heavy bound states.

Axial-vector FFs differ markedly: they vanish at $z \to 0$, peak sharply at $0.75 \lesssim z \lesssim 0.9$, and remain narrow under evolution, indicating reduced soft-gluon phase space and a strong preference for hard fragmentation. 
In all channels, bottomed FFs are suppressed by roughly three orders of magnitude compared to charmed ones, a mass effect overpowering the larger LDME normalization discussed in Section~\ref{ssec:FFs-inputs}. 
This confirms that the overall normalization is mainly driven by the perturbative coefficients.

Gluon-induced FFs display broader shapes, concentrated at smaller $z$. 
For scalar and tensor states, they decrease monotonically without forming a peak, whereas the axial-vector case develops a wide hump in $0.15 \lesssim z \lesssim 0.7$, whose position drifts upward with $\mu_F$. 
Since this channel is radiatively generated, the hump reflects the intrinsic DGLAP dynamics rather than the initial input. 
The systematic reduction of normalization with increasing quark mass persists also here, confirming the uniform suppression of bottomed configurations.

Comparing different quantum numbers, the axial-vector FFs are consistently smaller than the scalar and tensor ones, by more than one order of magnitude in both $Q$ and $g$ channels. 
This hierarchy follows NRQCD expectations~\cite{Weng:2020jao,Feng:2023agq}, where the antisymmetric spin-color coupling of the $1^{+-}$ state yields weaker overlap with gluon-induced production and smaller LDMEs~\cite{Bodwin:2002cfe,Ma:2015yka,Xu:2021mju}. 
The scalar and tensor cases, instead, benefit from enhanced compatibility with collinear selection rules.

As $\mu_F$ increases, all FFs gradually soften: heavy-quark ones retain their peak with moderate reduction, while gluon distributions broaden at small $z$, particularly in the axial-vector channel. 
The overall deformation remains mild even at $\mu_F=90$~GeV, showing that the essential fragmentation pattern is set by the initial inputs and preserved by evolution. 
Uncertainty bands, combining LDME and F-MHOU effects, stay moderate throughout, consistent with the conservative setup described in Section~\ref{ssec:FFs-inputs}.

\section{Hybrid factorization at $\NLL$ and beyond}
\label{sec:HE-resummation}

The first part of this section offers a concise overview of recent phenomenological advances in high-energy QCD resummation (see Section~\ref{sssec:HE-QCD}).
The second part presents a formal framework for our tetraquark-sensitive observables, formulated within the hybrid high-energy and collinear factorization scheme at $\NLLp$ accuracy (see Section~\ref{ssec:hybrid-factorization}).

\subsection{High-energy resummation in QCD: Status and prospects}
\label{sssec:HE-QCD}

Accurate predictions for high-energy observables depend on our ability to disentangle long-distance dynamics from short-distance effects in hadronic scatterings.
This separation allows nonperturbative and perturbative components to be factorized through the well-established collinear framework~\cite{Collins:1989gx,Sterman:1995fz}.

Certain regions of phase space, however, challenge this framework because large logarithmic corrections emerge.
These logarithms grow order by order in perturbation theory, compensating the smallness of the QCD running coupling and jeopardizing the convergence of the perturbative expansion.
In such regimes, standard collinear factorization must be supplemented by all-order resummations.

In the semi-hard regime of QCD~\cite{Gribov:1983ivg}, characterized by the scale hierarchy $\sqrt{s} \gg Q \gg \LQCD$, logarithms of the form $\ln(s/Q^2)$ appear at successive powers in perturbation theory and therefore require resummation~\cite{Celiberto:2017ius,Bolognino:2021bjd,Mohammed:2022gbk,Gatto:2025kfl}.

The \ac{BFKL} formalism~\cite{Fadin:1975cb,Kuraev:1976ge,Kuraev:1977fs,Balitsky:1978ic} provides the most natural framework for high-energy resummation.
It resums all terms $(\alpha_s \ln s)^n$ at \ac{LL} accuracy and $\alpha_s(\alpha_s \ln s)^n$ at NLL accuracy.

Within BFKL theory, any scattering amplitude can be expressed as a convolution of a universal Green's function with two singly off-shell, transverse-momentum-dependent emission vertices---known as impact factors---that describe the production of a forward particle from each incoming hadron.
In BFKL terminology, these are forward-production impact factors.
The Green's function satisfies an integral evolution equation whose kernel is known at NLO~\cite{Fadin:1998py,Ciafaloni:1998gs,Fadin:1998jv,Fadin:2000kx,Fadin:2000hu,Fadin:2004zq,Fadin:2005zj}, with ongoing progress toward higher orders~\cite{Caola:2021izf,Falcioni:2021dgr,DelDuca:2021vjq,Byrne:2022wzk,Fadin:2023roz,Byrne:2023nqx}.

The predictive power of NLL BFKL resummation remains constrained by the limited number of NLO-calculated off-shell emission functions, including
a) impact factors for incoming partons (quarks and gluons)~\cite{Fadin:1999de,Fadin:1999df}, which enable calculations of
b) forward-jet~\cite{Bartels:2001ge,Bartels:2002yj,Caporale:2011cc,Ivanov:2012ms,Colferai:2015zfa} and
c) forward light-hadron~\cite{Ivanov:2012iv} emissions.

Additional NLO results exist for
d) virtual-photon-to-light-vector-meson transitions~\cite{Ivanov:2004pp},
e) light-by-light scattering~\cite{Bartels:2000gt,Bartels:2001mv,Bartels:2002uz,Bartels:2004bi,Fadin:2001ap,Balitsky:2012bs}, and
f) forward Higgs-boson production, both in the infinite-top-mass limit~\cite{Hentschinski:2020tbi,Celiberto:2022fgx,Hentschinski:2022sko} and at finite top mass~\cite{Celiberto:2024bbv,Celiberto:2025ece} (see Ref.~\cite{DelDuca:2025vux} for first steps toward next-to-NLO accuracy).

At LO, several additional channels have been investigated, such as Drell--Yan pair production~\cite{Hentschinski:2012poz,Motyka:2014lya}, heavy-quark pair production~\cite{Celiberto:2017nyx,Bolognino:2019ccd,Bolognino:2019yls}, and forward $J/\psi$ production at low transverse momentum~\cite{Boussarie:2017oae} (see also Refs.~\cite{Boussarie:2015jar,Boussarie:2016gaq,Boussarie:2017xdy}).

Among the most sensitive probes of high-energy QCD at colliders are the so-called gold-plated channels: semi-inclusive processes particularly responsive to BFKL dynamics, including
Mueller--Navelet dijets at NLO~\cite{Mueller:1986ey,Colferai:2010wu,Ducloue:2013hia,Caporale:2013uva,Colferai:2015zfa,Caporale:2015uva,Ducloue:2015jba,Celiberto:2015yba,Celiberto:2015mpa,Celiberto:2016ygs,Celiberto:2016vva,Caporale:2018qnm,deLeon:2020myv,deLeon:2021ecb,Celiberto:2022gji,Egorov:2023duz,Egorov:2025ize,Baldenegro:2024ndr},
correlated dihadron emissions~\cite{Celiberto:2016hae,Celiberto:2016zgb,Celiberto:2017ptm,Celiberto:2017uae,Celiberto:2017ydk},
multi-jet configurations separated by large rapidity gaps~\cite{Caporale:2015vya,Caporale:2015int,Caporale:2016soq,Caporale:2016vxt,Caporale:2016xku,Celiberto:2016vhn,Caporale:2016djm,Caporale:2016pqe,Chachamis:2016qct,Chachamis:2016lyi,Caporale:2016lnh,Caporale:2016zkc,Caporale:2017jqj,Chachamis:2017vfa},
and more exclusive final states such as hadron-jet~\cite{Bolognino:2018oth,Bolognino:2019cac,Bolognino:2019yqj,Celiberto:2020wpk,Celiberto:2020rxb,Celiberto:2022kxx},
Higgs-jet~\cite{Celiberto:2020tmb,Celiberto:2021fjf,Celiberto:2021tky,Celiberto:2021txb,Celiberto:2021xpm},
heavy-light dijets~\cite{Bolognino:2021mrc,Bolognino:2021hxx}, and
heavy-flavored hadrons~\cite{Boussarie:2017oae,Celiberto:2017nyx,Bolognino:2019ouc,Bolognino:2019yls,Bolognino:2019ccd,Celiberto:2021dzy,Celiberto:2021fdp,Bolognino:2022wgl,Celiberto:2022dyf,Celiberto:2022grc,Bolognino:2022paj,Celiberto:2022qbh,Celiberto:2022keu,Celiberto:2022zdg,Celiberto:2022kza,Celiberto:2024omj,Celiberto:2025euy}.

Processes involving single forward particles provide especially clean probes of BFKL dynamics.
They are directly linked to the behavior of the proton's \ac{UGD} at small $x$, whose energy evolution follows the BFKL Green's function.
Representative examples include light-vector-meson leptoproduction at HERA~\cite{Anikin:2009bf,Anikin:2011sa,Besse:2013muy,Bolognino:2018rhb,Bolognino:2018mlw,Bolognino:2019bko,Bolognino:2019pba,Celiberto:2019slj,Luszczak:2022fkf} and future EIC projections~\cite{Bolognino:2021niq,Bolognino:2021gjm,Bolognino:2022uty,Celiberto:2022fam,Bolognino:2022ndh},
exclusive quarkonium photoproduction~\cite{Bautista:2016xnp,Garcia:2019tne,Hentschinski:2020yfm,Peredo:2023oym,Hentschinski:2025ovo},
inclusive Drell--Yan dileptons~\cite{Motyka:2014lya,Brzeminski:2016lwh,Motyka:2016lta,Celiberto:2018muu}, and
bottom-tagged jet emissions~\cite{Chachamis:2015ona,Chachamis:2013bwa,Chachamis:2009ks}.

The information encoded in the small-$x$ BFKL UGD has been pivotal for improving the theoretical description of collinear \ac{PDFs} incorporating small-$x$ resummation~\cite{Ball:2017otu,Abdolmaleki:2018jln,Bonvini:2019wxf,Silvetti:2022hyc,Silvetti:2023suu,Rinaudo:2024hdb,Celiberto:2025nnq}.
It also bridges model-based studies of twist-two gluon TMDs enhanced at low $x$~\cite{Bacchetta:2020vty,Celiberto:2021zww,Bacchetta:2021oht,Bacchetta:2021lvw,Bacchetta:2021twk,Bacchetta:2022esb,Bacchetta:2022crh,Bacchetta:2022nyv,Celiberto:2022omz,Bacchetta:2023zir,Bacchetta:2024fci,Bacchetta:2024uxb}.

Recent works~\cite{Hentschinski:2021lsh,Mukherjee:2023snp} refined the connection between small-$x$ dynamics and TMD factorization, while other studies~\cite{Boroun:2023goy,Boroun:2023ldq} related the UGD to dipole cross sections in color-glass frameworks.

Inclusive and differential Higgs and heavy-flavor distributions with small-$x$ resummation have also been investigated through the {\Hell} method~\cite{Bonvini:2018ixe,Silvetti:2022hyc}, based on the \ac{ABF} formalism~\cite{Ball:1995vc,Ball:1997vf,Altarelli:2001ji,Altarelli:2003hk,Altarelli:2005ni,Altarelli:2008aj,White:2006yh}.
This approach merges collinear factorization with high-energy resummation~\cite{Catani:1990xk,Catani:1990eg,Collins:1991ty,Catani:1993ww,Catani:1993rn,Catani:1994sq,Ball:2007ra,Caola:2010kv}, providing a complementary perspective on QCD at high energy.

A long-standing issue in BFKL descriptions of Mueller--Navelet jets arises from destabilizing NLL corrections.
Although formally of the same order as LL terms, they often carry opposite sign, leading to instabilities when exploring \ac{MHOUs} through variations of the characteristic energy scales.
Consequently, predictions can become unphysical at large jet rapidity separations, and azimuthal-angle correlations behave anomalously at both small and large rapidities.

Several strategies have been proposed to mitigate these effects.
The \ac{BLM} scale-setting prescription~\cite{Brodsky:1996sg,Brodsky:1997sd,Brodsky:1998kn,Brodsky:2002ka}, adapted to semi-hard reactions~\cite{Caporale:2015uva}, achieves partial stabilization---especially in azimuthal distributions--and modestly improves agreement with data.
However, its effectiveness is limited for observables involving light dihadron or hadron-jet emissions, since the BLM-optimized renormalization scales often exceed the natural ones of the processes~\cite{Celiberto:2017ius,Bolognino:2018oth,Celiberto:2020wpk}.
This leads to a strong suppression of total cross sections and thus poor statistics.

Encouraging signs of improved stability have recently been observed in final states sensitive to Higgs-boson production~\cite{Celiberto:2020tmb,Celiberto:2023rtu,Celiberto:2023uuk,Celiberto:2023eba,Celiberto:2023nym,Celiberto:2023dkr,Celiberto:2023rqp,Celiberto:2024mdt,Celiberto:2024bfu,Celiberto:2025edg}.
This trend first appeared in studies of semi-inclusive $\Lambda_c$ hyperons~\cite{Celiberto:2021dzy} and singly bottom-flavored hadrons~\cite{Celiberto:2021fdp} at the LHC.
Notably, this stability correlates with the characteristic behavior of VFNS collinear FFs, which govern singly heavy-flavored hadron production at high transverse momentum.

Subsequent analyses of vector quarkonia~\cite{Celiberto:2022dyf,Celiberto:2023fzz}, charmed $B$ mesons~\cite{Celiberto:2022keu,Celiberto:2024omj}, heavy tetraquarks~\cite{Celiberto:2023rzw,Celiberto:2024mab,Celiberto:2024mrq,Celiberto:2024beg,Celiberto:2025dfe,Celiberto:2025ziy,Celiberto:2025vra}, and rare $\Omega$ baryons~\cite{Celiberto:2025ogy} confirmed that this so-called \emph{natural stability} of QCD high-energy resummation~\cite{Celiberto:2022grc} is an intrinsic feature of final states sensitive to heavy flavor.

\subsection{Hybrid-factorization studies at $\NLL$ and beyond}
\label{ssec:hybrid-factorization}

\begin{figure}[!ht]
\centering

\includegraphics[width=0.575\textwidth]{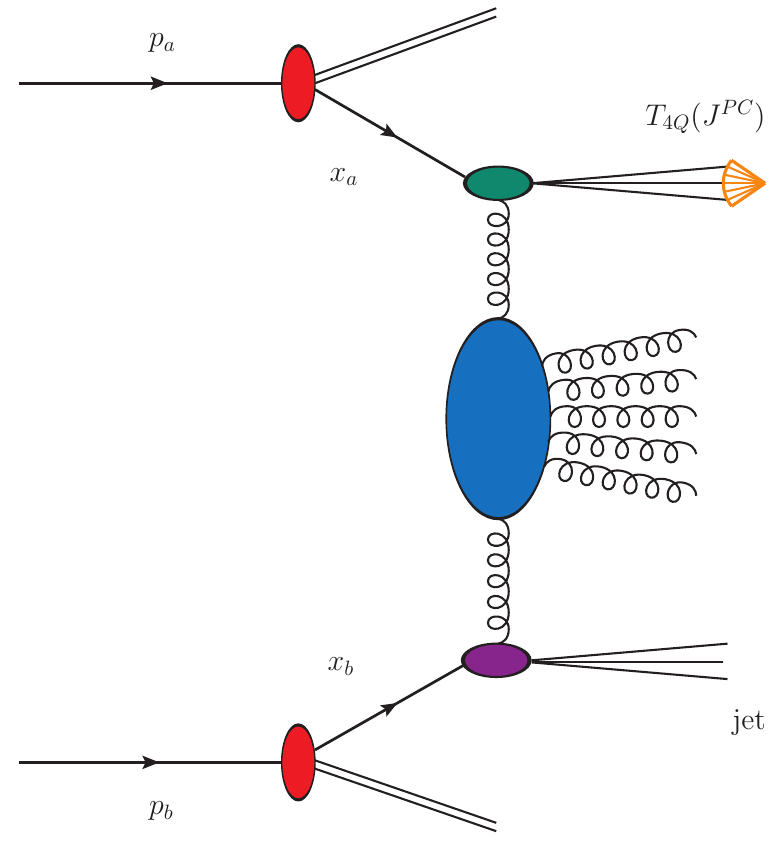}

\caption{Sketch of the semi-inclusive hadroproduction of a tetraquark-jet system within $\NLLp$ factorization (diagram generated with {\tt JaxoDraw 2.0}~\cite{Binosi:2008ig}).
Collinear PDFs are represented by red ovals, while the off-shell emission vertex associated with the tetraquark (jet) is shown as a green (violet) blob. 
Tetraquark emissions are indicated by orange arrows. 
The central blue oval denotes the BFKL Green's function.}
\label{fig:process}
\end{figure}

We focus on the process illustrated in Fig.~\ref{fig:process},
\begin{equation}
\label{process}
\setlength{\jot}{10pt} 
\begin{split}
    {\rm p}(p_a) \;+\; {\rm p}(p_2) &\;\rightarrow\; \TQQ(\vec \kappa_1, y_1) \;+\; {\cal X} \;+\; {\rm jet}(\vec \kappa_2, y_2) \; ,
\end{split}
\end{equation}
where a fully heavy tetraquark $\TQQ$ is produced in association with a light-flavored jet.
Both final-state objects carry large transverse momenta, $|\vec{\kappa}{1,2}| \gg \Lambda{\rm QCD}$, and are separated by a sizable rapidity gap $\DY \equiv y_1 - y_2$.

Such kinematic conditions ensure that the system probes the semi-hard regime, where high-energy resummation effects and collinear fragmentation dynamics coexist.
The transverse-momentum intervals are chosen sufficiently above the heavy-quark mass thresholds so that the VFNS collinear fragmentation becomes the dominant production mechanism for heavy hadrons.

The incoming proton momenta are decomposed along light-like Sudakov vectors, satisfying $p_a^2 = p_b^2 = 0$ and $2(p_a \!\cdot\! p_b) = s$.
Within this convention, the transverse momenta $\vec{\kappa}_{1,2}$ can be expressed as
\begin{equation}
\label{sudakov}
\kappa_{1,2} = x_{1,2} p_{1,2} - \frac{\kappa_{1,2\perp}^{\,2}}{x_{1,2} s}p_{2,1} + \kappa_{1,2\perp} \ , \qquad
\vec \kappa_{1,2}^{\,2} = -\kappa_{1,2\perp}^2\;.
\end{equation}
The longitudinal momentum fractions of the final-state objects, $x_{1,2}$, are related to their rapidities through
$y_{1,2} = \pm \frac{1}{2}\ln\!\left(\frac{x_{1,2}^2\,s}{\vec{\kappa}{1,2}^{\,2}}\right)$.
Differentiating, one obtains $\drv y{1,2} = \pm,\drv x_{1,2}/x_{1,2}$, and therefore
\begin{equation}
\label{DY_kin}
\DY = y_1 - y_2 \equiv \ln\!\left(\frac{x_1 x_2 s}{|\vec{\kappa}_1||\vec{\kappa}_2|}\right).
\end{equation}
These relations connect the measurable rapidity gap to the longitudinal fractions entering the partonic subprocess.

At LO in collinear QCD, the cross section for the process in Eq.~\eqref{process} reduces to a one-dimensional convolution of the proton PDFs, the hadron FFs, and the partonic hard factors.
Explicitly, one writes
\begin{equation}
\label{sigma_collinear}
\begin{split}
 \frac{\drv \sigma_{[p \;+\; p \;\;\to\;\; {\TQQ} \;+\; {\rm jet}]}^{\rm LO}}{\drv x_1 \drv x_2 \drv^2 \vec \kappa_1 \drv^2 \vec \kappa_2}
 &=\sum_{a,b} \int_0^1 \drv x_a \int_0^1 \drv x_b\ 
 f_a(x_a, \mu_F) f_b(x_b, \mu_F)
\int_{x_1}^1 \frac{\drv z}{z}D^{\TQQ}_{a}\left(\frac{x_1}{z}\right) 
\frac{\drv {\hat\sigma}_{a,b}}
{\drv x_1\drv x_2\drv z\,\drv ^2\vec \kappa_1\drv ^2\vec \kappa_2} \;.
\end{split}
\end{equation}
Here, the indices $(a,b)$ run over quarks, antiquarks, and the gluon; $f_{a,b}$ denote the proton PDFs, while $D_{a,b}^{\TQQ}$ are the corresponding FFs into the tetraquark state.
The variables $x_{a,b}$ represent the longitudinal momentum fractions of the incoming partons, $z$ is the fraction carried by the fragmenting parton in the final state, and $\drv\hat{\sigma}_{a,b}$ are the partonic hard-scattering cross sections.

In contrast, the derivation of the resummed cross section in the hybrid factorization framework proceeds in two stages.
First, one applies the high-energy factorization formalism dictated by the BFKL resummation, which systematically accounts for logarithms of the center-of-mass energy.
This is then matched to the collinear picture by incorporating proton PDFs and tetraquark FFs, providing a unified description valid across both regimes.
The resulting differential cross section naturally takes the form of a Fourier expansion in the azimuthal-angle difference, whose coefficients encapsulate the full dynamical content of the process.
Thus we have
\begin{equation}
 \label{dsigma_Fourier}
 \frac{\drv \sigma^\NLLp}{\drv y_1 \drv y_2 \drv \vec \kappa_1 \drv \vec \kappa_2 \drv \phi_1 \drv \phi_2} =
 \frac{1}{(2\pi)^2} \left[{\cal C}_0^\NLLp + 2 \sum_{n=1}^\infty \cos \left(n (\varphi - \pi)\right) \,
 {\cal C}_n^\NLLp \right]\,.
\end{equation}
Here, $\phi_{1,2}$ denote the azimuthal angles of the final-state objects, and $\varphi = \phi_1 - \phi_2$ represents their relative separation.
The corresponding azimuthal coefficients are computed within the BFKL framework and encode the resummation of high-energy logarithms at both LL and NLL accuracy.
All perturbative calculations are performed in the $\MSb$ renormalization scheme~\cite{PhysRevD.18.3998}, which ensures a consistent treatment of ultraviolet singularities and a systematic organization of higher-order corrections.
We then write
\begin{equation}
\label{Cn_NLLp_MSb}
\begin{split}
 \CnNLLp &= \int_0^{2\pi} \drv \varphi_1 \int_0^{2\pi} \drv \varphi_2\,
 \cos \left(n (\varphi - \pi)\right) \,
 \frac{\drv \sigma^\NLLp}{\drv y_1 \drv y_2\, \drv |\vec \kappa_1| \, \drv |\vec \kappa_2| \drv \varphi_1 \drv \varphi_2}\;
\\
 &= \; \frac{e^{\DY}}{s} 
 \int_{-\infty}^{+\infty} \drv \nu \, e^{{\DY} \bar \alpha_s(\mu_R)\chi^\NLO(n,\nu)}
\\
 &\times \; \alpha_s^2(\mu_R) \, 
 \biggl\{
 \R_H^\NLO(n,\nu,|\vec \kappa_1|, x_1)\,
 [\R_J^\NLO(n,\nu,|\vec \kappa_2|,x_2)]^*\,
\\ 
 &+ \,
 \left.
 \bar \alpha_s^2(\mu_R)
 \, \DY
 \frac{\beta_0}{4 N_c}\chi(n,\nu)f(\nu)
 \right\} \;.
\end{split}
\end{equation}
In this expression, $\bar{\alpha}_s(\mu_R) \equiv \alpha_s(\mu_R) N_c / \pi$, where $N_c$ is the number of QCD colors, and $\beta_0 = (11 N_c - 2 n_f)/3$ denotes the leading coefficient of the QCD $\beta$ function, with $n_f$ the number of active quark flavors.
We adopt a two-loop evolution for the strong coupling, initialized at $\alpha_s(M_Z) = 0.118$, and include the effect of a dynamically varying $n_f$ along the evolution.
The resummation kernel entering the exponent of Eq.~\eqref{Cn_NLLp_MSb} reads
\begin{eqnarray}
 \label{chi}
 \chi^\NLO(n,\nu) = \chi(n,\nu) + \bar\alpha_s \hat \chi(n,\nu) \;,
\end{eqnarray}
where
\begin{eqnarray}
 \label{kernel_LO}
 \chi\left(n,\nu\right) = -2\gamma_{\rm E} - 2 \, {\rm Re} \left\{ \psi\left(\frac{1+n}{2} + i \nu \right) \right\} \,. 
\end{eqnarray}
Here, $\chi_0(n,\nu)$ denotes the LO eigenvalue of the BFKL kernel, $\gamma_{\rm E}$ is the Euler--Mascheroni constant, and $\psi(z) \equiv \Gamma^\prime(z)/\Gamma(z)$ represents the logarithmic derivative of the Gamma function.
The function $\hat{\chi}(n,\nu)$ in Eq.~\eqref{chi} accounts for the NLO correction to the kernel,
\begin{equation}
\begin{split}
\label{chi_NLO}
\hat \chi\left(n,\nu\right) &= \bar\chi(n,\nu)+\frac{\beta_0}{8 N_c}\chi(n,\nu)
\left(-\chi(n,\nu)+\frac{10}{3}+2\ln\frac{\mu_R^2}{\mu_C^2}\right) \;,
\end{split}
\end{equation}
The $\TQQ$ mass is set to $m_{\TQQ} = 4 m_Q$, with $m_Q$ the constituent heavy-quark mass.
On the opposite side, the light-flavor jet carries no intrinsic heavy scale, so its transverse mass coincides with its transverse momentum, $|\vec{\kappa}_2|$.
The characteristic function $\bar{\chi}(n,\nu)$ appearing in the BFKL exponent is taken from Refs.~\cite{Kotikov:2000pm,Kotikov:2002ab} and reads
\begin{equation}
 \label{kernel_NLO}
 \bar \chi(n,\nu)\,=\, - \frac{1}{4}\left\{\frac{\pi^2 - 4}{3}\chi(n,\nu) - 6\zeta(3) - \frac{\drv^2 \chi}{\drv\nu^2} + \,2\,\Phi(n,\nu) + \,2\,\Phi(n,-\nu)
 \right.
\end{equation}
\[
 \left.
 +\; \frac{\pi^2\sinh(\pi\nu)}{2\,\nu\, \cosh^2(\pi\nu)}
 \left[
 \left(3+\left(1+\frac{n_f}{N_c^3}\right)\frac{11+12\nu^2}{16(1+\nu^2)}\right)
 \delta_{n0}
 -\left(1+\frac{n_f}{N_c^3}\right)\frac{1+4\nu^2}{32(1+\nu^2)}\delta_{n2}
\right]\right\} \, ,
\]
with
\begin{equation}
\label{kernel_NLO_phi}
 \Phi(n,\nu)\,=\,-\int_0^1 \drv x\,\frac{x^{-1/2+i\nu+n/2}}{1+x}\left\{\frac{1}{2}\left(\psi^\prime\left(\frac{n+1}{2}\right)-\zeta(2)\right)+\mbox{Li}_2(x)+\mbox{Li}_2(-x)\right.
\end{equation}
\[
\left.
 +\; \ln x\left[\psi(n+1)-\psi(1)+\ln(1+x)+\sum_{k=1}^\infty\frac{(-x)^k}{k+n}\right]+\sum_{k=1}^\infty\frac{x^k}{(k+n)^2}\left[1-(-1)^k\right]\right\}
\]
\[
 =\; \sum_{k=0}^\infty\frac{(-1)^{k+1}}{k+(n+1)/2+i\nu}\left\{\psi^\prime(k+n+1)-\psi^\prime(k+1)\right.
\]
\[
 \left.
 +\; (-1)^{k+1}\left[\beta_{\psi}(k+n+1)+\beta_{\psi}(k+1)\right]-\frac{\psi(k+n+1)-\psi(k+1)}{k+(n+1)/2+i\nu}\right\} \; ,
\]
where
\begin{equation}
\label{kernel_NLO_phi_beta_psi_dilog}
 \beta_{\psi}(z)=\frac{1}{4}\left[\psi^\prime\left(\frac{z+1}{2}\right)
 -\psi^\prime\left(\frac{z}{2}\right)\right] \;,
\qquad
\mbox{Li}_2(x) = \int^x_0 \drv y \,\frac{\ln(1-y)}{y} \; .
\end{equation}
The corresponding expressions for the singly off-shell emission functions are given by
\begin{equation}
\label{IFs}
\R_{H,J}^\NLO(n,\nu,|\vec \kappa_{1,2}|,x_{1,2}) =
\R_{H,J}(n,\nu,|\vec \kappa_{1,2}|,x_{1,2}) +
\alpha_s(\mu_R) \, \hat \R_{H,J}(n,\nu,|\vec \kappa_{1,2}|,x_{1,2}) \;.
\end{equation}
At LO, the functions describing the production of a forward hadron and a forward jet are given by
\begin{equation}
\label{LOHEF}
\begin{split}
\R_H(n,\nu,|\vec \kappa_1|,x_1) 
&= 2 \, \sqrt{\frac{C_F}{C_A}}
|\vec \kappa_1|^{2i\nu-1}\,\int_{x_1}^1\frac{\drv z}{z}
\left(\frac{z}{x_1} \right)
^{2 i\nu-1} 
 \left[\frac{C_A}{C_F}f_g(z)D_g^h\left(\frac{x_1}{z}\right)
 +\sum_{a=q,\bar q}f_a(z)D_a^h\left(\frac{x_1}{z}\right)\right] 
\end{split}
\end{equation}
and
\begin{equation}
 \label{LOJEF}
 \R_J(n,\nu,|\vec \kappa_2|,x_2) =  2 \sqrt{\frac{C_F}{C_A}}
 |\vec \kappa_2|^{2i\nu-1}\,\left[\frac{C_A}{C_F}f_g(x_2)
 +\sum_{b=q,\bar q}f_b(x_2)\right] \;,
\end{equation}
with $C_F \equiv (N_c^2-1)/(2N_c)$ being and $C_A \equiv N_c$ the QCD Casimir terms.
The $f(\nu)$ function embodies the logarithmic derivative of the LO impact factors
\begin{equation}
 f(\nu) = \frac{i}{2} \, \frac{\drv}{\drv \nu} \ln\left(\frac{\R_H}{\R_J^*}\right) + \ln\left(|\vec \kappa_1| |\vec \kappa_2|\right) \;.
\label{fnu}
\end{equation}

In Eq.~\eqref{Cn_NLLp_MSb}, the quantities $\hat \R_{H,J}$ represent the NLO corrections to the emission functions.
The forward-hadron contribution was computed in Ref.\cite{Ivanov:2012iv}, while for the forward jet we follow the approach of Refs.~\cite{Ivanov:2012iv,Ivanov:2012ms}.
For numerical implementation, we employ a jet selection function\footnote{Jet algorithms fall into two main classes: \emph{cone-type} and \emph{sequential-clustering} algorithms. See Refs.~\cite{Chekanov:2002rq,Salam:2010nqg} for a comprehensive overview. A notable example of the latter is the (anti-)$\kappa_\perp$ algorithm~\cite{Catani:1993hr,Cacciari:2008gp}.} evaluated in the \ac{SCA}\cite{Furman:1981kf,Aversa:1988vb}, adopting its cone-type version\cite{Colferai:2015zfa} with jet radius ${\cal R}=0.5$.

A full phenomenological comparison between our hybrid factorization predictions and fixed-order results would require a dedicated code for NLO two-particle production in hadronic collisions, which is presently unavailable.
As a practical alternative, we emulate the fixed-order behavior by truncating the azimuthal-coefficient expansion in Eq.~\eqref{Cn_NLLp_MSb} at ${\cal O}(\alpha_s^3)$, thus defining an effective high-energy fixed-order approximation ($\HENLOp$).
This prescription retains the leading-power asymptotic structure of NLO QCD while neglecting subleading terms suppressed by inverse powers of the partonic energy.
The $\MSb$ expression for the azimuthal coefficients at $\HENLOp$ reads
\begin{equation}
\label{Cn_HENLO_MSb}
 \CnHENLOp =
 \frac{e^{\DY}}{s}
 \int_{-\infty}^{+\infty} \drv \nu \,
 \alpha_s^2(\mu_R) \,
 \left[ 1 + \bar \alpha_s(\mu_R) \DY \chi(n,\nu) \right] \,
 \R_H^\NLO(n,\nu,|\vec \kappa_1|, x_1) \,[\R_J^\NLO(n,\nu,|\vec \kappa_2|,x_2)]^*
 \;,
\end{equation}
with the BFKL kernel being expanded and truncated at ${\cal O}(\alpha_s)$.
For the sake of comparison, we will also discuss predictions taken at LL:
\begin{equation}
\label{Cn_LL_MSb}
  \CnLL = \frac{e^{\DY}}{s} 
 \int_{-\infty}^{+\infty} \drv \nu \, e^{{\DY} \bar \alpha_s(\mu_R)\chi(n,\nu)} \, \alpha_s^2(\mu_R) \, \R_H(n,\nu,|\vec \kappa_1|, x_1)[\R_J(n,\nu,|\vec \kappa_2|,x_2)]^* \,.
\end{equation}

Equations~\eqref{Cn_NLLp_MSb} to~\eqref{Cn_LL_MSb} summarize the structure of our hybrid factorization framework.
Following the BFKL formalism, the hadronic cross section is written as a convolution in transverse momentum between the Green's function and two off-shell emission functions (impact factors).
These impact factors embed collinear PDFs and FFs, thereby merging high-energy resummation with collinear QCD dynamics in a consistent way.

The $\NLLp$ tag denotes full next-to-leading logarithmic resummation, built from NLO perturbative components, while the `$+$' superscript specifies that cross-terms from the two NLO impact-factor corrections are also included.

Renormalization and factorization scales follow the natural kinematic choice
\begin{equation}
\label{natural scales}
\mu_R = \mu_F = \mu_N = m_{1\perp} + m_{2\perp} \;,
\end{equation}
with $m_{i\perp}$ the transverse mass of each final-state particle.

Collinear PDFs are taken from the {\tt NNPDF4.0} NLO set~\cite{NNPDF:2021uiq,NNPDF:2021njg}, accessed through {\tt LHAPDF}~\cite{Buckley:2014ana}.
These distributions rely on global fits based on the replica method~\cite{Forte:2002fg}, now a standard in multidimensional proton-structure studies~\cite{Bacchetta:2017gcc,Scimemi:2019cmh,Bacchetta:2019sam,Bacchetta:2022awv,Bury:2022czx,Moos:2023yfa}; for a detailed comparison of inter-set correlations and uncertainty estimates, see Ref.~\cite{Ball:2021dab}.

All predictions in this review are consistently evaluated in the $\MSb$ renormalization scheme~\cite{PhysRevD.18.3998}.

\section{Tetraquark-jet emissions at the High-Luminosity LHC}
\label{sec:phenomenology}

Section~\ref{ssec:uncertainty} introduces the methodology adopted for a systematic evaluation of theoretical uncertainties affecting our predictions.
In Section~\ref{ssec:observables}, we define the observables under consideration and specify the kinematic domains and final-state selections consistent with realistic LHC detector acceptances.
Numerical results for rapidity spectra and azimuthal multiplicities are discussed in Sections~\ref{ssec:I} and~\ref{ssec:I-phi}, respectively.
All computations have been carried out using the {\Jethad} framework~\cite{Celiberto:2020wpk,Celiberto:2022rfj,Celiberto:2023fzz,Celiberto:2024mrq,Celiberto:2024swu,Celiberto:2025_P5Q_review,Celiberto:2025csa}, a hybrid \textsc{Python}/\textsc{Fortran} multimodular environment developed for the calculation, management, and post-processing of collider observables across diverse QCD frameworks.

\subsection{Uncertainties}
\label{ssec:uncertainty}

A robust phenomenological analysis demands a systematic and transparent treatment of theoretical uncertainties.
In this work, we disentangle the leading sources of error that affect our framework---both perturbative and nonperturbative---and quantify their individual and combined impact on collider-level observables.
This structured breakdown provides a clear picture of how each contribution shapes the final predictions.
Specifically, we account for the following effects:

\begin{itemize}

\item[a)]
\textbf{Perturbative H-MHOUs}.
These originate from the arbitrariness in choosing the renormalization and factorization scales that enter the hard matrix element of the partonic subprocess.
Their variation by factors of $1/2$ and $2$ around the central scales provides a standard estimate of missing higher-order corrections.

\item[b)]
\textbf{Perturbative F-MHOUs}.
These uncertainties trace back to the perturbative initial conditions of the FFs at the reference scale.
The \emph{evolution-ready} scale $Q_0$ is varied around its natural value, $5m_Q$, by a factor of $1/2$ to $2$, and the resulting envelope defines the corresponding uncertainty band.
This variation captures the effect of subleading terms not explicitly included in the evolution of the FFs.

\item[c)]
\textbf{Nonperturbative LDMEs}.
These parameters embody the long-distance dynamics governing hadron formation.
Their uncertainties are evaluated by scanning the relevant LDMEs within ranges supported by potential-model analyses.
The resulting bands quantify how model-dependent hadronization effects influence collider-level observables.

\item[d)]
\textbf{Proton PDFs}.
Since collinear PDFs are genuinely nonperturbative inputs extracted from global fits, they introduce an additional source of uncertainty.
Dedicated numerical tests for tetraquark-jet production, however, indicate that differences among PDF sets or replicas remain below the $1\%$ level.
Therefore, we adopt the central member of the {\tt NNPDF4.0} set~\cite{NNPDF:2021uiq,NNPDF:2021njg} and neglect the broader PDF-fit uncertainty, which is subdominant compared to MHOUs and LDME variations.

\item[e)]
\textbf{Phase-space numerical integration}.
The leading numerical uncertainty arises from multidimensional integrations over the final-state phase space.
These are performed using the native routines of {\Jethad}, with relative errors systematically kept below the $1\%$ level.
Subdominant effects originate from one-dimensional integrations over the partonic longitudinal momentum fractions $x$, which enter the convolution of PDFs and FFs in the LO and NLO tetraquark emission functions (see Eq.~\eqref{LOHEF}).
Extensive stability tests confirm that these contributions are negligible with respect to the multidimensional-integration uncertainty.

\end{itemize}

\subsection{Final-state observables}
\label{ssec:observables}

The first observable examined in our analysis is the rapidity distribution, expressed as the cross section differential in the rapidity distance $\DY = y_1 - y_2$ between the two final-state particles.  
It reads
\begin{equation}
\label{DY_distribution}
 \frac{\drv \sigma(\DY, s)}{\drv \DY} \, = \, C_0 \;,
\end{equation}
where $C_{n=0}$ denotes the azimuthal-angle averaged coefficient integrated over the final-state rapidity and momentum phase space at fixed $\DY$.  
Selecting the $n=0$ conformal spin removes all angular modulations, isolating the leading energy-dependent component of the cross section---the term most sensitive to resummation dynamics.
Rapidity-interval distributions provide a clean probe of the interplay between high-energy evolution and the collinear structure of hadrons, especially in semi-inclusive topologies involving a fully heavy tetraquark and a recoiling jet.

The kinematic selections mimic the acceptance and reconstruction thresholds of the CMS detector at the LHC.  
We require $|y_1| < 2.5$ for the tetraquark, corresponding to the barrel calorimeter coverage~\cite{Chatrchyan:2012xg}, and $|y_2| < 4.7$ for the jet, matching the end-cap region~\cite{Khachatryan:2016udy}.  
The transverse momentum of the $\TQQ(J^{PC})$ state is taken in the range $30$--$120$~GeV, while the associated jet spans $40$--$120$~GeV, in line with LHC analyses involving hadronic and jet observables~\cite{Khachatryan:2016udy,Khachatryan:2020mpd}.  
The use of asymmetric cuts for the observed transverse momenta enhances the visibility of high-energy resummation effects relative to fixed-order predictions~\cite{Celiberto:2015yba,Celiberto:2015mpa,Celiberto:2020wpk}.

The second key observable explored in our analysis is the azimuthal distribution.  
We will present a detailed study of this quantity for fully heavy tetraquark states.
The normalized angular multiplicities are defined as
\begin{equation}
\label{angular_multiplicity}
\frac{1}{\sigma} \frac{\drv \sigma(\varphi, s)}{\drv \varphi} = \frac{1}{\pi} \left[ \frac{1}{2} + \sum_{n=1}^\infty
\langle \cos(n \varphi) \rangle \, \cos (n \varphi) \right] \;,
\end{equation}
where $\varphi = \phi_1 - \phi_2 - \pi$, with $\phi_{1,2}$ the azimuthal angles of the two outgoing objects.  
The coefficients $C_n$ are integrated over the full rapidity and transverse-momentum phase space, and binned in the rapidity interval $\DY$. 
The $\langle \cos(n\varphi) \rangle = C_n/C_0$ moments quantify the azimuthal correlations of the system.

Initially introduced for light dijet studies~\cite{Marquet:2007xx,Ducloue:2013hia}, these observables have become sensitive probes of QCD dynamics in the high-energy regime.  
They encode the full azimuthal structure of partonic interactions and provide a direct measure of angular decorrelations driven by energy evolution.

Their differential dependence on $\varphi$ also makes them experimentally accessible, allowing robust comparison with data even under nonuniform azimuthal acceptance across $2\pi$.  
Recent analyses of dijet multiplicities~\cite{Celiberto:2022gji} demonstrated two main benefits: they mitigate the residual instabilities affecting resummed predictions for light final states~\cite{Bolognino:2018oth,Celiberto:2020wpk}, and they improve the consistency with CMS measurements at $\sqrt{s}=7$~TeV~\cite{Khachatryan:2016udy}.

For simplicity, we will focus on the fully charmed tensor resonance $\TQcTpp$, identified as the most plausible candidate for the $X(6900)$~\cite{LHCb:2020bwg}.  
However, numerical checks confirm that the corresponding fully bottomed state $\TQbTpp$ exhibits similar behavior within uncertainties.

\subsection{Rapidity-differential rates}
\label{ssec:I}

\begin{figure}[!t]
\centering

   \hspace{0.00cm}
   \includegraphics[scale=0.385,clip]{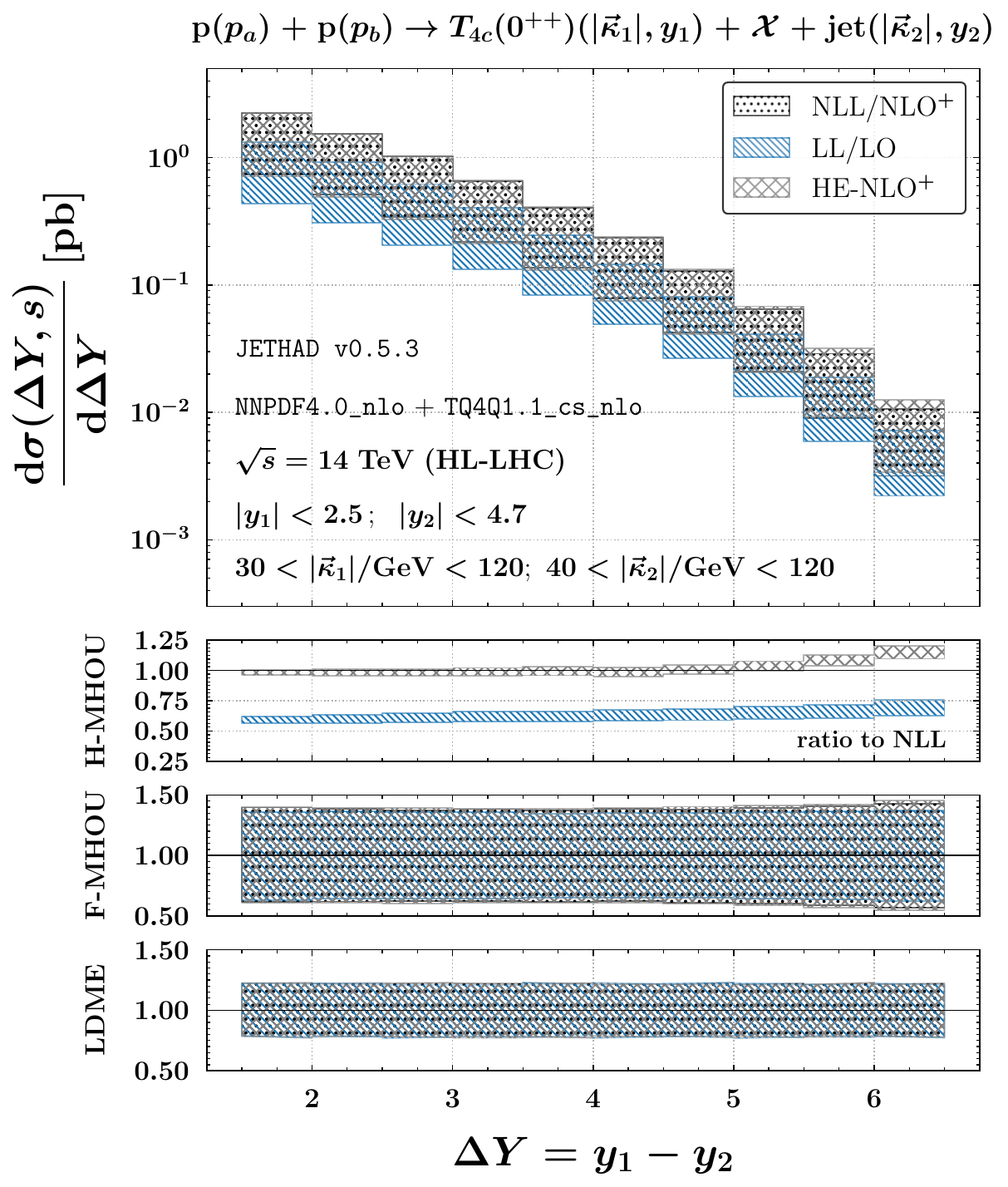}
   \hspace{-0.00cm}
   \includegraphics[scale=0.385,clip]{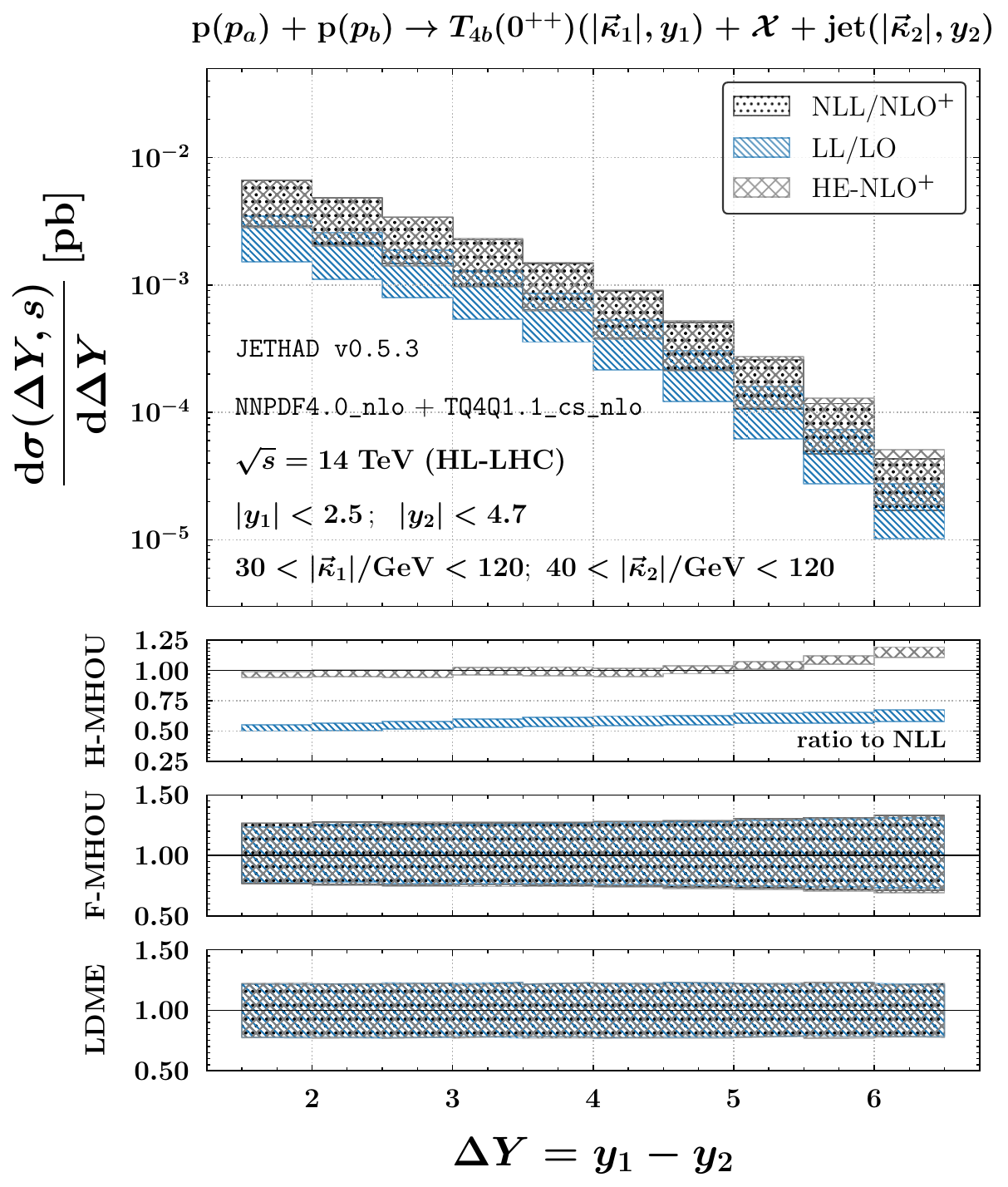}

\caption{
Rapidity spectrum of scalar tetraquarks $\TQcZpp$ (left) and $\TQbZpp$ (right) produced with a jet at $\sqrt{s}=14$ TeV.
Main panels show total uncertainties from combined H-MHOUs, F-MHOUs, LDME, and phase-space effects.
Ancillary panels display: a) $\LL$ and $\HENLOp$ ratios to the $\NLLp$ baseline, b) F-MHOUs envelopes, and c) LDME-induced variations.
}
\label{fig:I_TQ0}
\end{figure}

\begin{figure}[t]
\centering

   \hspace{0.00cm}
   \includegraphics[scale=0.385,clip]{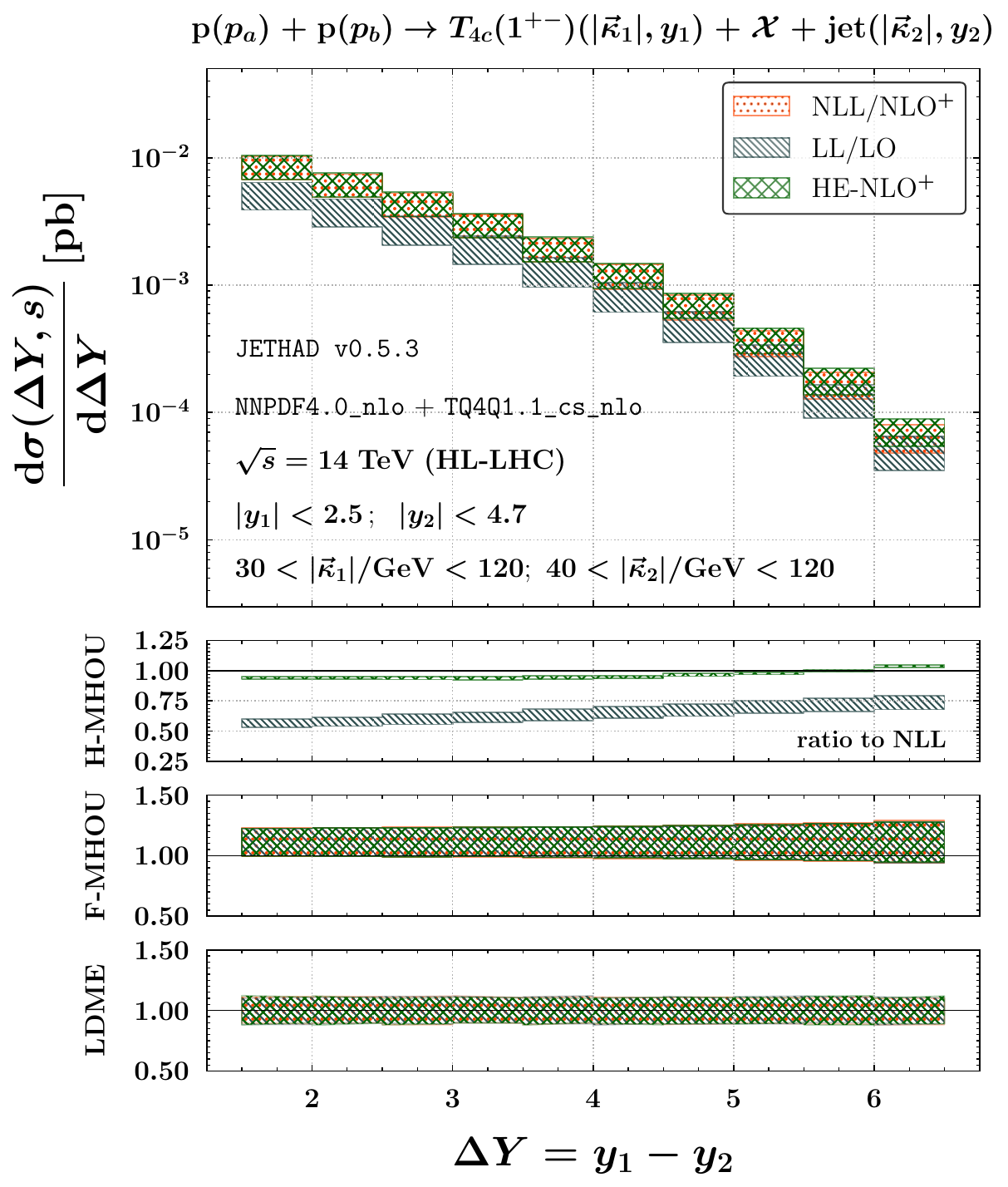}
   \hspace{-0.00cm}
   \includegraphics[scale=0.385,clip]{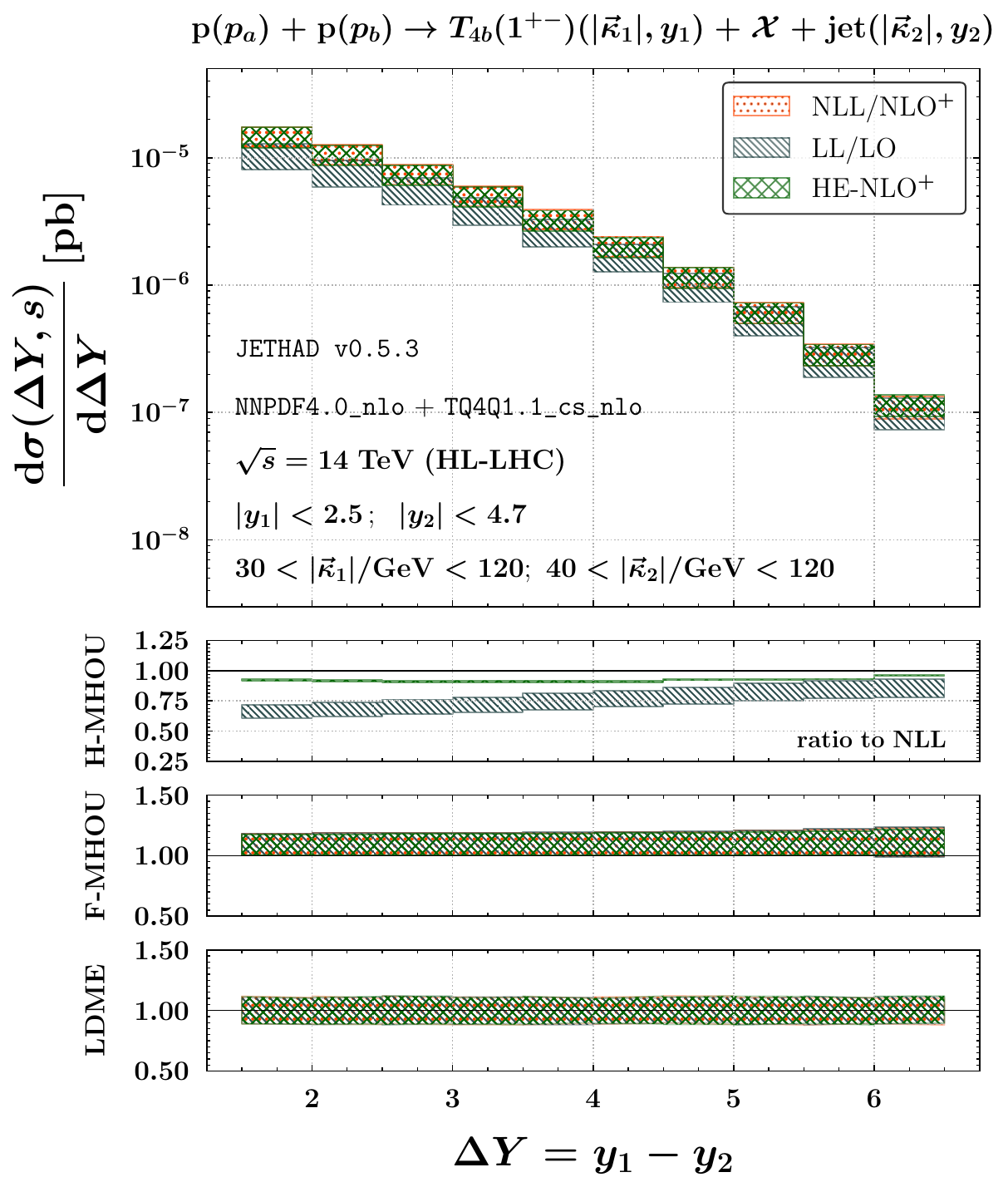}

\caption{
Rapidity spectrum of axial-vector tetraquarks $\TQcOpm$ (left) and $\TQbOpm$ (right) produced with a jet at $\sqrt{s}=14$ TeV.
Main panels show total uncertainties from combined H-MHOUs, F-MHOUs, LDME, and phase-space effects.
Ancillary panels display: a) $\LL$ and $\HENLOp$ ratios to the $\NLLp$ baseline, b) F-MHOUs envelopes, and c) LDME-induced variations.
}
\label{fig:I_TQ1}
\end{figure}

\begin{figure}[!t]
\centering

   \hspace{0.00cm}
   \includegraphics[scale=0.385,clip]{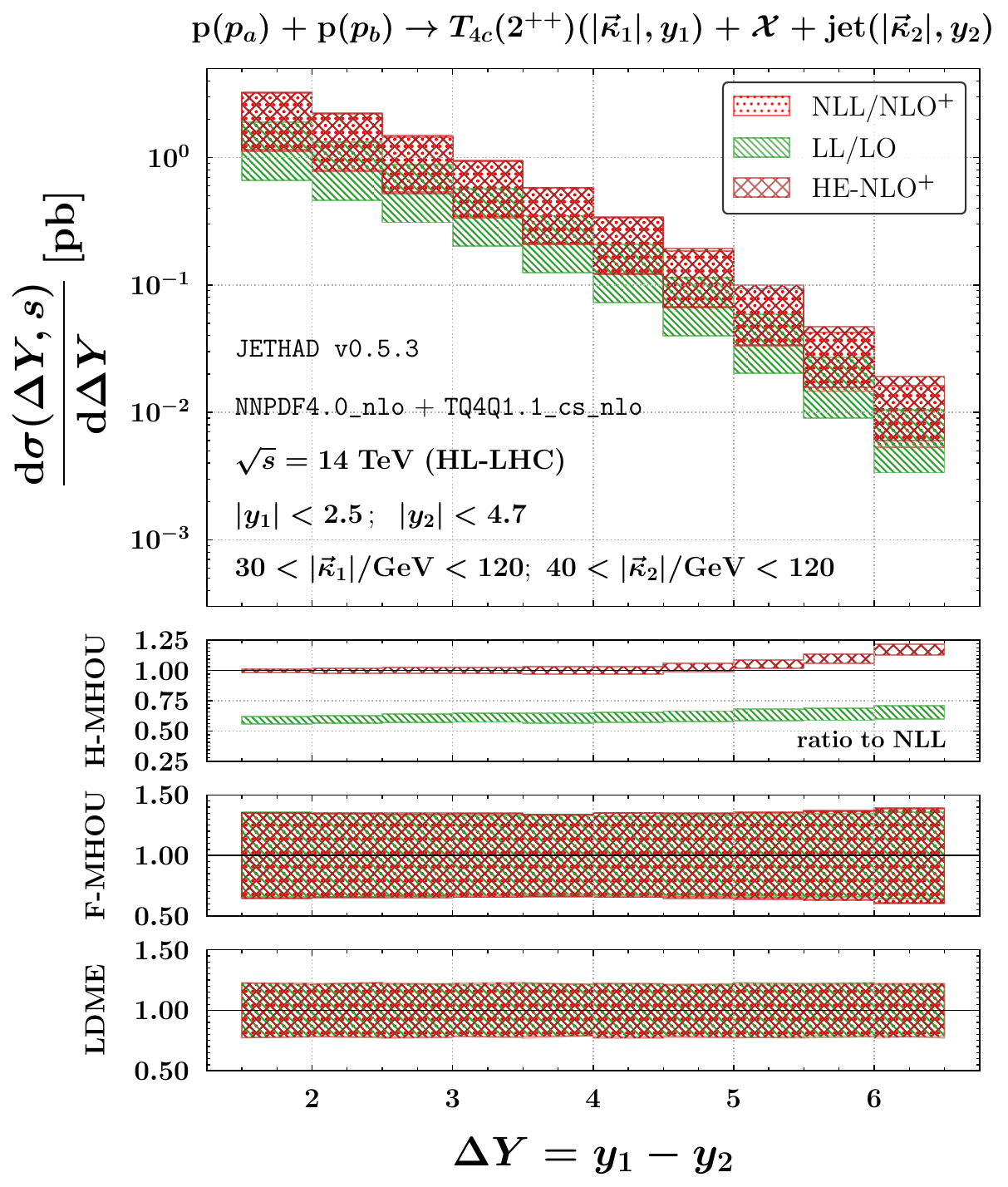}
   \hspace{-0.00cm}
   \includegraphics[scale=0.385,clip]{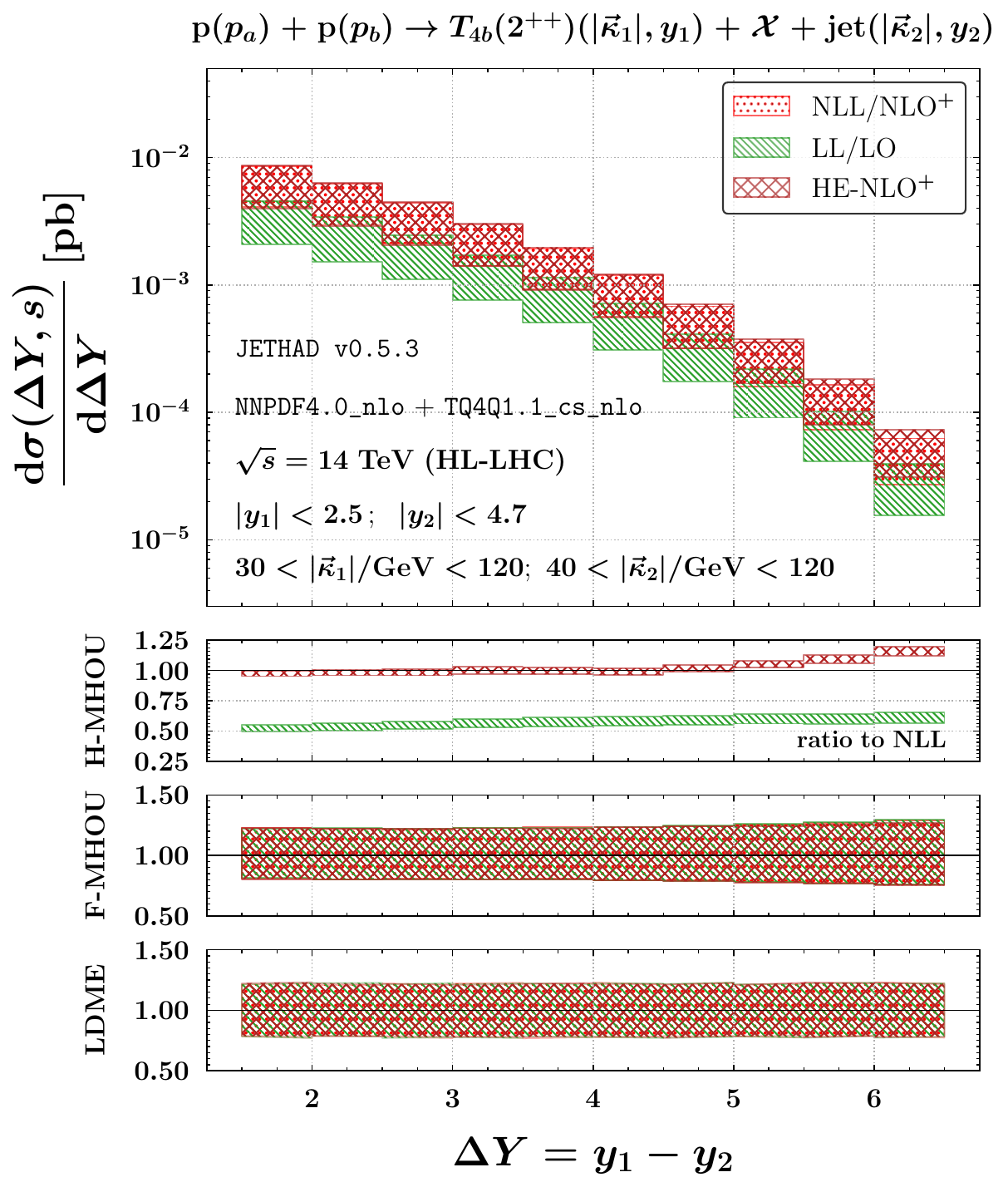}

\caption{
Rapidity spectrum of tensor tetraquarks $\TQcTpp$ (left) and $\TQbTpp$ (right) produced with a jet at $\sqrt{s}=14$ TeV.
Main panels show total uncertainties from combined H-MHOUs, F-MHOUs, LDME, and phase-space effects.
Ancillary panels display: a) $\LL$ and $\HENLOp$ ratios to the $\NLLp$ baseline, b) F-MHOUs envelopes, and c) LDME-induced variations.
}
\label{fig:I_TQ2}
\end{figure}

Predictions for the rapidity rates $\drv \sigma / \drv \DY$ of scalar ($0^{++}$), axial-vector ($1^{+-}$), and tensor ($2^{++}$) tetraquarks with charm and bottom content, produced in association with a light jet at $\sqrt{s}=14$~TeV, are shown in Figs.~\ref{fig:I_TQ0} to~\ref{fig:I_TQ2}.
Each main panel reports the absolute differential cross section, while the lower ones display ratios of $\LL$ and $\HENLOp$ to $\NLLp$ predictions.
Uncertainty bands combine H-MHOUs, F-MHOUs, LDME variations, and phase-space integration errors in quadrature; additional panels isolate the relative contributions of H-MHOUs, F-MHOUs, and LDMEs.
The bin width in $\DY$ is $0.5$ throughout.

Across all spin states, the cross section decreases as $\DY$ grows, reflecting the competition between the BFKL-driven rise of partonic coefficients and the suppression from collinear PDFs and FFs at large momentum fractions.
The falloff is moderate for $\DY \lesssim 3$, then gradually steeper toward larger intervals.
This universal trend, common to both charm and bottom sectors, demonstrates that the resummed component enhances the small-$x$ dynamics without destabilizing the perturbative series.

For scalar tetraquarks ($0^{++}$), we obtain the highest overall rates, from $10^{-2}$ pb to several pb in the charm case (left panel) and $10^{-5}$--$10^{-2}$ pb for bottom (right panel).
These values follow expectations from the compact $S$-wave structure and the absence of spin suppression.
Theoretical uncertainties remain moderate, with H-MHOUs below roughly 50\%.
LL results slightly overshoot $\NLLp$ at small $\DY$, while approaching it at larger intervals, indicating good convergence of the hybrid resummation.
Fixed-order $\HENLOp$ curves reproduce the $\NLLp$ pattern up to moderate $\DY$, confirming that genuine BFKL effects become relevant only at the upper end of the accessible range.

Axial-vector states ($1^{+-}$) exhibit cross sections roughly two orders of magnitude lower, ranging from $10^{-4}$ to $10^{-2}$ pb for charm and $10^{-7}$--$10^{-5}$ pb for bottom.
Their theoretical behavior is nevertheless clean and stable: uncertainty bands are the narrowest among all channels, owing to a single dominant heavy-quark fragmentation source and the strong evolution of the gluon FF, which compensates residual scale variations.
Ratios between $\HENLOp$ and $\NLLp$ remain close to unity, confirming a well-behaved perturbative hierarchy.
Although less abundant, the $1^{+-}$ channel provides the most transparent window on subleading resummation effects.

Tensor tetraquarks ($2^{++}$) feature cross sections comparable to or slightly above the scalar ones, with similarly stable uncertainty bands.
Their $\DY$ dependence mirrors that of the scalar case, displaying a smooth decline and limited scale sensitivity.
The $\HENLOp$ results lie consistently between $\LL$ and $\NLLp$, reinforcing the interpretation of the resummed predictions as a controlled extension of fixed order.

A global comparison across all spin channels shows that $\NLLp$ predictions offer a stable baseline, with resummation effects visible yet moderate.
Spin primarily influences normalization through FF suppression and the $\DY$ slope through the different scale evolution of the FFs.
Among the three, the axial-vector channel best isolates genuine logarithmic corrections, while scalar and tensor configurations provide higher rates and experimental robustness.
F-MHOUs and LDME variations contribute comparably to the total uncertainty, though F-MHOUs dominate slightly in most kinematic regions.
Overall, the results confirm that heavy-tetraquark production via collinear fragmentation remains perturbatively stable and sensitive to resummed high-energy dynamics already at HL-LHC energies.

\begin{figure}[!t]
\centering

   \hspace{0.00cm}
   \includegraphics[scale=0.385,clip]{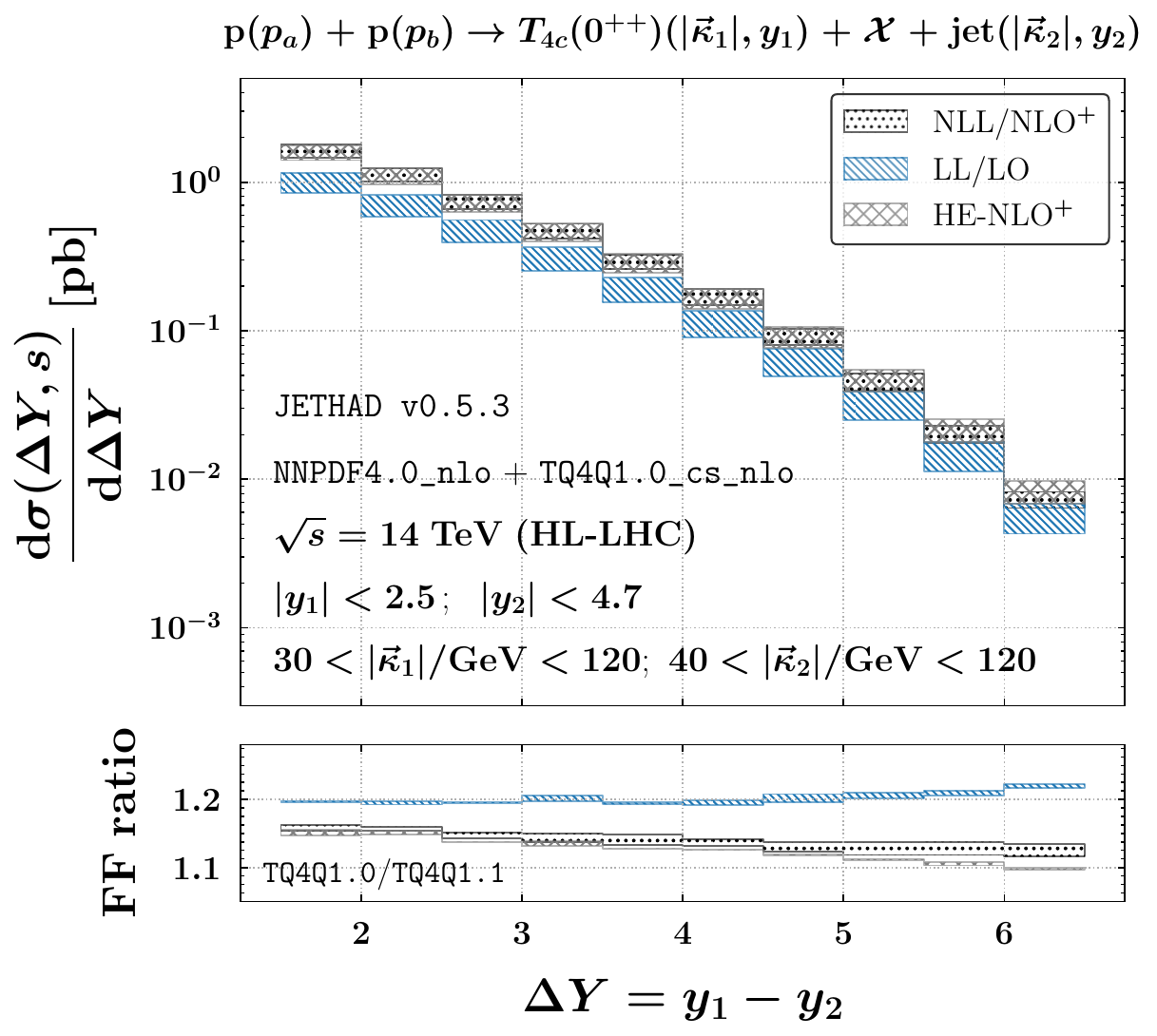}
   \hspace{-0.00cm}
   \includegraphics[scale=0.385,clip]{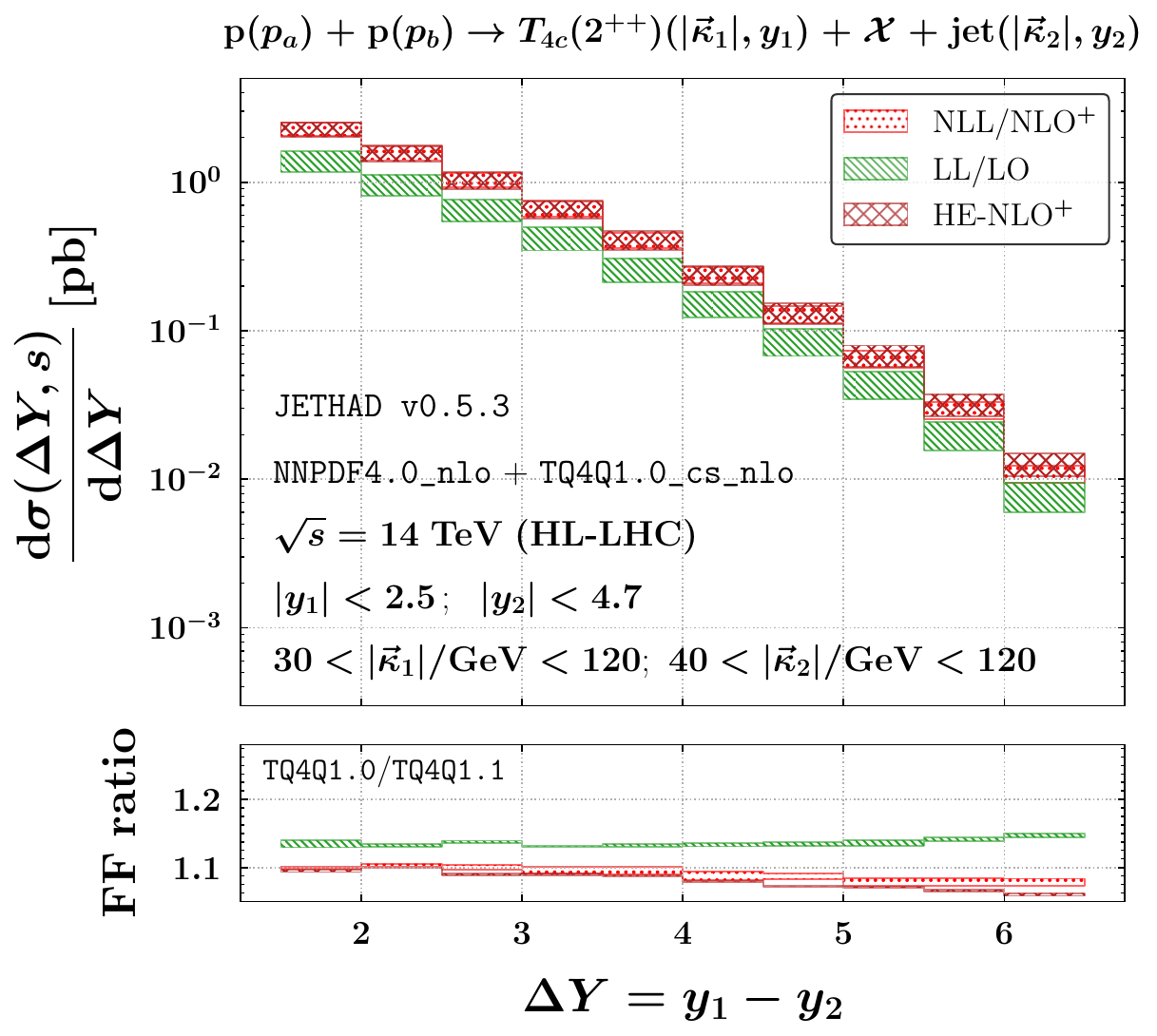}

\caption{
Rapidity spectrum of scalar and tensor tetraquarks, $\TQcZpp$ (left) and $\TQcTpp$ (right),produced with a jet at $\sqrt{s}=14$ TeV. 
Main-panel bands indicate the joint effect of H-MHOUs and phase-space integration.
Ancillary panels show ratios between results derived from the original {\tt TQ4Q1.0} FFs~\protect\cite{Celiberto:2024mab} and the new {\tt 1.1} version.
}
\label{fig:I_FF-comp}
\end{figure}

For comparison, Fig.~\ref{fig:I_FF-comp} displays the $\drv\sigma/\drv\DY$ spectra for scalar (left panel) and tensor (right panels) $\TQQ$ tetraquarks produced in association with a jet at $\sqrt{s}=14$~TeV, obtained using the earlier {\tt TQ4Q1.0} FFs~\cite{Celiberto:2024mab}.
Since the {\tt 1.0} release did not include dedicated uncertainty analyses for fragmentation---neither from scale variations in the short-distance coefficients (F-MHOUs) nor from LDMEs---the bands shown here reflect only phase-space and H-MHOU effects evaluated within the current setup.
Ancillary panels report the ratio between results based on the {\tt 1.0} and {\tt 1.1} functions, providing a direct comparison at the level of physical observables.

The overall pattern of the main panels remains consistent with that found using {\tt TQ4Q1.1}, confirming that the updated functions preserve the shape of the $\DY$ spectra within uncertainties.
However, the ratios indicate a systematic $\sim\!10$--$20\%$ reduction of the cross sections when moving from {\tt 1.0} to {\tt 1.1}, slightly more pronounced for the scalar channel.
This shift originates from the revised NRQCD-based heavy-quark inputs and the inclusion of LDME-driven normalization effects.
The remarkable stability of the ratios across the full $\DY$ range demonstrates that the upgrade to {\tt 1.1} yields a more conservative and theoretically consistent description, reducing model dependence without altering the physical behavior of the distributions.

\subsection{Azimuthal-angle distributions}
\label{ssec:I-phi}

\begin{figure}[!t]
\centering

 \hspace{0.00cm}
 \includegraphics[scale=0.370,clip]{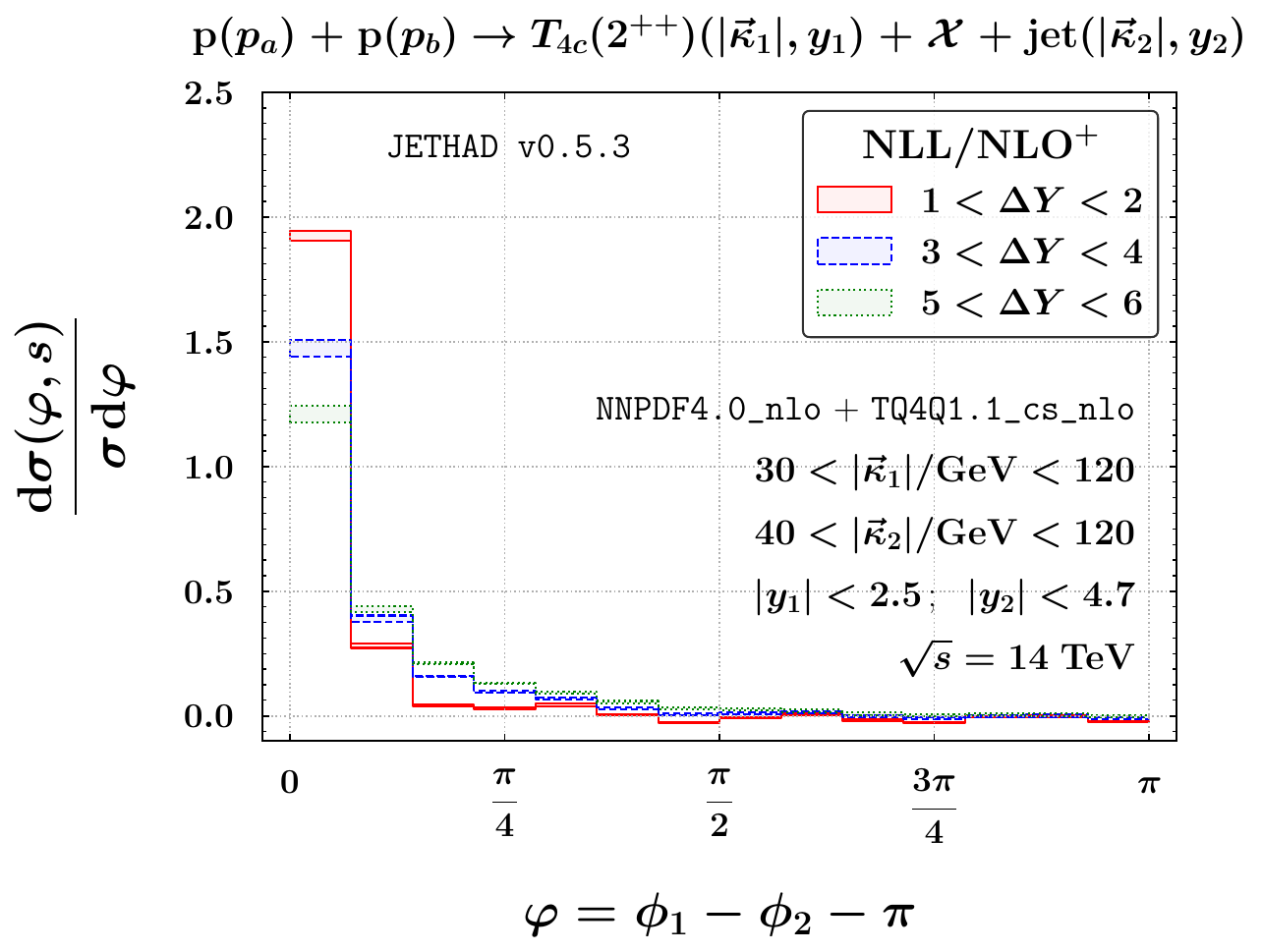}
 \hspace{-0.00cm}
 \includegraphics[scale=0.370,clip]{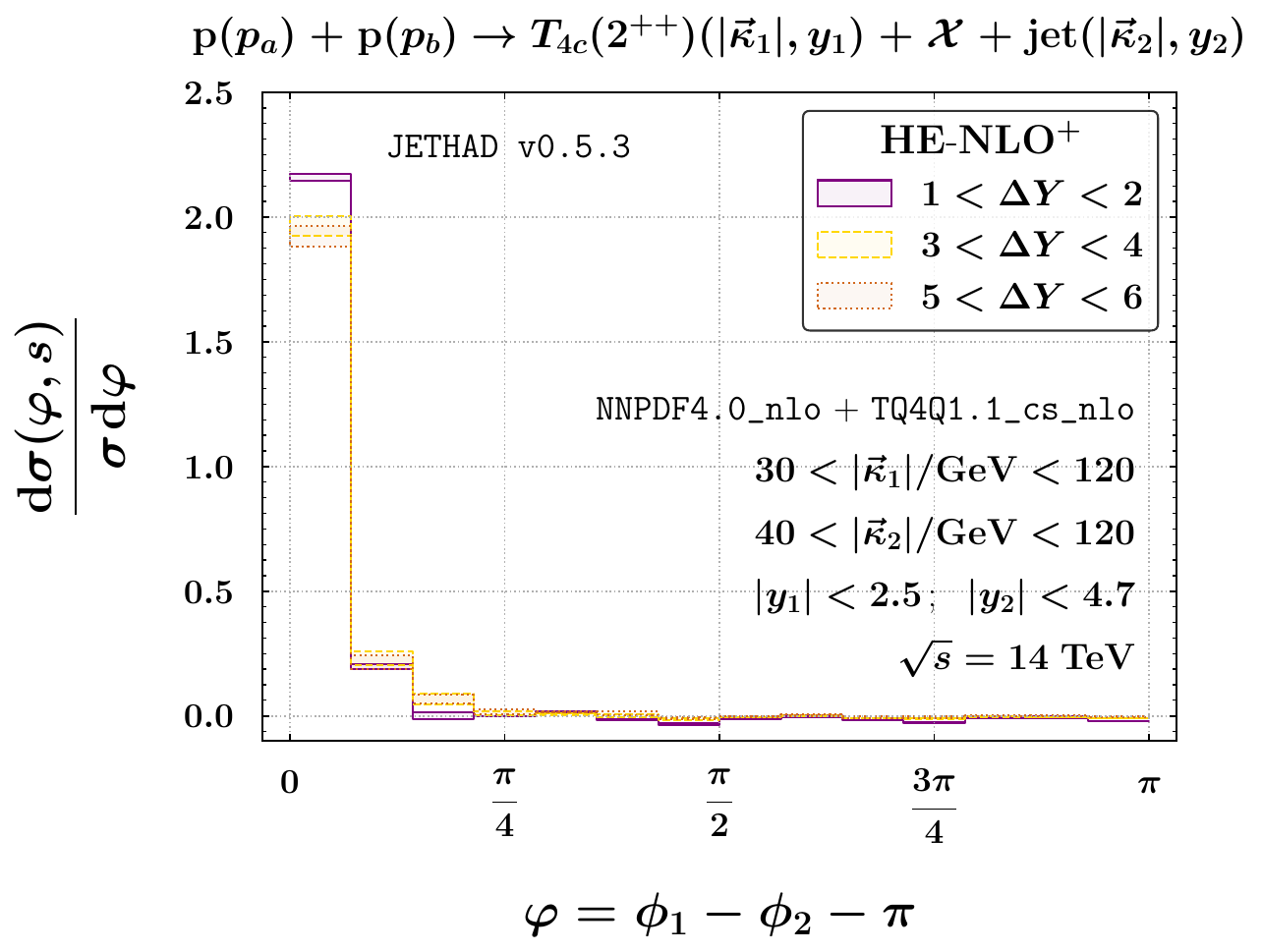}

\caption{
Azimuthal spectra of tensor tetraquarks $\TQcTpp$ produced with a jet at $\sqrt{s}=14$ TeV.
Left (right) panels show $\NLLp$ ($\HENLOp$) predictions obtained in the HyF framework.
Bands denote H-MHOU uncertainties.
}
\label{fig:I-phi_Tc2}
\end{figure}

Figure~\ref{fig:I-phi_Tc2} shows the normalized azimuthal spectra for the tensor tetraquark $\TQcTpp$ produced in association with a jet at $\sqrt{s}=14$~TeV within the HyF framework.
Predictions are presented at $\NLLp$ (left) and $\HENLOp$ (right) accuracy for three representative rapidity intervals, $1 < \Delta Y < 2$, $3 < \Delta Y < 4$, and $5 < \Delta Y < 6$.

Because the azimuthal multiplicities defined in Eq.~\eqref{angular_multiplicity} are symmetric under the transformation $-\varphi \to \varphi$, only the physical half-range $0 < \varphi < \pi$ is shown.
For ease of experimental comparison, all curves are averaged over equal-size $\varphi$ bins.
As these observables are normalized ratios of differential cross sections, most correlated uncertainties cancel out.
In particular, the impact of LDMEs and multidimensional phase-space integration is negligible, while F-MHOUs were numerically verified to have no visible effect.
Consequently, the uncertainty bands displayed in Fig.~\ref{fig:I-phi_Tc2} reflect only the contribution from H-MHOUs.

All spectra exhibit a pronounced peak at $\varphi \simeq 0$, corresponding to nearly back-to-back configurations of the final-state objects.
The peak height decreases as $\Delta Y$ increases, indicating stronger azimuthal decorrelation driven by soft-gluon radiation.
This broadening behavior reflects genuine BFKL-type dynamics, which enhance gluon emissions as the rapidity gap widens.
In contrast, $\HENLOp$ predictions display a narrower and more persistent peak, with only mild flattening at large $\Delta Y$, consistent with a DGLAP-like regime dominated by collinear radiation.
The distinction between the two accuracies therefore captures the classical ``BFKL-DGLAP'' interplay~\cite{Celiberto:2015yba,Celiberto:2015mpa,Celiberto:2020wpk,Celiberto:2022rfj,Celiberto:2022gji} that characterizes energy-flow correlations in semihard systems.

Uncertainty bands remain small over the full $\varphi$ range, confirming the perturbative stability of the HyF framework.
No negative or oscillating bins are observed, even in the large-$\varphi$ region, demonstrating that tensor-tetraquark fragmentation mitigates residual threshold effects.
Overall, this observable provides a robust probe of energy-momentum decorrelation in heavy exotic production and a clean window into high-energy QCD dynamics.

\section{Conslusions}
\label{sec:conclusions}

In this review, we have advanced the study of exotic matter formation in high-energy hadronic collisions by presenting the {\tt TQ4Q1.1} set of collinear FFs for fully heavy tetraquarks~\cite{Celiberto:2025_TQ4Q11}.
The analysis encompasses the three accessible spin-parity channels, $J^{PC} = 0^{++}$, $1^{+-}$, and $2^{++}$, and covers charmed and bottomed configurations.
These FFs are constructed within the leading-power single-parton fragmentation picture embedded in the NRQCD framework, where color-spin Fock-state components enter through their corresponding LDMEs.

SDCs for gluon and heavy-quark fragmentation were computed at the initial scale and evolved via the DGLAP equations in a VFNS using the threshold-matched {\HFNRevo} formalism~\cite{Celiberto:2025euy,Celiberto:2024mex,Celiberto:2024bxu,Celiberto:2024rxa,Celiberto:2025xvy}.
The {\tt 1.1} release introduces a systematic treatment of theoretical uncertainties, combining LDME variations, hard-scale dependencies (H-MHOUs), and fragmentation-scale inputs (F-MHOUs).
By analyzing each source separately and then merging them, we achieved a transparent breakdown of theoretical errors, enabling collider-level predictions with quantified reliability.

The {\psymJethad} multimodular interface~\cite{Celiberto:2020wpk,Celiberto:2022rfj,Celiberto:2023fzz,Celiberto:2024mrq,Celiberto:2024swu,Celiberto:2025_P5Q_review,Celiberto:2025csa} was used to compute hybrid-factorized cross sections for semi-inclusive tetraquark-jet production at $\sqrt{s} = 14$~TeV.
Working at full $\NLLp$ accuracy within the HyF formalism, which consistently merges collinear factorization and BFKL resummation, we analyzed rapidity-interval distributions and azimuthal multiplicities, two observables highly sensitive to the interplay between perturbative and nonperturbative dynamics.

The rapidity distributions display a monotonic decrease with increasing $\DY$ across all spin configurations, accompanied by well-behaved uncertainty bands that signal perturbative stability.
At large rapidity separations, the resummed $\NLLp$ signal becomes clearly distinguishable from the fixed-order background, particularly in scalar and tensor channels.
The axial-vector state, though suppressed in absolute rate, exhibits remarkable robustness and minimal uncertainty spread due to the strong $\mu_F$ evolution of the gluon FF and its reduced LDME sensitivity.
This ``natural stability''~\cite{Celiberto:2022grc} marks the $1^{+-}$ channel as an optimal benchmark for precision tests of high-energy resummation.

Azimuthal multiplicities provide an orthogonal probe of resummation dynamics at fixed $\DY$.
Being normalized ratios of cross sections, they are largely insensitive to nonperturbative modeling and phase-space effects, allowing for clean discrimination between $\NLLp$ and $\HENLOp$ predictions.
Their stability and positivity across the full angular range reinforce the suitability of heavy tetraquark production as a controlled laboratory for BFKL studies.

The {\tt TQ4Q1.1} functions thus acquire a dual role: precision tools for testing high-energy QCD through infrared-safe observables and diagnostic instruments for studying the color-spin composition of exotic bound states.
This synergy between structural and dynamical information establishes a unified framework for connecting exotic spectroscopy with the resummation frontier of QCD.

Looking forward, the same methodology may be validated using more accessible systems such as double $\Jpsi$ or $\Jpsi$-jet production, where forthcoming data could constrain the shape and normalization of heavy-quarkonium FFs and test the HyF resummation strategy.
Recent measurements by LHCb~\cite{LHCb:2023ybt} already provide a first window into this kinematic regime, and future analyses at higher transverse momenta may further consolidate the connection between tetraquark phenomenology and quarkonium production.

Further progress will focus on refining theoretical uncertainty quantification, in line with ongoing efforts to propagate model and scale variations through replica-based statistical approaches~\cite{Kassabov:2022orn,Harland-Lang:2018bxd,Ball:2021icz,McGowan:2022nag,NNPDF:2024dpb,Pasquini:2023aaf}.
Extending such strategies to fragmentation dynamics, particularly for F-MHOUs and LDMEs, will be essential for precision-era studies of exotic hadrons.
Parallel developments will aim at incorporating color-octet contributions into the NRQCD expansion, expected to modify both normalization and shape of the FFs, especially in scalar and tensor channels.

Another compelling direction concerns linking exotic-tetraquark structure to precision resummation and enabling data-driven benchmarking of FFs via global-fit and machine-learning analyses~\cite{Nocera:2017qgb,Bertone:2017xsf,Bertone:2017tyb,Bertone:2018ecm,Khalek:2021gxf,Khalek:2022vgy,Soleymaninia:2022qjf,Soleymaninia:2022alt}.

In parallel, new theoretical integrations will pursue the combination of high-energy and soft-gluon~\cite{Hatta:2020bgy,Hatta:2021jcd,Caucal:2022ulg,Taels:2022tza} resummations, jet-radius logarithms~\cite{Dasgupta:2014yra,Dasgupta:2016bnd,Banfi:2012jm,Banfi:2015pju,Liu:2017pbb}, and jet-angularity methods~\cite{Luisoni:2015xha,Caletti:2021oor,Reichelt:2021svh}.
Synergies with small-$x$ saturation physics~\cite{Gelis:2010nm,Kovchegov:2012mbw,Chirilli:2012jd,Boussarie:2014lxa,Benic:2016uku,Benic:2018hvb,Roy:2019hwr,Roy:2019cux,Beuf:2020dxl,Iancu:2021rup,Iancu:2023lel,vanHameren:2023oiq,Wallon:2023asa,Agostini:2024xqs,Altinoluk:2024zom,Altinoluk:2025dwd,Altinoluk:2025tms} and with angular-asymmetry studies in dijet and heavy-hadron production~\cite{Caucal:2021ent,Caucal:2022ulg,Taels:2022tza,Kotko:2015ura,vanHameren:2016ftb,Altinoluk:2020qet,Altinoluk:2021ygv,Boussarie:2021ybe,Caucal:2023nci,Cheung:2024qvw,Caucal:2025mth,Kang:2013hta,Ma:2014mri,Ma:2015sia,Ma:2018qvc,Stebel:2021bbn,Mantysaari:2021ryb,Mantysaari:2022kdm} will further expand the scope of hybrid approaches to heavy-flavor fragmentation.

From a phenomenological standpoint, the suppressed but clean environment of fully bottomed tetraquarks offers an ideal testing ground for the limits of fragmentation-based predictions.
The forthcoming high-luminosity LHC data will be pivotal in establishing whether their production can be experimentally accessed and, in turn, used to constrain FFs and the underlying color dynamics.
Meanwhile, the interplay between exotic hadron production, intrinsic heavy-quark components~\cite{Brodsky:1980pb,Brodsky:2015fna,Jimenez-Delgado:2014zga,Ball:2016neh,Hou:2017khm,Ball:2022qks,Guzzi:2022rca,NNPDF:2023tyk}, and mass-dependent QCD effects such as the \emph{dead-cone} phenomenon~\cite{Dokshitzer:1991fd,ALICE:2021aqk} will continue to shape our understanding of heavy-flavor dynamics in jets.

In summary, the {\tt TQ4Q1.1} release provides a coherent and uncertainty-quantified foundation for the collinear fragmentation of fully heavy tetraquarks.
It bridges the modeling of exotic bound states with the high-energy regime of QCD, connecting hadron-structure studies, precision resummation, and collider phenomenology.
This review thus sets the stage for an extended exploration of exotic-matter production, from tetraquarks to triply heavy baryons, within a unified, evolution-consistent framework that will accompany the forthcoming high-luminosity era of hadron physics~\cite{Chapon:2020heu,LHCspin:2025lvj,Anchordoqui:2021ghd,Feng:2022inv,Hentschinski:2022xnd,Accardi:2012qut,AbdulKhalek:2021gbh,Khalek:2022bzd,Acosta:2022ejc,AlexanderAryshev:2022pkx,LinearCollider:2025lya,LinearColliderVision:2025hlt,Brunner:2022usy,Arbuzov:2020cqg,Abazov:2021hku,Bernardi:2022hny,Amoroso:2022eow,Celiberto:2018hdy,Klein:2020nvu,2064676,MuonCollider:2022xlm,Aime:2022flm,MuonCollider:2022ded,Black:2022cth,Accettura:2023ked,InternationalMuonCollider:2024jyv,MuCoL:2024oxj,MuCoL:2025quu,InternationalMuonCollider:2025sys,Vignaroli:2023rxr,Dawson:2022zbb,Bose:2022obr,Begel:2022kwp,Abir:2023fpo,Accardi:2023chb,Gessner:2025acq,Altmann:2025feg}.

\section*{Funding}
\label{sec:funding}
\addcontentsline{toc}{section}{\nameref{sec:funding}}

This research was funded by Comunidad de Madrid, grant number 2022-T1/TIC-24176.

\section*{Data availability}
\label{sec:data_availability}
\addcontentsline{toc}{section}{\nameref{sec:data_availability}}

The original contributions presented in the study are included in the review. 
Further inquiries can be directed to the author.

\section*{Conflict of Interests}
\label{sec:coi}
\addcontentsline{toc}{section}{\nameref{sec:coi}}

The author declares no conflict of interest.



\printacronyms

\bibliographystyle{elsarticle-num}

\bibliography{references}

\end{document}